\documentclass[a4paper,11pt]{article}
\usepackage{jheppub} 
 \usepackage{graphics}
 \usepackage{float}
 \usepackage{tikz-cd} 
\usepackage{url}
\usepackage{float}
\usepackage{standalone}
\usepackage{bm}

 \usepackage{caption, subcaption}
 
 \usepackage{tikz}

\def\a{\alpha}
\def\b{\beta}
\def\c{\varepsilon}
\def\d{\delta}
\def\e{\epsilon}
\def\f{\phi}
\def\g{\gamma}
\def\h{\theta}
\def\k{\kappa}
\def\l{\lambda}
\def\m{\mu}
\def\n{\nu}
\def\p{\psi}
\def\q{\partial}
\def\r{\rho}
\def\s{\sigma}
\def\t{\tau}
\def\u{\upsilon}
\def\v{\varphi}
\def\w{\omega}
\def\x{\xi}
\def\y{\eta}
\def\z{\zeta}
\def\D{\Delta}
\def\G{\Gamma}
\def\H{\Theta}
\def\L{\zeta}
\def\F{\Phi}
\def\P{\Psi}
\def\S{\Sigma}

\def\aa{{\dot \a}}
\def\bb{{\dot \b}}
\def\ss{{\bar \s}}
\def\hh{{\bar \h}}
\def\CA{{\cal A}}
\def\CB{{\cal B}}
\def\CC{{\cal C}}
\def\CD{{\cal D}}
\def\CE{{\cal E}}
\def\CG{{\cal G}}
\def\CH{{\cal H}}
\def\CI{{\cal I}}
\def\CK{{\cal K}}
\def\CL{{\cal L}}
\def\CR{{\cal R}}
\def\CM{{\cal M}}
\def\CN{{\cal N}}
\def\CO{{\cal O}}
\def\CP{{\cal P}}
\def\CQ{{\cal Q}}
\def\CW{{\cal W}}

\def\capcup{\phantom{.}^{\frown}_{\smile}\phantom{.}}

\DeclareMathOperator{\Tr}{Tr}
\newcommand{\Slash}[1]{{\ooalign{\hfil/\hfil\crcr$#1$}}}

\def\o{\over}
\newcommand{\gsim}{ \mathop{}_{\textstyle \sim}^{\textstyle >} }
\newcommand{\lsim}{ \mathop{}_{\textstyle \sim}^{\textstyle <} }
\newcommand{\vev}[1]{ \left\langle {#1} \right\rangle }
\newcommand{\bra}[1]{ \langle {#1} | }
\newcommand{\ket}[1]{ | {#1} \rangle }
\newcommand{\EV}{ {\rm eV} }
\newcommand{\KEV}{ {\rm keV} }
\newcommand{\MEV}{ {\rm MeV} }
\newcommand{\GEV}{ {\rm GeV} }
\newcommand{\TEV}{ {\rm TeV} }
\def\diag{\mathop{\rm diag}\nolimits}
\def\Spin{\mathop{\rm Spin}}
\def\SO{\mathop{\rm SO}}
\def\O{\mathop{\rm O}}
\def\SU{\mathop{\rm SU}}
\def\U{\mathrm{U}}
\def\Sp{\mathop{\rm Sp}}
\def\SL{\mathop{\rm SL}}
\def\tr{\mathop{\rm tr}}
\def\rank{\mathop{\rm rank}}

\def\beq#1\eeq{\begin{align}#1\end{align}}

\title{On rank two theories with eight supercharges part II: Lefschetz pencils}

\author[]{Dan Xie}

\affiliation[]{Department of Mathematics, Tsinghua University, Beijing, 100084, China}

\abstract{The global Seiberg-Witten (SW) geometries for rank two theories with eight supercharges are studied. The theory is deformed generically so that there are only 
simplest $I_1$ or $\tilde{I}_1$ singularities on the Coulomb branch, which  geometrically gives the so-called Lefschetz pencils. The local singularity was shown to be determined 
by the conjugacy class of mapping class group (MCG);  The global study is then reduced to the questions about MCG: a) Find the factorization 
of the MCG element of the singular fiber into positive products of Dehn twists (which gives the $I_1$ singularity or $\tilde{I}_1$ singularity); 
b) Find the factorization of identity element in terms of Dehn twists. We solved above two MCG problems for most rank two theories.The results are very helpful in determining IR physics for all vacua of  4d SCFTs. 
Our approach is  combinatorial and many aspects can be straightforwardly generalized to the study of higher rank theory.  }

\begin{document} 
\maketitle
\flushbottom
%%%%%%%%%%%%%%%%%%%%%%%%%%%%%%%%%%%%%%%%%%%%%%%%%%%%%%%%%%%

\section{Introduction}
This is the second one of a series of papers in trying to classify rank two theories with eight supercharges.  
The basic idea is  to classify consistent Seiberg-Witten (SW) solution on the effective 4d Coulomb branch \footnote{See results in classifying 4d rank one SCFTs \cite{Argyres:2015ffa} and rank two SCFTs in \cite{Argyres:2022lah,Argyres:2022puv,Argyres:2022fwy}.}. In previous paper \cite{Xie:2022aad}, the local 
 of local singularities of SW solution are classified and the corresponding IR physical theories are identified by using the associated 
dual graph. 

The purpose of this paper is to study the global SW geometry for rank two theories, namely glue the local singularities 
consistently. There are a couple of  topological constraints: a) the product of local monodromy should be identity; b) there
are simple constraints on the sum of local invariants (see discussion in \cite{Xie:2022aad}) by assuming the geometry of the total space of the SW fiberation. However, the above constraints 
are  not very strong and a systematical study seems quite difficult. 

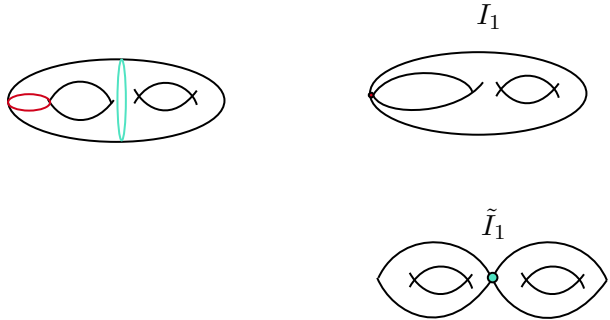
\begin{figure}
\begin{center}

\tikzset{every picture/.style={line width=0.75pt}} %set default line width to 0.75pt        

\begin{tikzpicture}[x=0.45pt,y=0.45pt,yscale=-1,xscale=1]
%uncomment if require: \path (0,964); %set diagram left start at 0, and has height of 964

%Shape: Ellipse [id:dp2732109334277837] 
\draw   (67,130.61) .. controls (67,111.5) and (107.29,96) .. (157,96) .. controls (206.71,96) and (247,111.5) .. (247,130.61) .. controls (247,149.72) and (206.71,165.22) .. (157,165.22) .. controls (107.29,165.22) and (67,149.72) .. (67,130.61) -- cycle ;
%Curve Lines [id:da1050906401575642] 
\draw    (101,132) .. controls (117,103.22) and (148,115.22) .. (153,133.22) ;
%Curve Lines [id:da3122526712301241] 
\draw    (100,129.22) .. controls (110,147.44) and (135,157.22) .. (155,130.22) ;
%Curve Lines [id:da8178070484931126] 
\draw    (172,133) .. controls (188,104.22) and (219,116.22) .. (224,134.22) ;
%Curve Lines [id:da31112229370476707] 
\draw    (172,121) .. controls (182,139.22) and (204,149.22) .. (224,122.22) ;
%Shape: Ellipse [id:dp7437130770712097] 
\draw   (368,124.61) .. controls (368,105.5) and (408.29,90) .. (458,90) .. controls (507.71,90) and (548,105.5) .. (548,124.61) .. controls (548,143.72) and (507.71,159.22) .. (458,159.22) .. controls (408.29,159.22) and (368,143.72) .. (368,124.61) -- cycle ;
%Curve Lines [id:da6017856742369068] 
\draw    (370,127) .. controls (386,98.22) and (448,105.22) .. (453,123.22) ;
%Curve Lines [id:da7648314124211668] 
\draw    (371,127) .. controls (381,145.22) and (442,142.22) .. (462,115.22) ;
%Curve Lines [id:da002621453922512007] 
\draw    (473,127) .. controls (489,98.22) and (520,110.22) .. (525,128.22) ;
%Curve Lines [id:da575627483536681] 
\draw    (473,115) .. controls (483,133.22) and (505,143.22) .. (525,116.22) ;
%Shape: Ellipse [id:dp09114965883187987] 
\draw  [color={rgb, 255:red, 208; green, 2; blue, 27 }  ,draw opacity=1 ] (67,132.11) .. controls (67,128.3) and (74.84,125.22) .. (84.5,125.22) .. controls (94.16,125.22) and (102,128.3) .. (102,132.11) .. controls (102,135.91) and (94.16,139) .. (84.5,139) .. controls (74.84,139) and (67,135.91) .. (67,132.11) -- cycle ;
%Shape: Ellipse [id:dp0788033866716038] 
\draw  [color={rgb, 255:red, 80; green, 227; blue, 194 }  ,draw opacity=1 ] (158,130.11) .. controls (158,111.39) and (159.57,96.22) .. (161.5,96.22) .. controls (163.43,96.22) and (165,111.39) .. (165,130.11) .. controls (165,148.83) and (163.43,164) .. (161.5,164) .. controls (159.57,164) and (158,148.83) .. (158,130.11) -- cycle ;
%Shape: Circle [id:dp7540616795101582] 
\draw  [fill={rgb, 255:red, 208; green, 2; blue, 27 }  ,fill opacity=1 ] (367,126) .. controls (367,124.9) and (367.9,124) .. (369,124) .. controls (370.1,124) and (371,124.9) .. (371,126) .. controls (371,127.1) and (370.1,128) .. (369,128) .. controls (367.9,128) and (367,127.1) .. (367,126) -- cycle ;
%Curve Lines [id:da24071301045486848] 
\draw    (375,279.22) .. controls (395,236.22) and (454,242.22) .. (470,281.22) ;
%Curve Lines [id:da721016691723871] 
\draw    (374,279.22) .. controls (387,310.22) and (432,334.22) .. (469,281.22) ;
%Curve Lines [id:da3483290540514863] 
\draw    (470,279.22) .. controls (490,236.22) and (549,242.22) .. (565,281.22) ;
%Curve Lines [id:da08554566318750068] 
\draw    (470,280.22) .. controls (483,311.22) and (528,334.22) .. (565,281.22) ;
%Curve Lines [id:da11405619391880739] 
\draw    (494,287) .. controls (510,258.22) and (541,270.22) .. (546,288.22) ;
%Curve Lines [id:da42327012996534985] 
\draw    (494,275) .. controls (504,293.22) and (526,303.22) .. (546,276.22) ;
%Curve Lines [id:da6806654236547269] 
\draw    (401,287) .. controls (417,258.22) and (448,270.22) .. (453,288.22) ;
%Curve Lines [id:da12733072762237563] 
\draw    (401,275) .. controls (411,293.22) and (433,303.22) .. (453,276.22) ;
%Shape: Circle [id:dp7001411219510569] 
\draw  [fill={rgb, 255:red, 80; green, 227; blue, 194 }  ,fill opacity=1 ] (466,279) .. controls (466,276.79) and (467.79,275) .. (470,275) .. controls (472.21,275) and (474,276.79) .. (474,279) .. controls (474,281.21) and (472.21,283) .. (470,283) .. controls (467.79,283) and (466,281.21) .. (466,279) -- cycle ;

\draw (454,47.4) node [anchor=north west][inner sep=0.75pt]    {$I_{1}$};
% Text Node
\draw (458,218.4) node [anchor=north west][inner sep=0.75pt]    {$\tilde{I}_{1}{}$};

\end{tikzpicture}
\end{center}
\caption{The simplest local degenerations. $I_1$ singularity: there is a non-separating vanishing cycle; $\tilde{I}_1$ singularity: there is a separating vanishing cycle: the Riemann surface is split into two parts after the degeneration.}
\label{intro1}
\end{figure}

It is certainly very helpful if one can write down the full SW families which encode the prescribed local singularities (See \cite{Argyres:2022lah}), which we will pursue elsewhere. 
We do not take that approach in this paper. Instead  a topological approach is taken so that the construction of global SW geometry can be carried out 
in a combinatorial way. The following two topological facts are important for us:
\begin{enumerate}

\item We will use  Mastumoto-Montesinos's (MM) theory \cite{matsumoto2011pseudo} for local singularity: the local degeneration is given by the conjugacy class of mapping class group.
 So there is a mapping class group element around each degeneration, which  completely characterizes the local singularity.

\item We consider the generic deformation of the theory so that
only the simplest possible singularity exists at the bulk: they are called $I_1$ or $\tilde{I_1}$ singularities, see figure. \ref{intro1} for the geometric illustration.  
The physical interpretation for them is that there is an extra massless particle associated with the vanishing cycle \cite{Seiberg:1994rs}.
One also need to add a singularity at $\infty$ which is also assumed to be split into $I_1$ and $\tilde{I_1}$ singularities. 
Therefore, one has a genus two SW fiberation with just $I_1$ or $\tilde{I}_1$ singularities, see figure. \ref{intro2}, and such fiberation is called Lefschetz pencils \cite{matsumoto1996lefschetz}.

\end{enumerate}

Now let's fixed a generic point at Coulomb branch, and one has an element in mapping class group by following a path around each singularity, see figure. \ref{intro2}.
The mapping class group element along $I_1$ or $\tilde{I}_1$ singularities is rather simple: it is given by the so-called Dehn twist along 
the vanishing cycle \cite{farb2011primer}.

 Once we have the  topological picture of global SW geometry shown in figure. \ref{intro2}, the classification is achieved by solving following  problems in genus two \textbf{mapping class group} $M_2$, which are generated by five Dehn twists, see figure. \ref{genustwodehn}:
\begin{enumerate}
\item The singular fiber at $\infty$ determines the UV theory, and so there is an associated mapping class group element.
Since it is now split into $I_1$ and $\tilde{I}_1$ singular fibers, which means that the corresponding mapping class group is given by 
the product of positive Dehn twists.  So the group theory question is  to find out the positive \textbf{factorization} of the mapping class group element associated with the UV theory, 
and this has been solved for  4d SCFTs, see table. [\ref{fac1},\ref{fac2},\ref{fac3},\ref{fac4}]. The candidate factorization for 4d asymptotical free theories, 5d and 6d KK theories  are also found.
\item  For the global SW geometry, since the base space is now compact, the product of Dehn twists should satisfy the topological condition 
\begin{equation*}
\tau_{i_1}\tau_{i_2}\ldots  \tau_{i_s}=1.
\end{equation*}
This amounts to find the \textbf{factorization} of the identity element in terms of Dehn twist.
Another topological constraint is the assumption of the total space being a rational surface \cite{Xie:2022aad}, which put the constraint on the number of $I_1$ and $\tilde{I}_1$ singularities \footnote{The choice of $(I_1, \tilde{I}_1)=(n,m)$ singularities are  $n+2m=20$, and so
the choices are $(n,m)=(20,0), (18, 1)$, $(16,2)$, etc.} on the Coulomb branch. Such factorization of identity element was found in this paper. 
\item The final step of finding global SW geometry is then to rearrange the factorization of identity so that one can get a desirable UV configuration at infinity (see step one), 
and this problem has been solved for most of 4d UV complete theories in this paper, see table. \ref{gl1} and \ref{gl2}.
\end{enumerate}
The factorization of mapping class group element associated with 4d SCFT is extremely useful physically: One can use the braid move and Hurwitz move 
to get configuration involving more complicated singularities, which would then determine \textbf{all} the IR configuration of the theory, so that one can solve 
this theory completely.

This paper is organized as follows: section two revisited the local singularities by using the classification of pseudo-periodic map; section three discusses 
the factorization of mapping class group elements in terms of Dehn twists, and the factorization for local singularity and global SW geometry for 4d SCFTs are given; section four gives 
several representation of mapping class group of genus two curve which would be useful for the further study, such as the UV singular fiber of 5d and 6d KK theory; finally a conclusion is given in section five.

\begin{figure}
\begin{center}

\tikzset{every picture/.style={line width=0.75pt}} %set default line width to 0.75pt        

\begin{tikzpicture}[x=0.45pt,y=0.45pt,yscale=-1,xscale=1]
%uncomment if require: \path (0,964); %set diagram left start at 0, and has height of 964

%Shape: Ellipse [id:dp6171997725805927] 
\draw   (140,177.61) .. controls (140,138.61) and (212.75,107) .. (302.5,107) .. controls (392.25,107) and (465,138.61) .. (465,177.61) .. controls (465,216.61) and (392.25,248.22) .. (302.5,248.22) .. controls (212.75,248.22) and (140,216.61) .. (140,177.61) -- cycle ;
%Straight Lines [id:da5776863552707197] 
\draw    (192,169.22) -- (202,177) ;
%Straight Lines [id:da3997780446044896] 
\draw    (193,177.22) -- (200,167.22) ;

%Straight Lines [id:da3769586936387894] 
\draw    (299,148.22) -- (309,156) ;
%Straight Lines [id:da038522714695941485] 
\draw    (300,156.22) -- (307,146.22) ;

%Straight Lines [id:da29380375222902533] 
\draw    (270,149.22) -- (280,157) ;
%Straight Lines [id:da7018154347255976] 
\draw    (271,157.22) -- (278,147.22) ;

%Straight Lines [id:da5395270805049477] 
\draw    (331,148.22) -- (341,156) ;
%Straight Lines [id:da7539267191665033] 
\draw    (332,156.22) -- (339,146.22) ;

%Straight Lines [id:da2869675831381453] 
\draw    (364,150.22) -- (374,158) ;
%Straight Lines [id:da08290309754777381] 
\draw    (365,158.22) -- (372,148.22) ;

%Straight Lines [id:da011067561401359116] 
\draw    (211,188.22) -- (221,196) ;
%Straight Lines [id:da44029259771240725] 
\draw    (212,196.22) -- (219,186.22) ;

%Straight Lines [id:da6213083236893719] 
\draw    (214,168.22) -- (224,176) ;
%Straight Lines [id:da2827049447932155] 
\draw    (215,176.22) -- (222,166.22) ;

%Straight Lines [id:da109769811774318] 
\draw    (191,185.22) -- (201,193) ;
%Straight Lines [id:da7071423425636538] 
\draw    (192,193.22) -- (199,183.22) ;

%Straight Lines [id:da8162010295422883] 
\draw [color={rgb, 255:red, 80; green, 227; blue, 194 }  ,draw opacity=1 ][fill={rgb, 255:red, 80; green, 227; blue, 194 }  ,fill opacity=1 ]   (296,184.22) -- (306,192) ;
%Straight Lines [id:da08073245390896266] 
\draw [color={rgb, 255:red, 80; green, 227; blue, 194 }  ,draw opacity=1 ][fill={rgb, 255:red, 80; green, 227; blue, 194 }  ,fill opacity=1 ]   (297,192.22) -- (304,182.22) ;

%Straight Lines [id:da39192146646753967] 
\draw [color={rgb, 255:red, 80; green, 227; blue, 194 }  ,draw opacity=1 ]   (332,181.22) -- (342,189) ;
%Straight Lines [id:da9383815407567846] 
\draw [color={rgb, 255:red, 80; green, 227; blue, 194 }  ,draw opacity=1 ]   (333,189.22) -- (340,179.22) ;

%Straight Lines [id:da18406813444927406] 
\draw    (52,292) -- (579,290.22) ;
%Shape: Ellipse [id:dp1718989642297417] 
\draw   (144,448.61) .. controls (144,409.61) and (216.75,378) .. (306.5,378) .. controls (396.25,378) and (469,409.61) .. (469,448.61) .. controls (469,487.61) and (396.25,519.22) .. (306.5,519.22) .. controls (216.75,519.22) and (144,487.61) .. (144,448.61) -- cycle ;
%Straight Lines [id:da623996166928884] 
\draw    (272,434.22) -- (282,442) ;
%Straight Lines [id:da5037522344839425] 
\draw    (273,442.22) -- (280,432.22) ;

%Straight Lines [id:da6598428461047732] 
\draw    (225,433.22) -- (235,441) ;
%Straight Lines [id:da9834204322943101] 
\draw    (226,441.22) -- (233,431.22) ;

%Straight Lines [id:da56094761557551] 
\draw    (351,433.22) -- (361,441) ;
%Straight Lines [id:da23994724521135868] 
\draw    (352,441.22) -- (359,431.22) ;

%Straight Lines [id:da4341855934745078] 
\draw    (404,437.22) -- (414,445) ;
%Straight Lines [id:da6367987544640259] 
\draw    (405,445.22) -- (412,435.22) ;

%Straight Lines [id:da6683200226147978] 
\draw [color={rgb, 255:red, 80; green, 227; blue, 194 }  ,draw opacity=1 ][fill={rgb, 255:red, 80; green, 227; blue, 194 }  ,fill opacity=1 ]   (310,441.22) -- (320,449) ;
%Straight Lines [id:da0304867445755157] 
\draw [color={rgb, 255:red, 80; green, 227; blue, 194 }  ,draw opacity=1 ][fill={rgb, 255:red, 80; green, 227; blue, 194 }  ,fill opacity=1 ]   (311,449.22) -- (318,439.22) ;

%Flowchart: Connector [id:dp48175919249521293] 
\draw  [fill={rgb, 255:red, 0; green, 0; blue, 0 }  ,fill opacity=1 ] (283,496.22) .. controls (283,495.11) and (283.9,494.22) .. (285,494.22) .. controls (286.1,494.22) and (287,495.11) .. (287,496.22) .. controls (287,497.32) and (286.1,498.22) .. (285,498.22) .. controls (283.9,498.22) and (283,497.32) .. (283,496.22) -- cycle ;
%Shape: Circle [id:dp15391652045332704] 
\draw   (218,438) .. controls (218,431.37) and (223.37,426) .. (230,426) .. controls (236.63,426) and (242,431.37) .. (242,438) .. controls (242,444.63) and (236.63,450) .. (230,450) .. controls (223.37,450) and (218,444.63) .. (218,438) -- cycle ;
%Shape: Circle [id:dp20697030443350384] 
\draw   (263,438) .. controls (263,431.37) and (268.37,426) .. (275,426) .. controls (281.63,426) and (287,431.37) .. (287,438) .. controls (287,444.63) and (281.63,450) .. (275,450) .. controls (268.37,450) and (263,444.63) .. (263,438) -- cycle ;
%Shape: Circle [id:dp7184015025981345] 
\draw   (301,444) .. controls (301,437.37) and (306.37,432) .. (313,432) .. controls (319.63,432) and (325,437.37) .. (325,444) .. controls (325,450.63) and (319.63,456) .. (313,456) .. controls (306.37,456) and (301,450.63) .. (301,444) -- cycle ;
%Shape: Circle [id:dp49402915042515083] 
\draw   (343,437) .. controls (343,430.37) and (348.37,425) .. (355,425) .. controls (361.63,425) and (367,430.37) .. (367,437) .. controls (367,443.63) and (361.63,449) .. (355,449) .. controls (348.37,449) and (343,443.63) .. (343,437) -- cycle ;
%Shape: Circle [id:dp2960735307080824] 
\draw   (396,440) .. controls (396,433.37) and (401.37,428) .. (408,428) .. controls (414.63,428) and (420,433.37) .. (420,440) .. controls (420,446.63) and (414.63,452) .. (408,452) .. controls (401.37,452) and (396,446.63) .. (396,440) -- cycle ;
%Curve Lines [id:da19730259629398406] 
\draw    (236,449) .. controls (244,460.22) and (241,498.22) .. (283,497.22) ;
%Curve Lines [id:da860721462238343] 
\draw    (276,450) .. controls (284,461.22) and (273,478.22) .. (286,497.22) ;
%Curve Lines [id:da4370489928886425] 
\draw    (407,453) .. controls (409,463.22) and (305,496.22) .. (287,496.22) ;
%Curve Lines [id:da7257895562242838] 
\draw    (314,456) .. controls (322,467.22) and (296,479.22) .. (285,496.22) ;
%Curve Lines [id:da9161431948525873] 
\draw    (352,448) .. controls (360,459.22) and (309,477.22) .. (287,496.22) ;

\draw (187,130.4) node [anchor=north west][inner sep=0.75pt]    {$\infty $};
% Text Node
\draw (213,400.4) node [anchor=north west][inner sep=0.75pt]    {$\tau _{1}$};
% Text Node
\draw (265,401.4) node [anchor=north west][inner sep=0.75pt]    {$\tau _{2}$};
% Text Node
\draw (308,405.4) node [anchor=north west][inner sep=0.75pt]    {$\sigma $};
% Text Node
\draw (353,401.4) node [anchor=north west][inner sep=0.75pt]    {$\tau _{3}$};
% Text Node
\draw (402,408.4) node [anchor=north west][inner sep=0.75pt]    {$\tau _{4}$};

\end{tikzpicture}
\end{center}
\caption{Up: there are only $I_1$ or $\tilde{I}_1$ singularities on the one dimensional slice of Coulomb branch; Notice that we should not mix the singularities at $\infty$ and the bulk ones;
Bottom: there is an associated mapping class group element called Dehn twist associated with each $I_1$ or $\tilde{I}_1$ singularity.}
\label{intro2}
\end{figure}

\newpage

\section{Mapping class group and genus two degeneration}

\subsection{Genus two degeneration revisited}
In the context of 4-dimensional $\mathcal{N}=2$ supersymmetric field theories, the Coulomb branch is a critical part of the moduli space \cite{Seiberg:1994rs,Seiberg:1994aj}. This branch is associated with the vacuum expectation values (VEVs) of scalar fields known as Coulomb branch operators. 

The Seiberg-Witten solution for Coulomb branch can be described as a bundle of abelian varieties over the moduli space. 
 In the majority of cases along the Coulomb branch (they are called generic vacua), the low-energy dynamics of the theory can be effectively modeled as a $U(1)^r$ abelian gauge theory, here $r$ represents the rank of the theory, and it corresponds to the number of massless photon fields in the theory. The complex structure of the abelian variety is identified with coupling constants of photons.

In contrast to the generic vacua, at certain special points along the Coulomb branch, the abelian variety undergoes a degeneration (i,e. it becomes singular). This results in additional degrees of freedom in the low-energy theory. Depending on the specifics of this degeneration, the low-energy behavior can become much more intricate and may exhibit features like interacting superconformal field theory (SCFT), infrared (IR) free abelian theories, or non-abelian gauge theories. The exact nature of the low-energy theory depends on the details of the degeneration of the abelian variety at these special points.

When the family of abelian varieties associated with a Coulomb branch solution can be described using the Jacobian of Riemann surfaces,
it implies that for each point on the Coulomb branch moduli space, one can associate a Riemann surfaces. This correspondence simplifies the study of the Coulomb branch, as it allows us to focus on the geometric properties of these curves, which is a more well-developed area in algebraic geometry.

In the special scenario where the rank of the theory is two, it is possible to represent all abelian varieties in terms of the Jacobian of genus two curves. This reduction is particularly significant because it streamlines the investigation of the Coulomb branch, effectively reducing it to the study of the properties of genus two curves and the associated Jacobians.
 It's useful to note that every genus two curve is indeed hyperelliptic \footnote{The equation for a genus two hyperelliptic curve is $y^2=x^5+\ldots$ or $y^2=x^6+\ldots$.}.

The first key step in analyzing rank two ($r=2$) Coulomb branch solutions  involves the classification of local degenerations of genus two curves. This classification is fundamental for comprehending how the theory behaves at specific points along the Coulomb branch. It's noteworthy that this classification has already been completed, with detailed information available in \cite{namikawa1973complete}.
To accomplish this classification, they make use of Hodge theory, a powerful mathematical framework within algebraic geometry.  Distinguishing the various degenerations in the rank two case requires considering three essential components:
\textbf{Monodromy Group}: This group captures the transformations on homology groups that occur as one traverses loops around singular points in the moduli space;
\textbf{Type of Modulus Point}: Specifying the type of modulus point is critical, as it signifies the nature of the singularity where the curve degenerates; 
\textbf{Additional Discrete Parameter $m$}: The inclusion of the discrete parameter $m$ serves to fine-tune the characterization of the degenerations, providing further details that refine the classification.

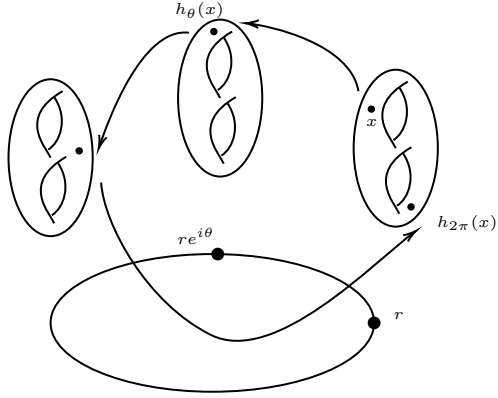
\begin{figure}
\begin{center}

\tikzset{every picture/.style={line width=0.75pt}} %set default line width to 0.75pt        

\begin{tikzpicture}[x=0.45pt,y=0.45pt,yscale=-1,xscale=1]
%uncomment if require: \path (0,799); %set diagram left start at 0, and has height of 799

%Shape: Ellipse [id:dp49351846711796177] 
\draw   (150,330.72) .. controls (150,298.84) and (210.22,273) .. (284.5,273) .. controls (358.78,273) and (419,298.84) .. (419,330.72) .. controls (419,362.6) and (358.78,388.44) .. (284.5,388.44) .. controls (210.22,388.44) and (150,362.6) .. (150,330.72) -- cycle ;
%Shape: Ellipse [id:dp5448975379896055] 
\draw   (402,184.22) .. controls (402,147.89) and (417.67,118.44) .. (437,118.44) .. controls (456.33,118.44) and (472,147.89) .. (472,184.22) .. controls (472,220.55) and (456.33,250) .. (437,250) .. controls (417.67,250) and (402,220.55) .. (402,184.22) -- cycle ;
%Curve Lines [id:da19093683411914986] 
\draw    (437,184.22) .. controls (433,175.44) and (409,150.44) .. (447,128.44) ;
%Curve Lines [id:da6436638318129626] 
\draw    (435,177.44) .. controls (445,171.44) and (452,151.44) .. (440,133.44) ;
%Curve Lines [id:da7602562900901861] 
\draw    (440,239.22) .. controls (436,230.44) and (412,205.44) .. (450,183.44) ;
%Curve Lines [id:da5537182294970349] 
\draw    (438,232.44) .. controls (448,226.44) and (455,206.44) .. (443,188.44) ;

%Shape: Circle [id:dp0634449584785447] 
\draw  [fill={rgb, 255:red, 0; green, 0; blue, 0 }  ,fill opacity=1 ] (414.5,330.72) .. controls (414.5,328.23) and (416.51,326.22) .. (419,326.22) .. controls (421.49,326.22) and (423.5,328.23) .. (423.5,330.72) .. controls (423.5,333.2) and (421.49,335.22) .. (419,335.22) .. controls (416.51,335.22) and (414.5,333.2) .. (414.5,330.72) -- cycle ;
%Shape: Ellipse [id:dp12353010808480624] 
\draw   (256,142.22) .. controls (256,105.89) and (271.67,76.44) .. (291,76.44) .. controls (310.33,76.44) and (326,105.89) .. (326,142.22) .. controls (326,178.55) and (310.33,208) .. (291,208) .. controls (271.67,208) and (256,178.55) .. (256,142.22) -- cycle ;
%Curve Lines [id:da9339662063090718] 
\draw    (291,142.22) .. controls (287,133.44) and (263,108.44) .. (301,86.44) ;
%Curve Lines [id:da06878002701827901] 
\draw    (289,135.44) .. controls (299,129.44) and (306,109.44) .. (294,91.44) ;
%Curve Lines [id:da748647090292899] 
\draw    (294,197.22) .. controls (290,188.44) and (266,163.44) .. (304,141.44) ;
%Curve Lines [id:da016218273559792218] 
\draw    (292,190.44) .. controls (302,184.44) and (309,164.44) .. (297,146.44) ;

%Shape: Ellipse [id:dp24164529350755715] 
\draw   (115,192.22) .. controls (115,155.89) and (130.67,126.44) .. (150,126.44) .. controls (169.33,126.44) and (185,155.89) .. (185,192.22) .. controls (185,228.55) and (169.33,258) .. (150,258) .. controls (130.67,258) and (115,228.55) .. (115,192.22) -- cycle ;
%Curve Lines [id:da5555459160409144] 
\draw    (150,192.22) .. controls (146,183.44) and (122,158.44) .. (160,136.44) ;
%Curve Lines [id:da301438646252443] 
\draw    (148,185.44) .. controls (158,179.44) and (165,159.44) .. (153,141.44) ;
%Curve Lines [id:da9389957093820688] 
\draw    (153,247.22) .. controls (149,238.44) and (125,213.44) .. (163,191.44) ;
%Curve Lines [id:da20552510792279222] 
\draw    (151,240.44) .. controls (161,234.44) and (168,214.44) .. (156,196.44) ;

%Shape: Circle [id:dp2518856516990885] 
\draw  [fill={rgb, 255:red, 0; green, 0; blue, 0 }  ,fill opacity=1 ] (284.5,273) .. controls (284.5,270.51) and (286.51,268.5) .. (289,268.5) .. controls (291.49,268.5) and (293.5,270.51) .. (293.5,273) .. controls (293.5,275.49) and (291.49,277.5) .. (289,277.5) .. controls (286.51,277.5) and (284.5,275.49) .. (284.5,273) -- cycle ;
%Shape: Circle [id:dp45386405034589294] 
\draw  [fill={rgb, 255:red, 0; green, 0; blue, 0 }  ,fill opacity=1 ] (171.5,186.47) .. controls (171.5,185.23) and (172.51,184.22) .. (173.75,184.22) .. controls (174.99,184.22) and (176,185.23) .. (176,186.47) .. controls (176,187.71) and (174.99,188.72) .. (173.75,188.72) .. controls (172.51,188.72) and (171.5,187.71) .. (171.5,186.47) -- cycle ;
%Shape: Circle [id:dp8821691486218962] 
\draw  [fill={rgb, 255:red, 0; green, 0; blue, 0 }  ,fill opacity=1 ] (414.5,151.47) .. controls (414.5,150.23) and (415.51,149.22) .. (416.75,149.22) .. controls (417.99,149.22) and (419,150.23) .. (419,151.47) .. controls (419,152.71) and (417.99,153.72) .. (416.75,153.72) .. controls (415.51,153.72) and (414.5,152.71) .. (414.5,151.47) -- cycle ;
%Shape: Circle [id:dp7783341207124342] 
\draw  [fill={rgb, 255:red, 0; green, 0; blue, 0 }  ,fill opacity=1 ] (283.5,86.47) .. controls (283.5,85.23) and (284.51,84.22) .. (285.75,84.22) .. controls (286.99,84.22) and (288,85.23) .. (288,86.47) .. controls (288,87.71) and (286.99,88.72) .. (285.75,88.72) .. controls (284.51,88.72) and (283.5,87.71) .. (283.5,86.47) -- cycle ;
%Shape: Circle [id:dp11688916915488423] 
\draw  [fill={rgb, 255:red, 0; green, 0; blue, 0 }  ,fill opacity=1 ] (447.5,233.47) .. controls (447.5,232.23) and (448.51,231.22) .. (449.75,231.22) .. controls (450.99,231.22) and (452,232.23) .. (452,233.47) .. controls (452,234.71) and (450.99,235.72) .. (449.75,235.72) .. controls (448.51,235.72) and (447.5,234.71) .. (447.5,233.47) -- cycle ;
%Curve Lines [id:da8324281893834764] 
\draw    (406,136.44) .. controls (399.25,96.33) and (331.96,81.48) .. (313.78,79.59) ;
\draw [shift={(312,79.44)}, rotate = 3.81] [color={rgb, 255:red, 0; green, 0; blue, 0 }  ][line width=0.75]    (10.93,-3.29) .. controls (6.95,-1.4) and (3.31,-0.3) .. (0,0) .. controls (3.31,0.3) and (6.95,1.4) .. (10.93,3.29)   ;
%Curve Lines [id:da011538412668397946] 
\draw    (265,87.44) .. controls (225.43,82.61) and (197.04,163.46) .. (190.61,182.6) ;
\draw [shift={(190,184.44)}, rotate = 288.43] [color={rgb, 255:red, 0; green, 0; blue, 0 }  ][line width=0.75]    (10.93,-3.29) .. controls (6.95,-1.4) and (3.31,-0.3) .. (0,0) .. controls (3.31,0.3) and (6.95,1.4) .. (10.93,3.29)   ;
%Curve Lines [id:da3342260732769443] 
\draw    (192,213) .. controls (194.19,247.72) and (229.96,314.32) .. (283.48,340.88) .. controls (336.2,367.04) and (418.28,280.31) .. (452.47,254.57) ;
\draw [shift={(454,253.44)}, rotate = 143.91] [color={rgb, 255:red, 0; green, 0; blue, 0 }  ][line width=0.75]    (10.93,-3.29) .. controls (6.95,-1.4) and (3.31,-0.3) .. (0,0) .. controls (3.31,0.3) and (6.95,1.4) .. (10.93,3.29)   ;

\draw (253,248.4) node [anchor=north west][inner sep=0.75pt]   [font=\tiny]   {$re^{i\theta }$};
% Text Node
\draw (433,319.4) node [anchor=north west][inner sep=0.75pt]    [font=\tiny]  {$r$};
% Text Node
\draw (410,157.4) node [anchor=north west][inner sep=0.75pt]    [font=\tiny]  {$x$};
% Text Node
\draw (251,59.4) node [anchor=north west][inner sep=0.75pt]  [font=\tiny]  {$h_{\theta }( x)$};
% Text Node
\draw (469,237.4) node [anchor=north west][inner sep=0.75pt]    [font=\tiny]  {$h_{2\pi }( x)$};

\end{tikzpicture}

\end{center}
\caption{There is a homeomorphism action $h$ associated with the loop around special vacua. The action acts on any point on Riemann surface, and in particular it induces an action on homology group.}
\label{MW}
\end{figure}

In the context of classifying degenerations of Riemann surfaces, the topological approach put forward by Matsumoto-Montesinios (MM) is highly valuable. Their theory, detailed in \cite{matsumoto2011pseudo}, is particularly useful for our purposes.
The key insight in MM's theory is that the conjugacy class of the mapping class group action serves as a complete determinant of the degeneration type, see figure. \ref{MW} for the description of mapping class group. This approach, aside from being systematic, is also combinatorial in nature, which greatly facilitates a comprehensive physical investigation \cite{Xie:2022aad}.
We will apply MM's theory specifically to the degeneration of genus two curves, aiming to recover the classification results presented in \cite{namikawa1973complete,Xie:2022aad}.

%To use Matsumoto-Montesinos's theory to study the low energy theory of $\mathcal{N}=2$ Coulomb branch, we make the following assumptions on SW solution: 

%$\bullet$: The low energy theory around the singular vacua is given by the abelian variety fiberation \cite{}, and 
%here we further require that abelian variety is given by the Jacobian of a Riemann surface; So using the Torrelli theorem, one might assume that locally the SW solution is given by the fiberation of a genus $g$ Riemann surface

%Notice that while there are a large number of theories whose SW solution can be described in above ways, 
%there are theories whose SW solution can not be described in above way. However, we do know that every abelian variety can be realized as the quotient of Jacobian variety, and so 
%Jacobian fibration might be served as the foundation for the general cases. 

In the study of Riemann surface degenerations, the conjugacy class of the mapping class group has a distinctive character known as a pseudo-periodic map of negative type \cite{matsumoto2011pseudo}. This map, denoted as $f$, can be classified using specific combinatorial data:
\begin{enumerate}

\item Admissible System of Cut Curves  ${\cal C}=\cup C_i$: The classification begins with an admissible system of cut curves, denoted as ${\cal C}=\cup C_i$. An admissible system is one where the irreducible component $B=\Sigma_g/{\cal C}$ satisfies certain conditions. Each component $B_i$ should have a non-negative Euler number $\chi_i$, which is calculated as $\chi_i=2-2g_i+n_i\geq 0$. Here, $n_i$ represents the number of boundary curves for an irreducible component $C_i$ , and $g_i$ is the genus of that component.

\item Finite Group Action on Oriented Graph $G_{\cal C}$: The map $f$ induces a finite group action on an oriented graph $G_{\cal C}$ defined in last step.

\item Screw Numbers for Annuli $C_i$: For each annulus $C_i$ in the system, the screw number is given. It's worth noting that these screw numbers must be negative in accordance with the classification.

\item Periodic Map Action: The action of $f$ on each irreducible component of $B$ is a periodic map. This periodicity is, in turn, determined by the valency data, denoted as $(n, g^{'}, {\sigma_1\over \lambda_1}+\frac{\sigma_2}{\lambda_2}+\ldots+\frac{\sigma_s}{\lambda_s})$. Here:
$n$ is the order of the map (i.e., $f^n=id$).
$g^{'}$ represents the genus of the base, defined by the covering map $f:\Sigma\to \Sigma^{'}$, 
$\sigma_i, \lambda_i$ are integral values that further specify the characteristics of the periodic map action.
\end{enumerate}

In summary, the classification of pseudo-periodic maps of negative type for Riemann surface degenerations relies on a systematic consideration of admissible cut curves, group actions on oriented graphs, screw numbers for annuli, and the valency data determining periodic map actions. This detailed combinatorial data provides a comprehensive understanding of the degeneration types.
The first two step gives rise to a weighted graph: each node has label representing genus and the number of internal cut curves; each edge represents a separating cut curve and a multiplicity from the finite group action. The third step gives an integer $K\geq -1$ along each weighted curve in weighted graph. Finally, one has a periodic map for each component in the cut system.
See figure. \ref{cut} for an example.

The three set of data of genus two degeneration given in \cite{namikawa1973complete} are recovered from MM's theory as follows: a) The monodromy group action is induced from the mapping group action \cite{farb2011primer}; b) The modulus point is given by the cut system (before the finite group action); c) The integral value $m$ is given by the 
screw number and the periodic map data on the boundaries of the annulus.

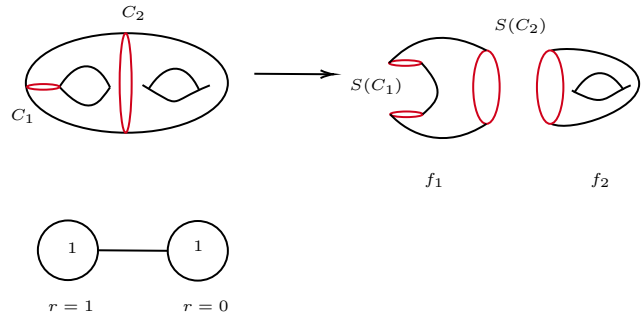
\begin{figure}
\begin{center}

\tikzset{every picture/.style={line width=0.75pt}} %set default line width to 0.75pt        

\begin{tikzpicture}[x=0.45pt,y=0.45pt,yscale=-1,xscale=1]
%uncomment if require: \path (0,799); %set diagram left start at 0, and has height of 799

%Shape: Ellipse [id:dp8346422299699846] 
\draw   (72,138.72) .. controls (72,115.68) and (109.61,97) .. (156,97) .. controls (202.39,97) and (240,115.68) .. (240,138.72) .. controls (240,161.76) and (202.39,180.44) .. (156,180.44) .. controls (109.61,180.44) and (72,161.76) .. (72,138.72) -- cycle ;
%Curve Lines [id:da01004509114736607] 
\draw    (100,142) .. controls (114,119.44) and (131,118.44) .. (142,142.44) ;
%Curve Lines [id:da8579761876965402] 
\draw    (100,142) .. controls (126,170.44) and (134,153.44) .. (142,142.44) ;

%Curve Lines [id:da9963047390571795] 
\draw    (174,144) .. controls (188,121.44) and (205,120.44) .. (216,144.44) ;
%Curve Lines [id:da8993367014196] 
\draw    (169,140.44) .. controls (204,161.44) and (188,155.72) .. (225,140.72) ;
%Shape: Ellipse [id:dp3162449614412318] 
\draw  [color={rgb, 255:red, 208; green, 2; blue, 27 }  ,draw opacity=1 ] (150,138.72) .. controls (150,115.68) and (152.24,97) .. (155,97) .. controls (157.76,97) and (160,115.68) .. (160,138.72) .. controls (160,161.76) and (157.76,180.44) .. (155,180.44) .. controls (152.24,180.44) and (150,161.76) .. (150,138.72) -- cycle ;
%Shape: Ellipse [id:dp6389484192091343] 
\draw  [color={rgb, 255:red, 208; green, 2; blue, 27 }  ,draw opacity=1 ] (73,142) .. controls (73,140.7) and (79.04,139.64) .. (86.5,139.64) .. controls (93.96,139.64) and (100,140.7) .. (100,142) .. controls (100,143.3) and (93.96,144.36) .. (86.5,144.36) .. controls (79.04,144.36) and (73,143.3) .. (73,142) -- cycle ;
%Straight Lines [id:da9577895186217642] 
\draw    (262,131) -- (326,131.42) ;
\draw [shift={(328,131.44)}, rotate = 180.38] [color={rgb, 255:red, 0; green, 0; blue, 0 }  ][line width=0.75]    (10.93,-3.29) .. controls (6.95,-1.4) and (3.31,-0.3) .. (0,0) .. controls (3.31,0.3) and (6.95,1.4) .. (10.93,3.29)   ;
%Curve Lines [id:da27876238923026453] 
\draw    (509,111.44) .. controls (533,106.44) and (582,117.44) .. (582,139.44) .. controls (582,161.44) and (526,180.44) .. (509,173) ;
%Shape: Ellipse [id:dp7007973300064821] 
\draw  [color={rgb, 255:red, 208; green, 2; blue, 27 }  ,draw opacity=1 ] (497,142.22) .. controls (497,125.22) and (501.92,111.44) .. (508,111.44) .. controls (514.08,111.44) and (519,125.22) .. (519,142.22) .. controls (519,159.22) and (514.08,173) .. (508,173) .. controls (501.92,173) and (497,159.22) .. (497,142.22) -- cycle ;
%Curve Lines [id:da931437814792998] 
\draw    (528,147) .. controls (542,124.44) and (557,126.44) .. (569,144.44) ;
%Curve Lines [id:da45061403993332805] 
\draw    (526,145) .. controls (555,160.44) and (552,150.44) .. (575,140.72) ;
%Shape: Ellipse [id:dp8420275016571473] 
\draw  [color={rgb, 255:red, 208; green, 2; blue, 27 }  ,draw opacity=1 ] (374,122) .. controls (374,120.7) and (380.04,119.64) .. (387.5,119.64) .. controls (394.96,119.64) and (401,120.7) .. (401,122) .. controls (401,123.3) and (394.96,124.36) .. (387.5,124.36) .. controls (380.04,124.36) and (374,123.3) .. (374,122) -- cycle ;
%Shape: Ellipse [id:dp886100868516245] 
\draw  [color={rgb, 255:red, 208; green, 2; blue, 27 }  ,draw opacity=1 ] (375,165) .. controls (375,163.7) and (381.04,162.64) .. (388.5,162.64) .. controls (395.96,162.64) and (402,163.7) .. (402,165) .. controls (402,166.3) and (395.96,167.36) .. (388.5,167.36) .. controls (381.04,167.36) and (375,166.3) .. (375,165) -- cycle ;
%Shape: Ellipse [id:dp5162179869702316] 
\draw  [color={rgb, 255:red, 208; green, 2; blue, 27 }  ,draw opacity=1 ] (444,142.22) .. controls (444,125.22) and (448.92,111.44) .. (455,111.44) .. controls (461.08,111.44) and (466,125.22) .. (466,142.22) .. controls (466,159.22) and (461.08,173) .. (455,173) .. controls (448.92,173) and (444,159.22) .. (444,142.22) -- cycle ;
%Curve Lines [id:da0693461507605293] 
\draw    (374,122) .. controls (391,104.44) and (415,90.44) .. (455,111.44) ;
%Curve Lines [id:da374740102754346] 
\draw    (375,165) .. controls (391,182.44) and (421,195.44) .. (455,173) ;
%Curve Lines [id:da5449122427684866] 
\draw    (402,165) .. controls (415,153.44) and (422,145.44) .. (401,122) ;
%Shape: Circle [id:dp6618611892792503] 
\draw   (82,279) .. controls (82,265.19) and (93.19,254) .. (107,254) .. controls (120.81,254) and (132,265.19) .. (132,279) .. controls (132,292.81) and (120.81,304) .. (107,304) .. controls (93.19,304) and (82,292.81) .. (82,279) -- cycle ;
%Straight Lines [id:da27207404724331496] 
\draw    (132,279) -- (190,279.44) ;
%Shape: Circle [id:dp8283795899562058] 
\draw   (190,279.44) .. controls (190,265.63) and (201.19,254.44) .. (215,254.44) .. controls (228.81,254.44) and (240,265.63) .. (240,279.44) .. controls (240,293.24) and (228.81,304.44) .. (215,304.44) .. controls (201.19,304.44) and (190,293.24) .. (190,279.44) -- cycle ;

\draw (57,158.4) node [anchor=north west][inner sep=0.75pt]  [font=\tiny]  {$C_{1}$};
% Text Node
\draw (149,71.4) node [anchor=north west][inner sep=0.75pt]   [font=\tiny]  {$C_{2}$};
% Text Node
\draw (104,270.4) node [anchor=north west][inner sep=0.75pt]    [font=\tiny] {$1$};
% Text Node
\draw (209,268.4) node [anchor=north west][inner sep=0.75pt]   [font=\tiny]  {$1$};
% Text Node
\draw (88,318.4) node [anchor=north west][inner sep=0.75pt]    [font=\tiny] {$r=1$};
% Text Node
\draw (200,318.4) node [anchor=north west][inner sep=0.75pt]    [font=\tiny] {$r=0$};
% Text Node
\draw (401,211.4) node [anchor=north west][inner sep=0.75pt]   [font=\tiny]  {$f_{1}$};
% Text Node
\draw (539,211.4) node [anchor=north west][inner sep=0.75pt]     [font=\tiny]{$f_{2}$};
% Text Node
\draw (459,81.4) node [anchor=north west][inner sep=0.75pt]   [font=\tiny]  {$S( C_{2})$};
% Text Node
\draw (339,132.4) node [anchor=north west][inner sep=0.75pt]    [font=\tiny] {$S( C_{1})$};

\end{tikzpicture}

\end{center}
\caption{Up: An admissible cut system for a genus two Riemann surface, and there are two irreducible components after the cutting; there is a screw number associated with each cutting curve, and a periodic map on each irreducible component.
 Bottom: A weighted graph for the above cut system: here one draw an edge for every separating curve (which will cut the Riemann surface into two separate components),
and one write the number of non-separating cutting curves for a vertex in the graph.}
\label{cut}
\end{figure}

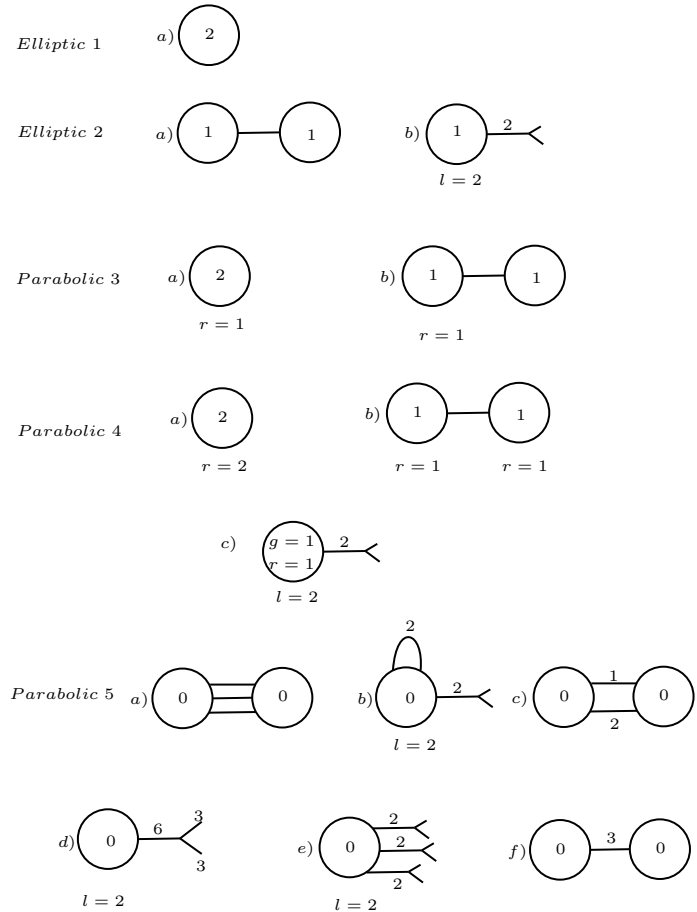
\begin{figure}
 \begin{center}

\tikzset{every picture/.style={line width=0.75pt}} %set default line width to 0.75pt        

\begin{tikzpicture}[x=0.45pt,y=0.45pt,yscale=-1,xscale=1]
%uncomment if require: \path (0,951); %set diagram left start at 0, and has height of 951

%Shape: Circle [id:dp7817928010957634] 
\draw   (176,72) .. controls (176,58.19) and (187.19,47) .. (201,47) .. controls (214.81,47) and (226,58.19) .. (226,72) .. controls (226,85.81) and (214.81,97) .. (201,97) .. controls (187.19,97) and (176,85.81) .. (176,72) -- cycle ;
%Shape: Circle [id:dp9176956848680582] 
\draw   (175,154) .. controls (175,140.19) and (186.19,129) .. (200,129) .. controls (213.81,129) and (225,140.19) .. (225,154) .. controls (225,167.81) and (213.81,179) .. (200,179) .. controls (186.19,179) and (175,167.81) .. (175,154) -- cycle ;
%Straight Lines [id:da19521025157372462] 
\draw    (225,154) -- (260,153.44) ;
%Shape: Circle [id:dp6919651868713496] 
\draw   (260,153.44) .. controls (260,139.63) and (271.19,128.44) .. (285,128.44) .. controls (298.81,128.44) and (310,139.63) .. (310,153.44) .. controls (310,167.24) and (298.81,178.44) .. (285,178.44) .. controls (271.19,178.44) and (260,167.24) .. (260,153.44) -- cycle ;
%Shape: Circle [id:dp7224470674155749] 
\draw   (382,155) .. controls (382,141.19) and (393.19,130) .. (407,130) .. controls (420.81,130) and (432,141.19) .. (432,155) .. controls (432,168.81) and (420.81,180) .. (407,180) .. controls (393.19,180) and (382,168.81) .. (382,155) -- cycle ;
%Straight Lines [id:da16217320361852283] 
\draw    (432,155) -- (467,154.44) ;
%Straight Lines [id:da8233007754960319] 
\draw    (467,154.44) -- (477,148.44) ;
%Straight Lines [id:da08726511916797552] 
\draw    (467,154.44) -- (479,163.44) ;
%Shape: Circle [id:dp7460432165845688] 
\draw   (362,274) .. controls (362,260.19) and (373.19,249) .. (387,249) .. controls (400.81,249) and (412,260.19) .. (412,274) .. controls (412,287.81) and (400.81,299) .. (387,299) .. controls (373.19,299) and (362,287.81) .. (362,274) -- cycle ;
%Straight Lines [id:da5930203032367205] 
\draw    (412,274) -- (447,273.44) ;
%Shape: Circle [id:dp7903984944518071] 
\draw   (447,273.44) .. controls (447,259.63) and (458.19,248.44) .. (472,248.44) .. controls (485.81,248.44) and (497,259.63) .. (497,273.44) .. controls (497,287.24) and (485.81,298.44) .. (472,298.44) .. controls (458.19,298.44) and (447,287.24) .. (447,273.44) -- cycle ;
%Shape: Circle [id:dp7429803675411172] 
\draw   (185,274) .. controls (185,260.19) and (196.19,249) .. (210,249) .. controls (223.81,249) and (235,260.19) .. (235,274) .. controls (235,287.81) and (223.81,299) .. (210,299) .. controls (196.19,299) and (185,287.81) .. (185,274) -- cycle ;
%Shape: Circle [id:dp8379035860903852] 
\draw   (187,393) .. controls (187,379.19) and (198.19,368) .. (212,368) .. controls (225.81,368) and (237,379.19) .. (237,393) .. controls (237,406.81) and (225.81,418) .. (212,418) .. controls (198.19,418) and (187,406.81) .. (187,393) -- cycle ;
%Shape: Circle [id:dp14033592921543703] 
\draw   (349,389) .. controls (349,375.19) and (360.19,364) .. (374,364) .. controls (387.81,364) and (399,375.19) .. (399,389) .. controls (399,402.81) and (387.81,414) .. (374,414) .. controls (360.19,414) and (349,402.81) .. (349,389) -- cycle ;
%Straight Lines [id:da8741895184964351] 
\draw    (399,389) -- (434,388.44) ;
%Shape: Circle [id:dp6432591035025335] 
\draw   (434,388.44) .. controls (434,374.63) and (445.19,363.44) .. (459,363.44) .. controls (472.81,363.44) and (484,374.63) .. (484,388.44) .. controls (484,402.24) and (472.81,413.44) .. (459,413.44) .. controls (445.19,413.44) and (434,402.24) .. (434,388.44) -- cycle ;
%Shape: Circle [id:dp6354823285378015] 
\draw   (246,505) .. controls (246,491.19) and (257.19,480) .. (271,480) .. controls (284.81,480) and (296,491.19) .. (296,505) .. controls (296,518.81) and (284.81,530) .. (271,530) .. controls (257.19,530) and (246,518.81) .. (246,505) -- cycle ;
%Straight Lines [id:da08861058074206307] 
\draw    (296,505) -- (331,504.44) ;
%Straight Lines [id:da6918994030377408] 
\draw    (331,504.44) -- (341,498.44) ;
%Straight Lines [id:da6389778452791505] 
\draw    (331,504.44) -- (343,513.44) ;
%Shape: Circle [id:dp1862645233578405] 
\draw   (154,628) .. controls (154,614.19) and (165.19,603) .. (179,603) .. controls (192.81,603) and (204,614.19) .. (204,628) .. controls (204,641.81) and (192.81,653) .. (179,653) .. controls (165.19,653) and (154,641.81) .. (154,628) -- cycle ;
%Straight Lines [id:da6822913252895975] 
\draw    (202,616.44) -- (240,616.44) ;
%Shape: Circle [id:dp535259885958569] 
\draw   (237,627.44) .. controls (237,613.63) and (248.19,602.44) .. (262,602.44) .. controls (275.81,602.44) and (287,613.63) .. (287,627.44) .. controls (287,641.24) and (275.81,652.44) .. (262,652.44) .. controls (248.19,652.44) and (237,641.24) .. (237,627.44) -- cycle ;
%Straight Lines [id:da49122143741674507] 
\draw    (204,628) -- (237,627.44) ;
%Straight Lines [id:da27681790001133555] 
\draw    (201,640) -- (240,639.44) ;
%Shape: Circle [id:dp35865619644947777] 
\draw   (340,627) .. controls (340,613.19) and (351.19,602) .. (365,602) .. controls (378.81,602) and (390,613.19) .. (390,627) .. controls (390,640.81) and (378.81,652) .. (365,652) .. controls (351.19,652) and (340,640.81) .. (340,627) -- cycle ;
%Straight Lines [id:da9232241904334081] 
\draw    (390,627) -- (425,626.44) ;
%Straight Lines [id:da2604462097306254] 
\draw    (425,626.44) -- (435,620.44) ;
%Straight Lines [id:da662775722973285] 
\draw    (425,626.44) -- (437,635.44) ;
%Shape: Circle [id:dp8606590705465278] 
\draw   (471,627) .. controls (471,613.19) and (482.19,602) .. (496,602) .. controls (509.81,602) and (521,613.19) .. (521,627) .. controls (521,640.81) and (509.81,652) .. (496,652) .. controls (482.19,652) and (471,640.81) .. (471,627) -- cycle ;
%Straight Lines [id:da21936009882403074] 
\draw    (519,615.44) -- (557,615.44) ;
%Shape: Circle [id:dp5637396865964348] 
\draw   (554,626.44) .. controls (554,612.63) and (565.19,601.44) .. (579,601.44) .. controls (592.81,601.44) and (604,612.63) .. (604,626.44) .. controls (604,640.24) and (592.81,651.44) .. (579,651.44) .. controls (565.19,651.44) and (554,640.24) .. (554,626.44) -- cycle ;
%Straight Lines [id:da45342128009242844] 
\draw    (518,639) -- (557,638.44) ;
%Shape: Circle [id:dp8877010435192394] 
\draw   (293,753) .. controls (293,739.19) and (304.19,728) .. (318,728) .. controls (331.81,728) and (343,739.19) .. (343,753) .. controls (343,766.81) and (331.81,778) .. (318,778) .. controls (304.19,778) and (293,766.81) .. (293,753) -- cycle ;
%Shape: Circle [id:dp6402873599878238] 
\draw   (468,755) .. controls (468,741.19) and (479.19,730) .. (493,730) .. controls (506.81,730) and (518,741.19) .. (518,755) .. controls (518,768.81) and (506.81,780) .. (493,780) .. controls (479.19,780) and (468,768.81) .. (468,755) -- cycle ;
%Shape: Circle [id:dp012163681931581127] 
\draw   (551,754.44) .. controls (551,740.63) and (562.19,729.44) .. (576,729.44) .. controls (589.81,729.44) and (601,740.63) .. (601,754.44) .. controls (601,768.24) and (589.81,779.44) .. (576,779.44) .. controls (562.19,779.44) and (551,768.24) .. (551,754.44) -- cycle ;
%Straight Lines [id:da4748843306484044] 
\draw    (518,755) -- (551,754.44) ;
%Shape: Circle [id:dp7865123407460154] 
\draw   (92,746) .. controls (92,732.19) and (103.19,721) .. (117,721) .. controls (130.81,721) and (142,732.19) .. (142,746) .. controls (142,759.81) and (130.81,771) .. (117,771) .. controls (103.19,771) and (92,759.81) .. (92,746) -- cycle ;
%Straight Lines [id:da6474447168954747] 
\draw    (142,746) -- (177,745.44) ;
%Straight Lines [id:da02958143505853228] 
\draw    (177,745.44) -- (195,731.44) ;
%Straight Lines [id:da5747088256010675] 
\draw    (177,745.44) -- (195,758.44) ;
%Straight Lines [id:da6144995884826782] 
\draw    (337,737) -- (372,736.44) ;
%Straight Lines [id:da46836236816081844] 
\draw    (372,736.44) -- (382,730.44) ;
%Straight Lines [id:da9478756338525812] 
\draw    (372,736.44) -- (384,745.44) ;
%Straight Lines [id:da9016213046834171] 
\draw    (343,756) -- (378,755.44) ;
%Straight Lines [id:da4570699641935838] 
\draw    (378,755.44) -- (388,749.44) ;
%Straight Lines [id:da9858815222384314] 
\draw    (378,755.44) -- (390,764.44) ;
%Straight Lines [id:da6242758103988597] 
\draw    (332,774) -- (367,773.44) ;
%Straight Lines [id:da5756096659241995] 
\draw    (367,773.44) -- (377,767.44) ;
%Straight Lines [id:da9740249536221303] 
\draw    (367,773.44) -- (379,782.44) ;
%Curve Lines [id:da6496017252170496] 
\draw    (354,604.44) .. controls (357,563.44) and (380,571.44) .. (377,604.44) ;

\draw (38,71.4) node [anchor=north west][inner sep=0.75pt]    [font=\tiny] {$Elliptic\ 1$};
% Text Node
\draw (195,64.4) node [anchor=north west][inner sep=0.75pt]    [font=\tiny] {$2$};
% Text Node
\draw (155,64.4) node [anchor=north west][inner sep=0.75pt]    [font=\tiny] {$a)$};
% Text Node

\draw (39,145.4) node [anchor=north west][inner sep=0.75pt]   [font=\tiny]  {$Elliptic\ 2$};
% Text Node
\draw (194,146.4) node [anchor=north west][inner sep=0.75pt]   [font=\tiny]  {$1$};
% Text Node
\draw (154,146.4) node [anchor=north west][inner sep=0.75pt]   [font=\tiny]  {$a)$};

\draw (280,148.4) node [anchor=north west][inner sep=0.75pt]   [font=\tiny]  {$1$};
% Text Node
\draw (401,145.4) node [anchor=north west][inner sep=0.75pt]   [font=\tiny]  {$1$};
% Text Node
\draw (361,145.4) node [anchor=north west][inner sep=0.75pt]   [font=\tiny]  {$b)$};
% Text Node

\draw (390,185.4) node [anchor=north west][inner sep=0.75pt]   [font=\tiny]  {$l=2$};
% Text Node
\draw (443,140.4) node [anchor=north west][inner sep=0.75pt]  [font=\tiny]  {$2$};
% Text Node
\draw (38,268.4) node [anchor=north west][inner sep=0.75pt]    [font=\tiny] {$Parabolic\ 3$};
% Text Node
\draw (39,396.4) node [anchor=north west][inner sep=0.75pt]   [font=\tiny]  {$Parabolic\ 4$};
% Text Node
\draw (33,616.4) node [anchor=north west][inner sep=0.75pt]   [font=\tiny]  {$Parabolic\ 5$};
% Text Node
\draw (381,266.4) node [anchor=north west][inner sep=0.75pt]  [font=\tiny]   {$1$};
% Text Node
\draw (341,266.4) node [anchor=north west][inner sep=0.75pt]  [font=\tiny]   {$b)$};
% Text Node

\draw (467,268.4) node [anchor=north west][inner sep=0.75pt]   [font=\tiny]  {$1$};
% Text Node
\draw (373,316.4) node [anchor=north west][inner sep=0.75pt]   [font=\tiny]  {$r=1$};
% Text Node
\draw (190,307.4) node [anchor=north west][inner sep=0.75pt]    [font=\tiny] {$r=1$};
% Text Node
\draw (204,266.4) node [anchor=north west][inner sep=0.75pt]   [font=\tiny]  {$2$};
% Text Node
\draw (164,266.4) node [anchor=north west][inner sep=0.75pt]   [font=\tiny]  {$a)$};
% Text Node

\draw (192,426.4) node [anchor=north west][inner sep=0.75pt]  [font=\tiny]   {$r=2$};
% Text Node
\draw (206,385.4) node [anchor=north west][inner sep=0.75pt]   [font=\tiny]  {$2$};
% Text Node
\draw (166,385.4) node [anchor=north west][inner sep=0.75pt]   [font=\tiny]  {$a)$};
% Text Node

\draw (368,381.4) node [anchor=north west][inner sep=0.75pt]   [font=\tiny]  {$1$};
% Text Node

\draw (328,381.4) node [anchor=north west][inner sep=0.75pt]   [font=\tiny]  {$b)$};
% Text Node

\draw (454,383.4) node [anchor=north west][inner sep=0.75pt]   [font=\tiny]  {$1$};
% Text Node
\draw (353,426.4) node [anchor=north west][inner sep=0.75pt]   [font=\tiny]  {$r=1$};
% Text Node
\draw (442,426.4) node [anchor=north west][inner sep=0.75pt]   [font=\tiny]  {$r=1$};
% Text Node
\draw (248,489.4) node [anchor=north west][inner sep=0.75pt]  [font=\tiny]   {$g=1$};
% Text Node
\draw (208,489.4) node [anchor=north west][inner sep=0.75pt]  [font=\tiny]   {$c)$};
% Text Node

\draw (254,535.4) node [anchor=north west][inner sep=0.75pt]   [font=\tiny]  {$l=2$};
% Text Node
\draw (307,490.4) node [anchor=north west][inner sep=0.75pt]  [font=\tiny]   {$2$};
% Text Node
\draw (248,508.4) node [anchor=north west][inner sep=0.75pt]  [font=\tiny]  {$r=1$};
% Text Node
\draw (173,620.4) node [anchor=north west][inner sep=0.75pt]   [font=\tiny]  {$0$};
% Text Node
\draw (133,620.4) node [anchor=north west][inner sep=0.75pt]   [font=\tiny]  {$a)$};
% Text Node

\draw (256,619.4) node [anchor=north west][inner sep=0.75pt]   [font=\tiny]  {$0$};
% Text Node
\draw (362,620.4) node [anchor=north west][inner sep=0.75pt]    [font=\tiny] {$0$};
% Text Node
\draw (322,620.4) node [anchor=north west][inner sep=0.75pt]    [font=\tiny] {$b)$};
% Text Node

\draw (352,659.4) node [anchor=north west][inner sep=0.75pt]  [font=\tiny]  {$l=2$};
% Text Node
\draw (401,612.4) node [anchor=north west][inner sep=0.75pt]  [font=\tiny]   {$2$};
% Text Node
\draw (490,619.4) node [anchor=north west][inner sep=0.75pt]   [font=\tiny]  {$0$};
% Text Node
\draw (450,619.4) node [anchor=north west][inner sep=0.75pt]   [font=\tiny]  {$c)$};
% Text Node

\draw (573,618.4) node [anchor=north west][inner sep=0.75pt]   [font=\tiny]  {$0$};
% Text Node
\draw (312,745.4) node [anchor=north west][inner sep=0.75pt]  [font=\tiny]   {$0$};
% Text Node
\draw (272,745.4) node [anchor=north west][inner sep=0.75pt]  [font=\tiny]   {$e)$};
% Text Node

\draw (487,747.4) node [anchor=north west][inner sep=0.75pt]  [font=\tiny]   {$0$};
% Text Node
\draw (447,747.4) node [anchor=north west][inner sep=0.75pt]  [font=\tiny]   {$f)$};
% Text Node

\draw (570,746.4) node [anchor=north west][inner sep=0.75pt]   [font=\tiny]  {$0$};
% Text Node
\draw (113,740.4) node [anchor=north west][inner sep=0.75pt]  [font=\tiny]   {$0$};
% Text Node
\draw (73,740.4) node [anchor=north west][inner sep=0.75pt]  [font=\tiny]   {$d)$};
% Text Node

\draw (93,790.4) node [anchor=north west][inner sep=0.75pt]  [font=\tiny]  {$l=2$};
% Text Node
\draw (152,730.4) node [anchor=north west][inner sep=0.75pt] [font=\tiny]  {$6$};
% Text Node
\draw (532,642.4) node [anchor=north west][inner sep=0.75pt]  [font=\tiny]   {$2$};
% Text Node
\draw (531,602.4) node [anchor=north west][inner sep=0.75pt] [font=\tiny]   {$1$};
% Text Node
\draw (529,738.4) node [anchor=north west][inner sep=0.75pt] [font=\tiny]   {$3$};
% Text Node
\draw (188,761.4) node [anchor=north west][inner sep=0.75pt]  [font=\tiny]  {$3$};
% Text Node
\draw (186,719.4) node [anchor=north west][inner sep=0.75pt]  [font=\tiny]  {$3$};
% Text Node
\draw (303,793.4) node [anchor=north west][inner sep=0.75pt]  [font=\tiny]  {$l=2$};
% Text Node
\draw (348,722.4) node [anchor=north west][inner sep=0.75pt]  [font=\tiny]   {$2$};
% Text Node
\draw (354,741.4) node [anchor=north west][inner sep=0.75pt] [font=\tiny]  {$2$};
% Text Node
\draw (351.5,777.12) node [anchor=north west][inner sep=0.75pt]  [font=\tiny]  {$2$};
% Text Node
\draw (362,560.4) node [anchor=north west][inner sep=0.75pt]  [font=\tiny]   {$2$};

\end{tikzpicture}
 
 \end{center}
 \caption{All possible weighted graphs for genus two Riemann surface. The number in the vertex represents the genus, and the number $r$ denotes the number of internal cut.}
  \label{weightedgenustwo}
 \end{figure}

Let's now revisit the classification of genus two degeneration by using MM's theory.  In fact, there are some missing items in \cite{namikawa1973complete}, but will be found here. 
There are a total of five classes, and the  combinatorial data is described as follows.

\textit{Remark}: The weighted graph 4a) with the periodic data   [$(g=2, k=0, r=2):~ord(f)=2,~(C_1, C_2),~~\frac{1}{2}+\frac{1}{2}+(1)+(1)$] seems missing in \cite{namikawa1973complete}.

\textbf{Weighted graph}: The weighted graph for genus two degeneration is described in figure. \ref{weightedgenustwo}, and we identify them with the names in \cite{namikawa1973complete,Xie:2023lko}. 
They are found by first imposing the non-negative Euler number constraint (This is just the constraint in Deligne-Mumford theory), and then find the finite group actions.
%\begin{enumerate}
%\item Elliptic 1: the weighted graph is  a genus two curve without any cutting curve, and 
%the data is  the periodic map which is classified in table. \ref{}. The stable model is  a smooth genus two curve, namely type M in \cite{Xie:2023lko}, \cite{Li:2023ffx}.
%\item Elliptic 2: the weighted graph is  two genus one curves connected by one edge, and the quotient of it.   The stable model is just  two genus one curve connected by a double point, namely type N in \cite{}.
%\item Parabolic 3: The stable model consists of: a) Genus two curve with one cut; b): a genus one curve with one cut, and a genus one curve without cut, and they are connected by a single double point, namely type B in \cite{}.
%\item Parabolic 4: the weighted graph is shown in figure. \ref{}. The stable model consists of: a) Genus two curve with two cuts; b): Two genus one curves with one cut, and they are connected by a double point, namely type C in \cite{}.
%\item Parabolic 5: the weighted graph is shown in figure. \ref{}. The stable model is two rational curves meeting at three points, namely type D in \cite{}.
%\end{enumerate}

\textbf{Periodic Maps}: The list of periodic maps for curves with genus one and two is given in table. \ref{rank1}. Additionally, we'll explore periodic maps on curves that include boundaries.
The fundamental component in this context is labeled as $(g, k, r)$, where $k$ represents the count of boundary edges, and $r$ signifies the number of internal cutting curves. We employ bold letters to indicate data associated with the boundary curves, while data enclosed within brackets pertains to the internal curves.
See table. \ref{periodicboun} for the full data.

 \begin{table}[H]
\begin{center}
\begin{tabular}{|c|c|}
\hline
Order & $Data$   \\ \hline
$n=6$ & $(\frac{1}{6}+\frac{1}{3}+\frac{1}{2}, \frac{5}{6}+\frac{2}{3}+\frac{1}{2})$ \\ \hline
$n=4$ & $(\frac{1}{4}+\frac{1}{4}+\frac{1}{2}, \frac{3}{4}+\frac{3}{4}+\frac{1}{2})$ \\ \hline
$n=3$ & $(\frac{1}{3}+\frac{1}{3}+\frac{1}{3}, \frac{2}{3}+\frac{2}{3}+\frac{2}{3})$ \\ \hline
$n=2$ & $\frac{1}{2}+\frac{1}{2}+\frac{1}{2}+\frac{1}{2}$ \\ \hline
\end{tabular}
\begin{tabular}{|c|c|}
\hline
Order & $Data$   \\ \hline
$n=10$ & $(\frac{1}{10}+\frac{2}{5}+\frac{1}{2}, \frac{9}{10}+\frac{3}{5}+\frac{1}{2})$,   $(\frac{3}{10}+\frac{1}{5}+\frac{1}{2}, \frac{7}{10}+\frac{4}{5}+\frac{1}{2})$, \\ \hline
$n=8$ & $(\frac{1}{8}+\frac{3}{8}+\frac{1}{2}, \frac{7}{8}+\frac{5}{8}+\frac{1}{2})$ \\ \hline
$n=6$ & $(\frac{1}{6}+\frac{1}{6}+\frac{2}{3}, \frac{5}{6}+\frac{5}{6}+\frac{1}{3}),~~(\frac{1}{3}+\frac{2}{3}+\frac{1}{2}+\frac{1}{2})$ \\ \hline
$n=5$ & $(\frac{1}{5}+\frac{1}{5}+\frac{3}{5}, \frac{4}{5}+\frac{4}{5}+\frac{2}{5}),~~(\frac{1}{5}+\frac{1}{5}+\frac{2}{5}, \frac{4}{5}+\frac{4}{5}+\frac{3}{5})$ \\ \hline
$n=4$ & $\frac{1}{4}+\frac{3}{4}+\frac{1}{2}+\frac{1}{2}$ \\ \hline
$n=3$ & $\frac{1}{3}+\frac{1}{3}+\frac{2}{3}+\frac{2}{3}$ \\ \hline
$n=2$ & $\frac{1}{2}+\frac{1}{2}+\frac{1}{2}+\frac{1}{2}+\frac{1}{2}+\frac{1}{2}$ \\ \hline
$g^{'}=1,n=2$ & $\frac{1}{2}+\frac{1}{2}$ \\ \hline
\end{tabular}
\end{center}
\caption{The periodic maps for genus $g=1$ an $g=2$. The genus $g^{'}=0$ is ignored.}
\label{rank1}
\end{table}%

\begin{table}[H]
\begin{center}
\begin{tabular}{|c|l|}
\hline
Type & $Data$   \\ \hline
$(g=1, k=1, r=0)$ & $(1):~f=id,~~~~~~(2):~\bm{\frac{5}{6}}+\frac{2}{3}+\frac{1}{2},~~(3):~\bm{\frac{1}{6}}+\frac{1}{3}+\frac{1}{2}$, \\ 
~&$(4):~\bm{\frac{1}{4}}+\frac{1}{4}+\frac{1}{2},~~(5):~\bm{\frac{3}{4}}+\frac{3}{4}+\frac{1}{2},~~(6):~\bm{\frac{1}{3}}+\frac{1}{3}+\frac{1}{3},$\\
~&$(7):~\bm{\frac{2}{3}}+\frac{2}{3}+\frac{2}{3},~~(8):~\bm{\frac{1}{2}}+\frac{1}{2}+\frac{1}{2}+\frac{1}{2}$\\ \hline
$(g=1, k=1, r=1)$ & $(1): f=id $ \\ 
& $(2):~ord(f)=2:~Amp(C_1),~~\bm{\frac{1}{2}}+\frac{1}{2}+(1)$, \\  \hline
$(g=2, k=0, r=1)$ &  $(1):~f=id,~~~~~~~(2):~(\frac{1}{4})+(\frac{1}{4})+\frac{1}{2},~~~~~~(3):~(\frac{3}{4})+(\frac{3}{4})+\frac{1}{2}$ \\ 
  $f(C_1)=C_1$& $(4):~(\frac{1}{3})+(\frac{1}{3})+\frac{1}{3},~~(5):~(\frac{2}{3})+(\frac{2}{3})+\frac{2}{3}, ~~~(6):~(\frac{1}{2})+(\frac{1}{2})+\frac{1}{2}+\frac{1}{2}$ \\ \hline
  $Amp(C_1)$ & $(1):~\frac{1}{4}+\frac{1}{4}+(\frac{1}{2}),~~(2):~\frac{3}{4}+\frac{3}{4}+(\frac{1}{2}),~(3):~\frac{5}{6}+(\frac{2}{3})+\frac{1}{2},$ \\ 
  ~&$(4):~\frac{1}{6}+(\frac{1}{3})+\frac{1}{2},~(5):~\frac{1}{2}+\frac{1}{2}+\frac{1}{2}+\frac{1}{2}+(1)$ \\ \hline
  $(g=2, k=0, r=2)$ &  $(1):~f=id,~~~~~~~ $ \\ 
   &  $(2):~ord(f)=2,~(C_1, C_2),~~\frac{1}{2}+\frac{1}{2}+(1)+(1)$    \\
    & $(3):~ord(f)=2,~Amp(C_1),~f(C_2)=C_2,~~(\frac{1}{2})+(\frac{1}{2})+(1)$    \\ 
    &  $ (4):~ord(f)=2,~Amp(C_1),~Amp(C_2),~~\frac{1}{2}+\frac{1}{2}+(1)+(1)$,  \\
 & $(5):~ord(f)=4,~Amp(C_1,C_2),~~\frac{3}{4}+\frac{1}{4}+(1)$ \\ \hline
   $(g=0, k=3, r=0)$ &  $(1):~f=id,~~~~~~~ $ \\ 
   &$(2):~ord(f)=2,~(\partial_1, \partial_2),~f(\partial_3)=\partial_3,~\frac{1}{2}+\bm{\frac{1}{2}}+\bm{1}$ \\ 
   &$(3):~ord(f)=3,~(\partial_1, \partial_2,\partial_3),~~\frac{1}{3}+\frac{2}{3}+\bm{1}$ \\ \hline
\end{tabular}
\end{center}
\caption{The periodic maps for curves with boundaries and internal cuts.}
\label{periodicboun}
\end{table}%

\textbf{Dual graph}: One can attach a dual graph (star-shaped) for the periodic maps, and 
these sub-graphs are glued together to form a full dual graph for each degeneration. The
dual graph is closely related to 3d mirror of the IR theory, and is used in an essential way 
in \cite{Xie:2023lko} to determine the IR theory. Here we give a short review of the results in \cite{Xie:2023lko}.

 First, one can define a dual graph for 
a period map as follows. The data for a periodic map is  $(n ,g^{'}, {\sigma_1 \over \lambda_1}+\ldots+{\sigma_l\over \lambda_l})$, here $n$ is the order of the map, $g^{'}$ is 
the genus of the base defined by the covering map $f:g\to g^{'}$, and ${\sigma_i\over \lambda_i}$ gives the data for the ramification point of the covering map. These data are constrained by Hurwitz formula
\begin{equation*}
2g-2=n[ (2g^{'}-2) +\sum_i(1-\frac{1}{\lambda_i})]
\end{equation*}
The dual graph is constructed from the combinatorial data $(n ,g^{'}, {\sigma_1 \over \lambda_1}+\ldots+{\sigma_l\over \lambda_l})$ as follows:
\begin{enumerate}
\item First given a valency data ${\sigma\over \lambda}$ ($m\lambda =n$), one attach a linear chain of spheres with following nonzero multiplicities $a_0>a_1>a_2\ldots>a_s=1$:
\begin{equation*}
a_0=\lambda,~~a_1=\sigma,~~{a_{i+1}+a_{i-1}\over a_i}=\lambda_i \in Z
\end{equation*}
Given $a_i$ and $a_{i-1}$, the above formula uniquely determines the integer $a_{i+1}$. 
Since $n=\lambda m$,  the final chain of spheres for the valency data ${\sigma\over \lambda}$ are 
\begin{equation*}
ma_0-ma_1-ma_2-\ldots-ma_{s-1}-m
\end{equation*}
So one get a star-shaped dual graph, with the central node having genus $g^{'}$ and all the other node having genus zero.

\item Then one can glue the dual graph together as follows.
Let's first assume $C_i$ to be non-amphidrome. Then $m^{(1)}=m^{(2)}=m$,  and one obtain  two sequences of integers $a_0>a_1>\ldots >a_u=1$ and 
$b_0>b_1>\ldots > b_v=1$. Graphically, one get two quiver tails from above sequence of integers.  Define an integer 
\begin{equation}
K=-s(C_i)-\delta^{(1)}/\lambda^{(1)}-\delta^{(2)}/\lambda^{(2)}
\end{equation}
where $\delta^{(j)}$ are integers such that $\sigma^{(j)}\delta^{(j)}=1 (mod \lambda^{(j)}$),~$0\leq \delta^{(j)}<\lambda^{(j)}-1$. If $\lambda^{(j)} =1$, one set $\delta^{(j)}=0$.  $K$ satisfies condition $K\geq -1$, as $s(C_i)<0, 0\leq \delta^{(j)}/\lambda^{(j)} < 1$. The gluing 
for the two quiver tails is defined as follows
\begin{enumerate}
\item If $K\geq 1$, then the glued tail looks as follows
\begin{equation*}
(ma_0, ma_1,\ldots, ma_u,\underbrace{m,m,\ldots,m}_{K-1},mb_v,\ldots,mb_1,mb_0)
\end{equation*} 
\item  If $K= 0$, then the glued tail looks as follows
\begin{equation*}
(ma_0, ma_1,\ldots, ma_{u-1},m,mb_{v-1},\ldots,mb_1,mb_0)
\end{equation*} 
\item Finally, if $K=-1$, then one can find $u_0<u$ and $v_0<v$ so that $a_{u_0}=b_{v_0}$, and 
$(a_{u_0-1}+b_{v_0-1})/a_{u_0}$ is an integer greater than one. Then the quiver tail looks like
\begin{equation*}
(ma_0,ma_1,\ldots, ma_{u_0},mb_{v_0-1},\ldots, mb_1,mb_0)
\end{equation*}
\end{enumerate}

Let's now assume $C_i$ to be amphidrome, then $C_i^{'}, C_i^{''}$ has valency data $(2m, \lambda, \sigma)$. Similarly, one has 
a sequence of integers $a_0>a_1>\ldots>a_u=1$, from which one can get a quiver tail. Then $K=-s(C_i)/2-\delta/\lambda$ is a non-negative integer where $\delta \sigma=1(mod \lambda)$. The glued quiver tail now has $u+K+2$ spheres, and it is a 
Dynkin diagram of $D$ type
\begin{equation*}
(2m a_0, 2ma_1,\ldots, 2m a_u, \underbrace{2m,\ldots, 2m}_K \text{(the tree part)},~m,m~\text{(the terminal part}))
\end{equation*}
\end{enumerate}

One can get a 3d quiver gauge theory from the dual graph: the multiplicities gave the gauge group $U(n_i)$ and the 
edges give the bi-fundamental matter. The quiver gauge theory determines wether the degeneration is allowed or not: it is allowed if the
Higgs branch dimension of it is equal to two (after contracting $-1$ curve and peeling off the quiver tail), see details in \cite{Xie:2023lko}, and this is not 
always possible. If it is an allowed degeneration, 
the (modified) quiver gives the 3d mirror for the IR  theory, which can then be used to determine the IR theory.

\textbf{Example}: Let's now give an example showing how to find the classification for a given weighted graph.  The weighted graph is taken to be $(g=2, r=2)$ (4(a) in figure. \ref{weightedgenustwo}), which means that there are two non-separating cutting curves for the genus two Riemann surface.  There are following situations 
that one need to consider (see table. \ref{periodicboun}): 
\begin{enumerate}
\item The action of $f$ on two cutting curves $C_1, C_2$ is non-amphidrome, and $f(C_1)=C_1, f(C_2)=C_2$. After the cut, one has a sphere with four marked points representing four boundaries, see figure, \ref{cut}. The periodic map fixed all the boundary curves, i.e. $f(C_1^{'})=C_1^{'}$, etc, and so 
the order of periodic map is one.  
\item The action of $f$ is  $f(C_1)=C_2, f(C_2)=C_1$. The periodic map on the fourth punctured sphere now acts as $f(C_1^{'})=C_2^{'}, f(C_1^{''})=C_2^{''}$. So the periodic map  has order two, 
and $(C_1^{'}, C_2^{'})$, $(C_1^{''}, C_2^{''})$ are in  $Z_2$ orbits.  The periodic map is then $f:\Sigma \to \Sigma^{'}$ with both $\Sigma$ and $\Sigma^{'}$ genus zero curve. To satisfy the Hurwitz formula, there must be two fixed points on $\Sigma$ whose valency data is $\frac{1}{2}$. 
Therefore the valency data on the punctured sphere is $(1)+(1)+\frac{1}{2}+\frac{1}{2}$, with the $1$ in the bracket indicating that the boundaries of two cutting curves are in the $Z_2$ orbit. 
\item  The action of $f$ on $C_1$ is amphidrome: $Amp(C_1), f(C_2)=C_2$. The periodic map on the fourth punctured sphere now acts as $f(C_1^{'})=C_1^{''}, f(C_2^{'})=C_2^{'}, f(C_2^{''})=C_2^{''}$.  This means that $C_1, C_1^{'}$ are in a $Z_2$ orbit, while $C_2^{''}, C_2^{'}$ are fixed points. 
So the periodic map on the fourth punctured sphere has order two, and the valency data is $(\textbf{1})+(\frac{1}{2})+(\frac{1}{2})$.
\item  The action of $f$ on $C_1$ and $C_2$ is amphidrome: $Amp(C_1), Amp(C_2)$. The periodic map on the fourth punctured sphere now acts as $f(C_1^{'})=C_1^{''}, f(C_2^{'})=C_2^{''}$.  This means that $C_1^{'}, C_1^{''}$ are in a $Z_2$ orbit, while $C_2^{'}, C_2^{''}$ are another $Z_2$ orbit. 
So the periodic map on the fourth punctured sphere has order two, and the valency data is $(\textbf{1})+(\textbf{1})+\frac{1}{2}+\frac{1}{2}$.
\item The action of $f$ on two cutting curves $C_1, C_2$ is non-amphidrome,  i.e. $Am(C_1, C_2)$. The  boundaries $C_1^{'}, C_1^{''}, C_2^{'}, C_2^{''}$ are in a $Z_4$ orbit. So the periodic map on the fourth punctured sphere has order four, 
and the valency data on the punctured sphere is $(\textbf{1})+(\textbf{1})+\frac{3}{4}+\frac{1}{4}$.
\end{enumerate}

\begin{figure}
\begin{center}

\tikzset{every picture/.style={line width=0.75pt}} %set default line width to 0.75pt        

\begin{tikzpicture}[x=0.50pt,y=0.50pt,yscale=-1,xscale=1]
%uncomment if require: \path (0,787); %set diagram left start at 0, and has height of 787

%Shape: Ellipse [id:dp9217056413675766] 
\draw   (120,135.5) .. controls (120,100.43) and (163.09,72) .. (216.25,72) .. controls (269.41,72) and (312.5,100.43) .. (312.5,135.5) .. controls (312.5,170.57) and (269.41,199) .. (216.25,199) .. controls (163.09,199) and (120,170.57) .. (120,135.5) -- cycle ;
%Shape: Ellipse [id:dp935860416872493] 
\draw   (162,132) .. controls (162,124.82) and (172.63,119) .. (185.75,119) .. controls (198.87,119) and (209.5,124.82) .. (209.5,132) .. controls (209.5,139.18) and (198.87,145) .. (185.75,145) .. controls (172.63,145) and (162,139.18) .. (162,132) -- cycle ;
%Shape: Ellipse [id:dp7690357647954402] 
\draw   (223,133) .. controls (223,125.82) and (233.63,120) .. (246.75,120) .. controls (259.87,120) and (270.5,125.82) .. (270.5,133) .. controls (270.5,140.18) and (259.87,146) .. (246.75,146) .. controls (233.63,146) and (223,140.18) .. (223,133) -- cycle ;
%Shape: Ellipse [id:dp04123473410935119] 
\draw  [color={rgb, 255:red, 208; green, 2; blue, 27 }  ,draw opacity=1 ] (118.5,133) .. controls (118.5,131.34) and (127.68,130) .. (139,130) .. controls (150.32,130) and (159.5,131.34) .. (159.5,133) .. controls (159.5,134.66) and (150.32,136) .. (139,136) .. controls (127.68,136) and (118.5,134.66) .. (118.5,133) -- cycle ;
%Shape: Ellipse [id:dp20926237352812838] 
\draw  [color={rgb, 255:red, 208; green, 2; blue, 27 }  ,draw opacity=1 ] (270.5,134) .. controls (270.5,132.34) and (279.68,131) .. (291,131) .. controls (302.32,131) and (311.5,132.34) .. (311.5,134) .. controls (311.5,135.66) and (302.32,137) .. (291,137) .. controls (279.68,137) and (270.5,135.66) .. (270.5,134) -- cycle ;
%Shape: Arc [id:dp18642284133348164] 
\draw  [draw opacity=0] (585.25,113.66) .. controls (585.28,85.77) and (620.81,63.1) .. (664.7,63) .. controls (706.32,62.91) and (740.52,83.14) .. (744.08,109.02) -- (664.82,113.59) -- cycle ; \draw   (585.25,113.66) .. controls (585.28,85.77) and (620.81,63.1) .. (664.7,63) .. controls (706.32,62.91) and (740.52,83.14) .. (744.08,109.02) ;  
%Shape: Ellipse [id:dp8944425404498453] 
\draw   (587,113) .. controls (587,110.24) and (596.74,108) .. (608.75,108) .. controls (620.76,108) and (630.5,110.24) .. (630.5,113) .. controls (630.5,115.76) and (620.76,118) .. (608.75,118) .. controls (596.74,118) and (587,115.76) .. (587,113) -- cycle ;
%Shape: Ellipse [id:dp8678577679132623] 
\draw   (701,109) .. controls (701,106.24) and (710.74,104) .. (722.75,104) .. controls (734.76,104) and (744.5,106.24) .. (744.5,109) .. controls (744.5,111.76) and (734.76,114) .. (722.75,114) .. controls (710.74,114) and (701,111.76) .. (701,109) -- cycle ;
%Shape: Arc [id:dp530103108091175] 
\draw  [draw opacity=0] (631.71,165.08) .. controls (635.69,161.05) and (638.5,151.12) .. (638.5,139.5) .. controls (638.5,126.24) and (634.83,115.18) .. (629.95,112.58) -- (627.75,139.5) -- cycle ; \draw   (631.71,165.08) .. controls (635.69,161.05) and (638.5,151.12) .. (638.5,139.5) .. controls (638.5,126.24) and (634.83,115.18) .. (629.95,112.58) ;  
%Shape: Ellipse [id:dp009747075204184164] 
\draw   (589,164) .. controls (589,161.24) and (598.74,159) .. (610.75,159) .. controls (622.76,159) and (632.5,161.24) .. (632.5,164) .. controls (632.5,166.76) and (622.76,169) .. (610.75,169) .. controls (598.74,169) and (589,166.76) .. (589,164) -- cycle ;
%Shape: Arc [id:dp5959123256345] 
\draw  [draw opacity=0] (699.68,166.46) .. controls (694.34,162.52) and (690.5,151.49) .. (690.5,138.5) .. controls (690.5,123.98) and (695.3,111.91) .. (701.62,109.46) -- (704,138.5) -- cycle ; \draw   (699.68,166.46) .. controls (694.34,162.52) and (690.5,151.49) .. (690.5,138.5) .. controls (690.5,123.98) and (695.3,111.91) .. (701.62,109.46) ;  
%Shape: Ellipse [id:dp6489927280418647] 
\draw   (699,166) .. controls (699,163.24) and (708.74,161) .. (720.75,161) .. controls (732.76,161) and (742.5,163.24) .. (742.5,166) .. controls (742.5,168.76) and (732.76,171) .. (720.75,171) .. controls (708.74,171) and (699,168.76) .. (699,166) -- cycle ;
%Shape: Arc [id:dp8438280364392976] 
\draw  [draw opacity=0] (742.39,167.75) .. controls (744.3,202.87) and (711.6,230.65) .. (669.3,229.83) .. controls (627.55,229.02) and (592.04,200.65) .. (589.17,166.14) -- (665.73,166.16) -- cycle ; \draw   (742.39,167.75) .. controls (744.3,202.87) and (711.6,230.65) .. (669.3,229.83) .. controls (627.55,229.02) and (592.04,200.65) .. (589.17,166.14) ;  
%Straight Lines [id:da6228803401692122] 
\draw    (392,132) -- (509.5,132.98) ;
\draw [shift={(511.5,133)}, rotate = 180.48] [color={rgb, 255:red, 0; green, 0; blue, 0 }  ][line width=0.75]    (10.93,-3.29) .. controls (6.95,-1.4) and (3.31,-0.3) .. (0,0) .. controls (3.31,0.3) and (6.95,1.4) .. (10.93,3.29)   ;

% Text Node
\draw (136,106.4) node [anchor=north west][inner sep=0.75pt]  [font=\tiny]  {$C_{1}$};
% Text Node
\draw (285,111.4) node [anchor=north west][inner sep=0.75pt]  [font=\tiny]  {$C_{2}$};
% Text Node
\draw (604,78.4) node [anchor=north west][inner sep=0.75pt]  [font=\tiny]  {$C_{1}^{'}$};
% Text Node
\draw (709,78.4) node [anchor=north west][inner sep=0.75pt]  [font=\tiny]  {$C_{2}^{'}$};
% Text Node
\draw (610,179.4) node [anchor=north west][inner sep=0.75pt]  [font=\tiny]  {$C_{1}^{''}$};
% Text Node
\draw (714,176.4) node [anchor=north west][inner sep=0.75pt]  [font=\tiny]  {$C_{2}^{''}$};

\end{tikzpicture}

\end{center}
\caption{Genus two curve has two non-separating cutting curves, and one get a fourth punctured sphere after the cutting.}
\label{cut}
\end{figure}
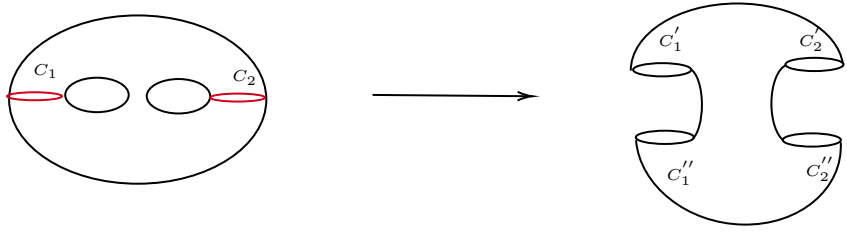

\subsection{IR theory}
 Let's now use the result of last subsection to classify the possible degeneration for rank two theories. The subtly is that one can not get 
 sensible physical interpretation for all the degenerations. To determine the IR theory, we use the link between the dual graph and 3d mirror, 
 and the basic assumption is that one  should get a consistent 3d mirror for a physical sensible degeneration, otherwise 
 we will not consider it.

\subsubsection{4d $\mathcal{N}=2$ SCFTs}
 The basic assumption to get a SCFT is following: a): one get consistent 3d mirror from the dual graph ; b): there is no variable link number $K$: the link number is truncated dual to 
 the consistency of 3d mirror; This excludes the weighted graph with separating curves, since if the gluing is possible, the link number is never 
 truncated; for the internal cutting curve, one has to consider the amphidrome cutting.
 
\begin{figure}[H]
\begin{center}

\tikzset{every picture/.style={line width=0.75pt}} %set default line width to 0.75pt        

\begin{tikzpicture}[x=0.45pt,y=0.45pt,yscale=-1,xscale=1]
%uncomment if require: \path (0,964); %set diagram left start at 0, and has height of 964

%Shape: Circle [id:dp8071340569819327] 
\draw   (85,70) .. controls (85,56.19) and (96.19,45) .. (110,45) .. controls (123.81,45) and (135,56.19) .. (135,70) .. controls (135,83.81) and (123.81,95) .. (110,95) .. controls (96.19,95) and (85,83.81) .. (85,70) -- cycle ;
%Straight Lines [id:da35628061605647465] 
\draw    (135,71) -- (171,71.22) ;
%Straight Lines [id:da26348406831492066] 
\draw    (170,71) -- (187,88.22) ;
%Straight Lines [id:da36519621330574203] 
\draw    (171,71) -- (190,58.22) ;
%Straight Lines [id:da8956521638886012] 
\draw    (261,70) -- (320,70.21) ;
\draw [shift={(322,70.22)}, rotate = 180.21] [color={rgb, 255:red, 0; green, 0; blue, 0 }  ][line width=0.75]    (10.93,-3.29) .. controls (6.95,-1.4) and (3.31,-0.3) .. (0,0) .. controls (3.31,0.3) and (6.95,1.4) .. (10.93,3.29)   ;
%Shape: Circle [id:dp19073457009081918] 
\draw   (380,75) .. controls (380,61.19) and (391.19,50) .. (405,50) .. controls (418.81,50) and (430,61.19) .. (430,75) .. controls (430,88.81) and (418.81,100) .. (405,100) .. controls (391.19,100) and (380,88.81) .. (380,75) -- cycle ;
%Straight Lines [id:da8311281642298418] 
\draw    (430,76) -- (466,76.22) ;
%Straight Lines [id:da5769219855861216] 
\draw    (261,172.22) -- (279,187.22) ;
%Straight Lines [id:da3589010521765861] 
\draw    (284,210) -- (259,231.22) ;
%Straight Lines [id:da7140127211780909] 
\draw    (309,197) -- (335,197.22) ;
%Straight Lines [id:da08992758743232798] 
\draw    (430,204) -- (444,214.22) ;
%Straight Lines [id:da24535678818942808] 
\draw    (431,192) -- (446,180.22) ;
%Straight Lines [id:da9556089580227871] 
\draw    (241,290.22) -- (259,305.22) ;
%Straight Lines [id:da6263450991245825] 
\draw    (264,328) -- (239,349.22) ;
%Straight Lines [id:da2590394125630595] 
\draw    (289,315) -- (315,315.22) ;
%Straight Lines [id:da7744587828834741] 
\draw    (334,326) -- (348,336.22) ;
%Straight Lines [id:da6241179834268451] 
\draw    (335,308) -- (350,296.22) ;
%Straight Lines [id:da2246068220614319] 
\draw    (269,407.22) -- (287,422.22) ;
%Straight Lines [id:da24550694941001194] 
\draw    (292,445) -- (267,466.22) ;
%Straight Lines [id:da7165674794644767] 
\draw    (317,432) -- (343,432.22) ;
%Straight Lines [id:da8915349755308906] 
\draw    (395,441) -- (409,451.22) ;
%Straight Lines [id:da7503042907202158] 
\draw    (397,430) -- (412,418.22) ;
%Straight Lines [id:da16289566258036237] 
\draw    (371,318.72) -- (428,318.72) ;
\draw [shift={(430,318.72)}, rotate = 180] [color={rgb, 255:red, 0; green, 0; blue, 0 }  ][line width=0.75]    (10.93,-3.29) .. controls (6.95,-1.4) and (3.31,-0.3) .. (0,0) .. controls (3.31,0.3) and (6.95,1.4) .. (10.93,3.29)   ;
%Straight Lines [id:da21315230591223644] 
\draw    (489,331) -- (503,341.22) ;
%Straight Lines [id:da7778697511794472] 
\draw    (490,313) -- (505,301.22) ;
%Curve Lines [id:da13701194339362566] 
\draw    (472,309.72) .. controls (469,287.72) and (482,252.72) .. (485,308.72) ;
%Straight Lines [id:da4560115548717554] 
\draw    (239,522.22) -- (257,537.22) ;
%Straight Lines [id:da7900028236554717] 
\draw    (262,560) -- (237,581.22) ;
%Straight Lines [id:da022197691342486392] 
\draw    (279,547.72) -- (313,547.22) ;
%Straight Lines [id:da9126112638980213] 
\draw    (332,558) -- (346,568.22) ;
%Straight Lines [id:da5841103165993071] 
\draw    (333,540) -- (348,528.22) ;
%Straight Lines [id:da4934501659488044] 
\draw    (369,550.72) -- (426,550.72) ;
\draw [shift={(428,550.72)}, rotate = 180] [color={rgb, 255:red, 0; green, 0; blue, 0 }  ][line width=0.75]    (10.93,-3.29) .. controls (6.95,-1.4) and (3.31,-0.3) .. (0,0) .. controls (3.31,0.3) and (6.95,1.4) .. (10.93,3.29)   ;
%Straight Lines [id:da4557168034137804] 
\draw    (487,563) -- (501,573.22) ;
%Straight Lines [id:da6019405801268467] 
\draw    (488,545) -- (503,533.22) ;
%Straight Lines [id:da5288763965453467] 
\draw    (447,549) -- (473,549.22) ;
%Straight Lines [id:da014523678621143121] 
\draw    (448,556) -- (474,556.22) ;
%Straight Lines [id:da9906232917336224] 
\draw    (269,643.22) -- (287,658.22) ;
%Straight Lines [id:da8075618203593864] 
\draw    (292,681) -- (267,702.22) ;
%Straight Lines [id:da8072358358226361] 
\draw    (317,668) -- (343,668.22) ;
%Straight Lines [id:da6443250947216057] 
\draw    (382,677) -- (396,687.22) ;
%Straight Lines [id:da36135401894895725] 
\draw    (382,666) -- (397,654.22) ;
%Straight Lines [id:da7680061156770912] 
\draw    (236,744.22) -- (254,759.22) ;
%Straight Lines [id:da8635221306191252] 
\draw    (259,782) -- (234,803.22) ;
%Straight Lines [id:da9528365035055095] 
\draw    (276,769.72) -- (310,769.22) ;
%Straight Lines [id:da758168330862816] 
\draw    (329,780) -- (343,790.22) ;
%Straight Lines [id:da31321732785247036] 
\draw    (330,762) -- (345,750.22) ;
%Straight Lines [id:da9734837315302283] 
\draw    (366,772.72) -- (423,772.72) ;
\draw [shift={(425,772.72)}, rotate = 180] [color={rgb, 255:red, 0; green, 0; blue, 0 }  ][line width=0.75]    (10.93,-3.29) .. controls (6.95,-1.4) and (3.31,-0.3) .. (0,0) .. controls (3.31,0.3) and (6.95,1.4) .. (10.93,3.29)   ;
%Straight Lines [id:da8666457481541718] 
\draw    (484,785) -- (498,795.22) ;
%Straight Lines [id:da033469459265931256] 
\draw    (485,767) -- (500,755.22) ;
%Straight Lines [id:da07344791642409798] 
\draw    (444,775) -- (470,775.22) ;
%Straight Lines [id:da6696536174029688] 
\draw    (434,787) -- (455,809.72) ;
%Straight Lines [id:da5655122694384827] 
\draw    (472,804.72) -- (480,789.72) ;
%Straight Lines [id:da6743416375221376] 
\draw    (230,843.22) -- (248,858.22) ;
%Straight Lines [id:da5425544260140822] 
\draw    (244,869) -- (218,868.72) ;
%Straight Lines [id:da5958740407629646] 
\draw    (270,868.72) -- (304,868.22) ;
%Straight Lines [id:da7854897805850176] 
\draw    (323,879) -- (337,889.22) ;
%Straight Lines [id:da29799748856623887] 
\draw    (324,861) -- (339,849.22) ;
%Straight Lines [id:da05395721868612213] 
\draw    (250,879.72) -- (226,900.72) ;

\draw    (10,910.72) -- (726,910.72) ;

\draw (105,63.4) node [anchor=north west][inner sep=0.75pt]     [font=\tiny]{$1$};
% Text Node
\draw (97,114.4) node [anchor=north west][inner sep=0.75pt]     [font=\tiny]{$l=2$};
% Text Node
\draw (150,51.4) node [anchor=north west][inner sep=0.75pt]     [font=\tiny]{$2$};
% Text Node
\draw (400,68.4) node [anchor=north west][inner sep=0.75pt]    [font=\tiny] {$1$};
% Text Node
\draw (445,56.4) node [anchor=north west][inner sep=0.75pt]    [font=\tiny] {$1$};
% Text Node
\draw (61,176.4) node [anchor=north west][inner sep=0.75pt]     [font=\tiny]{$\frac{5}{6} +\frac{2}{3} +\frac{1}{2}$};
% Text Node
\draw (282,187) node [anchor=north west][inner sep=0.75pt]    [font=\tiny][align=left] {12};
% Text Node
\draw (229,157.4) node [anchor=north west][inner sep=0.75pt]    [font=\tiny] {$4,8$};
% Text Node
\draw (235,225.4) node [anchor=north west][inner sep=0.75pt]   [font=\tiny]  {$6$};
% Text Node
\draw (340,189.4) node [anchor=north west][inner sep=0.75pt] [font=\tiny]{$10,8,6,4,2$};
% Text Node
\draw (446,163.4) node [anchor=north west][inner sep=0.75pt]    [font=\tiny] {$1$};
% Text Node
\draw (450,209.4) node [anchor=north west][inner sep=0.75pt]  [color={rgb, 255:red, 208; green, 2; blue, 27 }  ,opacity=1 ]   [font=\tiny]{$1$};
% Text Node
\draw (61,297.4) node [anchor=north west][inner sep=0.75pt]     [font=\tiny]{$\frac{1}{6} +\frac{1}{3} +\frac{1}{2}$};
% Text Node
\draw (262,305) node [anchor=north west][inner sep=0.75pt]    [font=\tiny][align=left] {12};
% Text Node
\draw (218,275.4) node [anchor=north west][inner sep=0.75pt]     [font=\tiny]{$4$};
% Text Node
\draw (215,343.4) node [anchor=north west][inner sep=0.75pt]     [font=\tiny]{$6$};
% Text Node
\draw (320,307.4) node [anchor=north west][inner sep=0.75pt]    [font=\tiny] {$2$};
% Text Node
\draw (357,283.4) node [anchor=north west][inner sep=0.75pt]    [font=\tiny] {$1$};
% Text Node
\draw (359,330.4) node [anchor=north west][inner sep=0.75pt]  [color={rgb, 255:red, 0; green, 0; blue, 0 }  ,opacity=1 ]  [font=\tiny] {$1$};
% Text Node
\draw (56,415.4) node [anchor=north west][inner sep=0.75pt]     [font=\tiny]{$\frac{3}{4} +\frac{3}{4} +\frac{1}{2}$};
% Text Node
\draw (290,422) node [anchor=north west][inner sep=0.75pt]    [font=\tiny][align=left] {8};
% Text Node
\draw (215,392.4) node [anchor=north west][inner sep=0.75pt]    [font=\tiny] {$2,4,6$};
% Text Node
\draw (243,460.4) node [anchor=north west][inner sep=0.75pt]    [font=\tiny] {$4$};
% Text Node
\draw (348,424.4) node [anchor=north west][inner sep=0.75pt]    [font=\tiny] {$6,4,2$};
% Text Node
\draw (415,400.4) node [anchor=north west][inner sep=0.75pt]    [font=\tiny] {$1$};
% Text Node
\draw (416,448.4) node [anchor=north west][inner sep=0.75pt]  [color={rgb, 255:red, 208; green, 2; blue, 27 }  ,opacity=1 ]   [font=\tiny]{$1$};
% Text Node
\draw (473,313.4) node [anchor=north west][inner sep=0.75pt]    [font=\tiny] {$2$};
% Text Node
\draw (512,288.4) node [anchor=north west][inner sep=0.75pt]    [font=\tiny] {$1$};
% Text Node
\draw (514,335.4) node [anchor=north west][inner sep=0.75pt]  [color={rgb, 255:red, 208; green, 2; blue, 27 }  ,opacity=1 ]   [font=\tiny]{$1$};
% Text Node
\draw (59,529.4) node [anchor=north west][inner sep=0.75pt]    [font=\tiny] {$\frac{1}{4} +\frac{1}{4} +\frac{1}{2}$};
% Text Node
\draw (260,537) node [anchor=north west][inner sep=0.75pt]   [font=\tiny] [align=left] {8};
% Text Node
\draw (216,507.4) node [anchor=north west][inner sep=0.75pt]    [font=\tiny] {$2$};
% Text Node
\draw (213,575.4) node [anchor=north west][inner sep=0.75pt]    [font=\tiny] {$4$};
% Text Node
\draw (318,539.4) node [anchor=north west][inner sep=0.75pt]    [font=\tiny] {$2$};
% Text Node
\draw (355,515.4) node [anchor=north west][inner sep=0.75pt]    [font=\tiny] {$1$};
% Text Node
\draw (357,562.4) node [anchor=north west][inner sep=0.75pt]  [color={rgb, 255:red, 0; green, 0; blue, 0 }  ,opacity=1 ]  [font=\tiny] {$1$};
% Text Node
\draw (477,545.4) node [anchor=north west][inner sep=0.75pt]    [font=\tiny] {$2$};
% Text Node
\draw (510,520.4) node [anchor=north west][inner sep=0.75pt]    [font=\tiny] {$1$};
% Text Node
\draw (512,567.4) node [anchor=north west][inner sep=0.75pt]  [color={rgb, 255:red, 208; green, 2; blue, 27 }  ,opacity=1 ]  [font=\tiny] {$1$};
% Text Node
\draw (436,545.4) node [anchor=north west][inner sep=0.75pt]    [font=\tiny] {$2$};
% Text Node
\draw (56,651.4) node [anchor=north west][inner sep=0.75pt]    [font=\tiny] {$\frac{2}{3} +\frac{2}{3} +\frac{2}{3}$};
% Text Node
\draw (292,659) node [anchor=north west][inner sep=0.75pt]   [font=\tiny] [align=left] {6};
% Text Node
\draw (233,631.4) node [anchor=north west][inner sep=0.75pt]    [font=\tiny] {$2,4$};
% Text Node
\draw (348,660.4) node [anchor=north west][inner sep=0.75pt]   [font=\tiny]  {$4,2$};
% Text Node
\draw (404,637.4) node [anchor=north west][inner sep=0.75pt]   [font=\tiny]  {$1$};
% Text Node
\draw (403,687.4) node [anchor=north west][inner sep=0.75pt]  [color={rgb, 255:red, 208; green, 2; blue, 27 }  ,opacity=1 ] [font=\tiny]  {$1$};
% Text Node
\draw (233,691.4) node [anchor=north west][inner sep=0.75pt]    [font=\tiny] {$2,4$};
% Text Node
\draw (59,744.4) node [anchor=north west][inner sep=0.75pt]    [font=\tiny] {$\frac{1}{3} +\frac{1}{3} +\frac{1}{3}$};
% Text Node
\draw (257,759) node [anchor=north west][inner sep=0.75pt]   [font=\tiny] [align=left] {6};
% Text Node
\draw (213,729.4) node [anchor=north west][inner sep=0.75pt]     [font=\tiny]{$2$};
% Text Node
\draw (210,797.4) node [anchor=north west][inner sep=0.75pt]    [font=\tiny] {$2$};
% Text Node
\draw (315,761.4) node [anchor=north west][inner sep=0.75pt]    [font=\tiny] {$2$};
% Text Node
\draw (352,737.4) node [anchor=north west][inner sep=0.75pt]    [font=\tiny] {$1$};
% Text Node
\draw (354,784.4) node [anchor=north west][inner sep=0.75pt]  [color={rgb, 255:red, 0; green, 0; blue, 0 }  ,opacity=1 ]   [font=\tiny]{$1$};
% Text Node
\draw (474,767.4) node [anchor=north west][inner sep=0.75pt]     [font=\tiny]{$2$};
% Text Node
\draw (507,742.4) node [anchor=north west][inner sep=0.75pt]    [font=\tiny] {$1$};
% Text Node
\draw (509,789.4) node [anchor=north west][inner sep=0.75pt]  [color={rgb, 255:red, 208; green, 2; blue, 27 }  ,opacity=1 ]   [font=\tiny]{$1$};
% Text Node
\draw (433,767.4) node [anchor=north west][inner sep=0.75pt]    [font=\tiny] {$2$};
% Text Node
\draw (458,804.4) node [anchor=north west][inner sep=0.75pt]    [font=\tiny] {$2$};
% Text Node
\draw (53,843.4) node [anchor=north west][inner sep=0.75pt]    [font=\tiny] {$\frac{1}{2} +\frac{1}{2} +\frac{1}{2} +\frac{1}{2}$};
% Text Node
\draw (251,858) node [anchor=north west][inner sep=0.75pt]   [font=\tiny][align=left] {4};
% Text Node
\draw (207,828.4) node [anchor=north west][inner sep=0.75pt]     [font=\tiny]{$2$};
% Text Node
\draw (202,862.4) node [anchor=north west][inner sep=0.75pt]   [font=\tiny]  {$2$};
% Text Node
\draw (309,860.4) node [anchor=north west][inner sep=0.75pt]    [font=\tiny] {$2$};
% Text Node
\draw (346,836.4) node [anchor=north west][inner sep=0.75pt]   [font=\tiny]  {$1$};
% Text Node
\draw (348,883.4) node [anchor=north west][inner sep=0.75pt]  [color={rgb, 255:red, 208; green, 2; blue, 27 }  ,opacity=1 ]   [font=\tiny]{$1$};
% Text Node
\draw (206,895.4) node [anchor=north west][inner sep=0.75pt]    [font=\tiny] {$2$};

\end{tikzpicture}

\tikzset{every picture/.style={line width=0.75pt}} %set default line width to 0.75pt        

\begin{tikzpicture}[x=0.45pt,y=0.45pt,yscale=-1,xscale=1]
%uncomment if require: \path (0,964); %set diagram left start at 0, and has height of 964

%Shape: Circle [id:dp8071340569819327] 
\draw   (85,70) .. controls (85,56.19) and (96.19,45) .. (110,45) .. controls (123.81,45) and (135,56.19) .. (135,70) .. controls (135,83.81) and (123.81,95) .. (110,95) .. controls (96.19,95) and (85,83.81) .. (85,70) -- cycle ;
%Straight Lines [id:da39147409646347464] 
\draw    (312,186.22) -- (329,198.22) ;
%Straight Lines [id:da08837461543191805] 
\draw    (327,214) -- (311,228.22) ;
%Straight Lines [id:da9888659192046503] 
\draw    (363,203.22) -- (347,203.22) ;
%Straight Lines [id:da9714858620624287] 
\draw    (403,183) -- (387,197.22) ;
%Straight Lines [id:da38562207258154757] 
\draw    (406,220.22) -- (391,210.22) ;
%Straight Lines [id:da2878250902936317] 
\draw    (325,342.22) -- (342,354.22) ;
%Straight Lines [id:da10171440397276243] 
\draw    (340,370) -- (324,384.22) ;
%Straight Lines [id:da8444644862500605] 
\draw    (376,359.22) -- (360,359.22) ;
%Straight Lines [id:da1894487078988285] 
\draw    (421,335) -- (405,349.22) ;
%Straight Lines [id:da42173451355456715] 
\draw    (421,379.22) -- (406,369.22) ;

\draw (105,63.4) node [anchor=north west][inner sep=0.75pt]     [font=\tiny]{$2$};
% Text Node
\draw (97,114.4) node [anchor=north west][inner sep=0.75pt]     [font=\tiny]{$r=1$};
% Text Node
\draw (65,189.4) node [anchor=north west][inner sep=0.75pt]    [font=\tiny] {$\frac{3}{4} +\frac{3}{4} +\left(\frac{1}{2}\right)$};
% Text Node
\draw (331,196.4) node [anchor=north west][inner sep=0.75pt]   [font=\tiny]  {$4$};
% Text Node
\draw (264,168.4) node [anchor=north west][inner sep=0.75pt]   [font=\tiny]  {$1,2,3$};
% Text Node
\draw (262,227.4) node [anchor=north west][inner sep=0.75pt]    [font=\tiny] {$1,2,3$};
% Text Node
\draw (371,195.4) node [anchor=north west][inner sep=0.75pt]    [font=\tiny] {$2$};
% Text Node
\draw (410,169.4) node [anchor=north west][inner sep=0.75pt]    [font=\tiny] {$1$};
% Text Node
\draw (412,212.4) node [anchor=north west][inner sep=0.75pt]  [color={rgb, 255:red, 208; green, 2; blue, 27 }  ,opacity=1 ]  [font=\tiny] {$1$};
% Text Node
\draw (73,323.4) node [anchor=north west][inner sep=0.75pt]     [font=\tiny]{$\frac{5}{6} +\frac{1}{2} +\left(\frac{2}{3}\right)$};
% Text Node
\draw (344,352.4) node [anchor=north west][inner sep=0.75pt]    [font=\tiny] {$6$};
% Text Node
\draw (251,322.4) node [anchor=north west][inner sep=0.75pt]    [font=\tiny] {$1,2,3,4,5$};
% Text Node
\draw (305,382.4) node [anchor=north west][inner sep=0.75pt]    [font=\tiny] {$3$};
% Text Node
\draw (375,351.4) node [anchor=north west][inner sep=0.75pt]    [font=\tiny] {$4,2$};
% Text Node
\draw (423,325.4) node [anchor=north west][inner sep=0.75pt]    [font=\tiny] {$1$};
% Text Node
\draw (429,373.4) node [anchor=north west][inner sep=0.75pt]  [color={rgb, 255:red, 208; green, 2; blue, 27 }  ,opacity=1 ]  [font=\tiny] {$1$};

\end{tikzpicture}

\end{center}
\caption{Upper: dual graph for SCFTs from weighted graph 2b). Here one first contract $-1$ curve and then peel off the red tail. These give the rank two $H_i$ type theories.
Bottom:  dual graph for SCFTs from weighted graph 3a), and one also peel off the red tail. The 3d mirror for those theories were found in \cite{Xie:2012hs}.}
\label{scft1}
\end{figure}
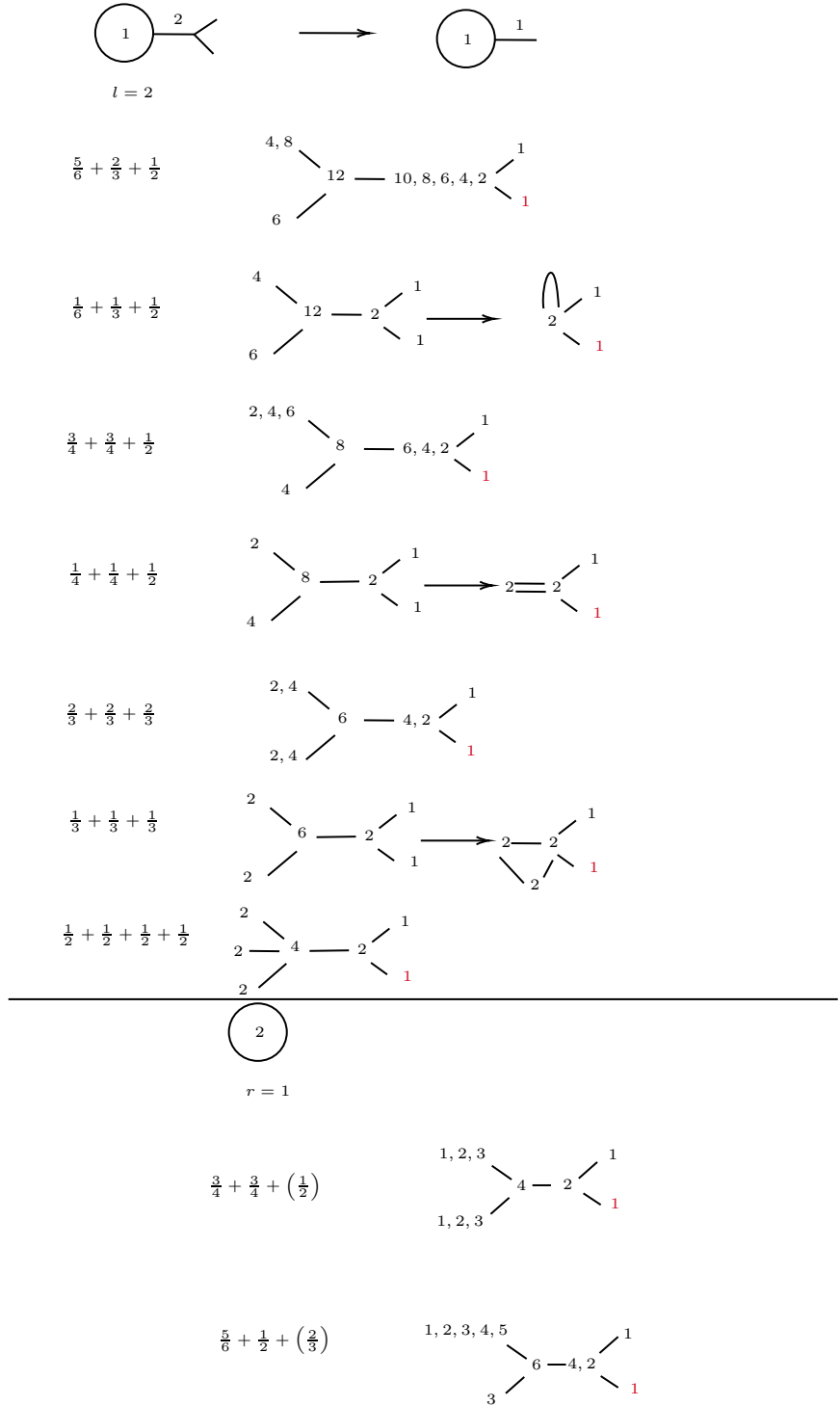

One find the following possibilities by looking at the weighted graph in figure.\ref{weightedgenustwo}:
 \begin{itemize}
\item  The weighted graph 1a) in figure. \ref{weightedgenustwo} has no cut curve and so there is no variable link number K. The only constraint would be that the modification of the mirror quiver is possible.
By looking at the list of genus two periodic map, we find that
 the degeneration $(\frac{9}{10}+\bm{\frac{3}{5}}+\frac{1}{2}),~(\frac{3}{10}+\bm{\frac{1}{5}}+\frac{1}{2})$,~$(\frac{5}{6}+\frac{5}{6}+\bm{\frac{1}{3}}),~(\frac{2}{3}+\bm{\frac{1}{3}}+\frac{1}{2}+\frac{1}{2})$ are not good.
 The bad valency data is in put in bold letter. Others give SCFT, see \cite{Xie:2022aad}.
\item The weighted graph 2b) in figure. \ref{weightedgenustwo} would give SCFT.  The basic data for the periodic map is $(g=1, r=0, k=1)$.
 The corresponding dual graph are listed in figure. \ref{scft1}. Here we need to first do the contraction of $-1$ curve \cite{Xie:2023lko}, and then use the modification procedure to get 
 the 3d mirror.
\item The weighted graph is 3a) in figure. \ref{weightedgenustwo},  and the cut curve is taken to be amphidrome. The basic building block of the periodic map is $(g=2, r=1)$,
 See figure. \ref{scft1} for the dual graphs.
\end{itemize}

\subsubsection{Non-abelian gauge theory}
Let's now consider the degeneration whose low energy theory is IR free non-abelian gauge theory. The rank two gauge group could be $A_2, A_1\times A_1, B_2(=C_2), G_2$. 
One can find non-abelian gauge groups from weighted graph as follows \cite{Xie:2023lko}: 1) if there is a weight $n$ edges connecting two components in the weighted graph, the gauge group would be $SU(n)$; 2) The internal cut is amphidrome with weight $m$ ($m$  cutting curves will be mapped to each other by the mapping group action), the 
gauge group would be $Sp(2m)$. 

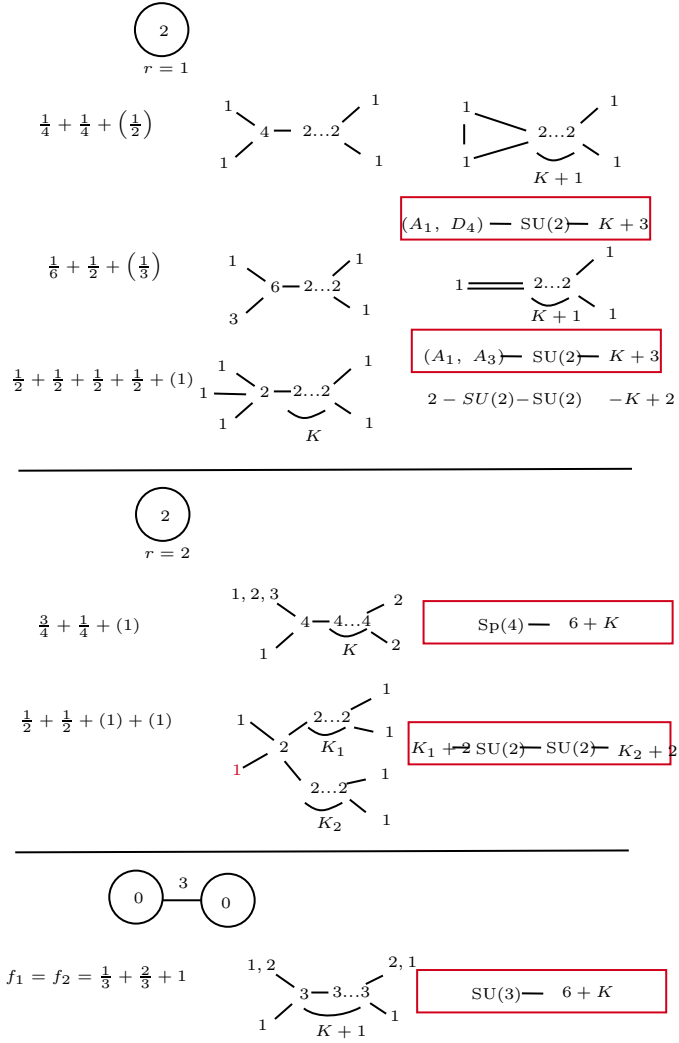
\begin{figure}[htbp]
\begin{center}

\tikzset{every picture/.style={line width=0.75pt}} %set default line width to 0.75pt        

\begin{tikzpicture}[x=0.40pt,y=0.40pt,yscale=-1,xscale=1]
%uncomment if require: \path (0,1110); %set diagram left start at 0, and has height of 1110

%Shape: Circle [id:dp8071340569819327] 
\draw   (150,67) .. controls (150,53.19) and (161.19,42) .. (175,42) .. controls (188.81,42) and (200,53.19) .. (200,67) .. controls (200,80.81) and (188.81,92) .. (175,92) .. controls (161.19,92) and (150,80.81) .. (150,67) -- cycle ;
%Straight Lines [id:da39147409646347464] 
\draw    (245,145.22) -- (262,157.22) ;
%Straight Lines [id:da08837461543191805] 
\draw    (260,173) -- (244,187.22) ;
%Straight Lines [id:da9888659192046503] 
\draw    (296,162.22) -- (280,162.22) ;
%Straight Lines [id:da9714858620624287] 
\draw    (360,140) -- (344,154.22) ;
%Straight Lines [id:da38562207258154757] 
\draw    (361,184.22) -- (346,174.22) ;
%Straight Lines [id:da2878250902936317] 
\draw    (255,292.22) -- (272,304.22) ;
%Straight Lines [id:da10171440397276243] 
\draw    (270,320) -- (254,334.22) ;
%Straight Lines [id:da8444644862500605] 
\draw    (304,309.22) -- (288,309.22) ;
%Straight Lines [id:da1894487078988285] 
\draw    (351,285) -- (335,299.22) ;
%Straight Lines [id:da42173451355456715] 
\draw    (351,329.22) -- (336,319.22) ;
%Straight Lines [id:da06408380551045645] 
\draw    (41,483) -- (615,481.22) ;
%Straight Lines [id:da5777178280978403] 
\draw    (245,391.22) -- (262,403.22) ;
%Straight Lines [id:da0396133935220675] 
\draw    (260,419) -- (244,433.22) ;
%Straight Lines [id:da7583396862674683] 
\draw    (296,408.22) -- (280,408.22) ;
%Straight Lines [id:da740375692751855] 
\draw    (352,387) -- (336,401.22) ;
%Straight Lines [id:da20276323531967377] 
\draw    (352,429.22) -- (337,419.22) ;
%Straight Lines [id:da2432968184405384] 
\draw    (254,411) -- (224,410.22) ;
%Straight Lines [id:da14553529677143362] 
\draw    (468,146.22) -- (516,162.22) ;
%Straight Lines [id:da42988489283773745] 
\draw    (518,173.22) -- (467,188.22) ;
%Straight Lines [id:da6920246561704639] 
\draw    (583,141) -- (567,155.22) ;
%Straight Lines [id:da4289706437812959] 
\draw    (584,185.22) -- (569,175.22) ;
%Straight Lines [id:da7964462960104184] 
\draw    (458,175.22) -- (458,156.22) ;
%Curve Lines [id:da18319825959776548] 
\draw    (526,183.22) .. controls (539,193.22) and (544,192.22) .. (561,183.22) ;
%Straight Lines [id:da5822336539540398] 
\draw    (462,306.22) -- (513,306.22) ;
%Straight Lines [id:da9816997846750064] 
\draw    (514,311.22) -- (461,311.22) ;
%Straight Lines [id:da24440757222446852] 
\draw    (579,284) -- (563,298.22) ;
%Straight Lines [id:da022843450352642902] 
\draw    (580,328.22) -- (565,318.22) ;
%Curve Lines [id:da4039808828374549] 
\draw    (521,320.22) .. controls (534,330.22) and (539,329.22) .. (556,320.22) ;
%Curve Lines [id:da878715186999595] 
\draw    (293,426.22) .. controls (306,436.22) and (311,435.22) .. (328,426.22) ;
%Shape: Circle [id:dp2593585801500409] 
\draw   (151,524.22) .. controls (151,510.41) and (162.19,499.22) .. (176,499.22) .. controls (189.81,499.22) and (201,510.41) .. (201,524.22) .. controls (201,538.03) and (189.81,549.22) .. (176,549.22) .. controls (162.19,549.22) and (151,538.03) .. (151,524.22) -- cycle ;
%Straight Lines [id:da44425019668908816] 
\draw    (283,608.22) -- (300,620.22) ;
%Straight Lines [id:da2607323025937318] 
\draw    (298,636) -- (282,650.22) ;
%Straight Lines [id:da34699623837079874] 
\draw    (332,625.22) -- (316,625.22) ;
%Straight Lines [id:da4370056938868401] 
\draw    (383,609.22) -- (367,617.22) ;
%Straight Lines [id:da9407965875347213] 
\draw    (386,645.22) -- (371,635.22) ;
%Straight Lines [id:da0394618390708279] 
\draw    (258,720.22) -- (280,737.22) ;
%Straight Lines [id:da4458410219145055] 
\draw    (290,757) -- (305,775.22) ;
%Straight Lines [id:da7091785435114661] 
\draw    (311,720.22) -- (294,732.22) ;
%Straight Lines [id:da28479713912757876] 
\draw    (371,698.22) -- (352,708.22) ;
%Straight Lines [id:da7287781573641372] 
\draw    (373,729.22) -- (355,725.22) ;
%Straight Lines [id:da9147294873010643] 
\draw    (274,750) -- (251,763.22) ;
%Straight Lines [id:da14065014545708254] 
\draw    (366,774.22) -- (348,778.22) ;
%Straight Lines [id:da32955157350373643] 
\draw    (366,805.22) -- (351,793.22) ;
%Straight Lines [id:da04788895822406247] 
\draw    (554,250) -- (574,250.22) ;
%Straight Lines [id:da187445319224252] 
\draw    (482,250) -- (502,250.22) ;
%Shape: Rectangle [id:dp20506659653458592] 
\draw  [color={rgb, 255:red, 208; green, 2; blue, 27 }  ,draw opacity=1 ] (399,225) -- (632,225) -- (632,265) -- (399,265) -- cycle ;
%Straight Lines [id:da8045815604570676] 
\draw    (564,375) -- (584,375.22) ;
%Straight Lines [id:da5423139599241087] 
\draw    (492,375) -- (512,375.22) ;
%Shape: Rectangle [id:dp9501608256569372] 
\draw  [color={rgb, 255:red, 208; green, 2; blue, 27 }  ,draw opacity=1 ] (409,350) -- (642,350) -- (642,390) -- (409,390) -- cycle ;
%Straight Lines [id:da2697585525624717] 
\draw    (518,628) -- (538,628.22) ;
%Shape: Rectangle [id:dp3091024192851972] 
\draw  [color={rgb, 255:red, 208; green, 2; blue, 27 }  ,draw opacity=1 ] (420,606) -- (653,606) -- (653,646) -- (420,646) -- cycle ;
%Curve Lines [id:da4999481653551341] 
\draw    (332,632.22) .. controls (345,642.22) and (350,641.22) .. (367,632.22) ;
%Straight Lines [id:da5852250183063614] 
\draw    (577,744) -- (593,744.22) ;
%Straight Lines [id:da9419108472048157] 
\draw    (510,744) -- (530,744.22) ;
%Shape: Rectangle [id:dp9355538540696129] 
\draw  [color={rgb, 255:red, 208; green, 2; blue, 27 }  ,draw opacity=1 ] (406,720.22) -- (651,720.22) -- (651,760) -- (406,760) -- cycle ;
%Curve Lines [id:da5059022339188586] 
\draw    (312,725.22) .. controls (325,735.22) and (330,734.22) .. (347,725.22) ;
%Curve Lines [id:da5182941719970149] 
\draw    (309,796.22) .. controls (322,806.22) and (327,805.22) .. (344,796.22) ;
%Straight Lines [id:da7023229590578681] 
\draw    (447,744) -- (463,744.22) ;
%Straight Lines [id:da7779996114358174] 
\draw    (38,843) -- (612,841.22) ;
%Shape: Circle [id:dp8870110334631234] 
\draw   (126,885.22) .. controls (126,871.41) and (137.19,860.22) .. (151,860.22) .. controls (164.81,860.22) and (176,871.41) .. (176,885.22) .. controls (176,899.03) and (164.81,910.22) .. (151,910.22) .. controls (137.19,910.22) and (126,899.03) .. (126,885.22) -- cycle ;
%Straight Lines [id:da6191841804074283] 
\draw    (176,888.22) -- (212,888.22) ;
%Shape: Circle [id:dp04854881780038145] 
\draw   (212,888.22) .. controls (212,874.41) and (223.19,863.22) .. (237,863.22) .. controls (250.81,863.22) and (262,874.41) .. (262,888.22) .. controls (262,902.03) and (250.81,913.22) .. (237,913.22) .. controls (223.19,913.22) and (212,902.03) .. (212,888.22) -- cycle ;
%Straight Lines [id:da06872870451283553] 
\draw    (282,958.22) -- (299,970.22) ;
%Straight Lines [id:da9265043271229907] 
\draw    (297,986) -- (281,1000.22) ;
%Straight Lines [id:da5485446978605488] 
\draw    (331,975.22) -- (315,975.22) ;
%Straight Lines [id:da8913765575043606] 
\draw    (382,959.22) -- (366,967.22) ;
%Straight Lines [id:da818488445985029] 
\draw    (385,995.22) -- (370,985.22) ;
%Curve Lines [id:da44595471377255147] 
\draw    (308,990.22) .. controls (321,1000.22) and (344,998.22) .. (361,989.22) ;
%Straight Lines [id:da9616071835205975] 
\draw    (512,976) -- (532,976.22) ;
%Shape: Rectangle [id:dp3115643542714268] 
\draw  [color={rgb, 255:red, 208; green, 2; blue, 27 }  ,draw opacity=1 ] (414,954) -- (647,954) -- (647,994) -- (414,994) -- cycle ;

\draw (170,60.4) node [anchor=north west][inner sep=0.75pt]   [font=\tiny]  {$2$};
% Text Node
\draw (155,95.4) node [anchor=north west][inner sep=0.75pt]    [font=\tiny] {$r=1$};
% Text Node
\draw (56,140.4) node [anchor=north west][inner sep=0.75pt]    [font=\tiny] {$\frac{1}{4} +\frac{1}{4} +\left(\frac{1}{2}\right)$};
% Text Node
\draw (264,155.4) node [anchor=north west][inner sep=0.75pt]   [font=\tiny]  {$4$};
% Text Node
\draw (230,131.4) node [anchor=north west][inner sep=0.75pt]   [font=\tiny]  {$1$};
% Text Node
\draw (226,185.4) node [anchor=north west][inner sep=0.75pt]   [font=\tiny]  {$1$};
% Text Node
\draw (304,154.4) node [anchor=north west][inner sep=0.75pt]   [font=\tiny]  {$2...2$};
% Text Node
\draw (368,125.4) node [anchor=north west][inner sep=0.75pt]   [font=\tiny]  {$1$};
% Text Node
\draw (371,183.4) node [anchor=north west][inner sep=0.75pt]  [color={rgb, 255:red, 0; green, 0; blue, 0 }  ,opacity=1 ]  [font=\tiny] {$1$};
% Text Node
\draw (64,274.4) node [anchor=north west][inner sep=0.75pt]    [font=\tiny] {$\frac{1}{6} +\frac{1}{2} +\left(\frac{1}{3}\right)$};
% Text Node
\draw (274,302.4) node [anchor=north west][inner sep=0.75pt]  [font=\tiny]   {$6$};
% Text Node
\draw (234,277.4) node [anchor=north west][inner sep=0.75pt]  [font=\tiny]   {$1$};
% Text Node
\draw (235,332.4) node [anchor=north west][inner sep=0.75pt]   [font=\tiny]  {$3$};
% Text Node
\draw (305,301.4) node [anchor=north west][inner sep=0.75pt]    [font=\tiny] {$2...2$};
% Text Node
\draw (353,275.4) node [anchor=north west][inner sep=0.75pt]   [font=\tiny]  {$1$};
% Text Node
\draw (359,323.4) node [anchor=north west][inner sep=0.75pt]  [color={rgb, 255:red, 0; green, 0; blue, 0 }  ,opacity=1 ]  [font=\tiny] {$1$};
% Text Node
\draw (31,384.4) node [anchor=north west][inner sep=0.75pt]   [font=\tiny]  {$\frac{1}{2} +\frac{1}{2} +\frac{1}{2} +\frac{1}{2} +( 1)$};
% Text Node
\draw (264,401.4) node [anchor=north west][inner sep=0.75pt]   [font=\tiny]  {$2$};
% Text Node
\draw (224,376.4) node [anchor=north west][inner sep=0.75pt]    [font=\tiny] {$1$};
% Text Node
\draw (225,431.4) node [anchor=north west][inner sep=0.75pt]  [font=\tiny]   {$1$};
% Text Node
\draw (295,400.4) node [anchor=north west][inner sep=0.75pt]  [font=\tiny]   {$2...2$};
% Text Node
\draw (361,371.4) node [anchor=north west][inner sep=0.75pt]    [font=\tiny] {$1$};
% Text Node
\draw (362,430.4) node [anchor=north west][inner sep=0.75pt]  [color={rgb, 255:red, 0; green, 0; blue, 0 }  ,opacity=1 ]  [font=\tiny] {$1$};
% Text Node
\draw (207,402.4) node [anchor=north west][inner sep=0.75pt]  [color={rgb, 255:red, 0; green, 0; blue, 0 }  ,opacity=1 ] [font=\tiny]  {$1$};
% Text Node
\draw (453,132.4) node [anchor=north west][inner sep=0.75pt]   [font=\tiny]  {$1$};
% Text Node
\draw (453,184.4) node [anchor=north west][inner sep=0.75pt]   [font=\tiny]  {$1$};
% Text Node
\draw (524,155.4) node [anchor=north west][inner sep=0.75pt]  [font=\tiny]   {$2...2$};
% Text Node
\draw (591,126.4) node [anchor=north west][inner sep=0.75pt]   [font=\tiny]  {$1$};
% Text Node
\draw (594,184.4) node [anchor=north west][inner sep=0.75pt]  [color={rgb, 255:red, 0; green, 0; blue, 0 }  ,opacity=1 ] [font=\tiny]  {$1$};
% Text Node
\draw (519,198.62) node [anchor=north west][inner sep=0.75pt]   [font=\tiny]  {$K+1$};
% Text Node
\draw (446,300.4) node [anchor=north west][inner sep=0.75pt]   [font=\tiny]  {$1$};
% Text Node
\draw (520,298.4) node [anchor=north west][inner sep=0.75pt]   [font=\tiny]  {$2...2$};
% Text Node
\draw (587,269.4) node [anchor=north west][inner sep=0.75pt]    [font=\tiny] {$1$};
% Text Node
\draw (590,327.4) node [anchor=north west][inner sep=0.75pt]  [color={rgb, 255:red, 0; green, 0; blue, 0 }  ,opacity=1 ] [font=\tiny]  {$1$};
% Text Node
\draw (517,328.62) node [anchor=north west][inner sep=0.75pt]  [font=\tiny]   {$K+1$};
% Text Node
\draw (306,441.62) node [anchor=north west][inner sep=0.75pt]   [font=\tiny]  {$K$};
% Text Node
\draw (171,517.62) node [anchor=north west][inner sep=0.75pt]  [font=\tiny]   {$2$};
% Text Node
\draw (156,552.62) node [anchor=north west][inner sep=0.75pt]   [font=\tiny]  {$r=2$};
% Text Node
\draw (55,616.4) node [anchor=north west][inner sep=0.75pt]  [font=\tiny]   {$\frac{3}{4} +\frac{1}{4} +( 1)$};
% Text Node
\draw (302,618.4) node [anchor=north west][inner sep=0.75pt]   [font=\tiny]  {$4$};
% Text Node
\draw (237,591.4) node [anchor=north west][inner sep=0.75pt]  [font=\tiny]   {$1,2,3$};
% Text Node
\draw (263,648.4) node [anchor=north west][inner sep=0.75pt]   [font=\tiny]  {$1$};
% Text Node
\draw (333,617.4) node [anchor=north west][inner sep=0.75pt] [font=\tiny]    {$4...4$};
% Text Node
\draw (389,597.4) node [anchor=north west][inner sep=0.75pt]  [font=\tiny]   {$2$};
% Text Node
\draw (387,639.4) node [anchor=north west][inner sep=0.75pt]  [color={rgb, 255:red, 0; green, 0; blue, 0 }  ,opacity=1 ] [font=\tiny]  {$2$};
% Text Node
\draw (39,706.4) node [anchor=north west][inner sep=0.75pt]    [font=\tiny] {$\frac{1}{2} +\frac{1}{2} +( 1) +( 1)$};
% Text Node
\draw (282,735.4) node [anchor=north west][inner sep=0.75pt] [font=\tiny]    {$2$};
% Text Node
\draw (242,710.4) node [anchor=north west][inner sep=0.75pt] [font=\tiny]    {$1$};
% Text Node
\draw (314,707.4) node [anchor=north west][inner sep=0.75pt]   [font=\tiny]  {$2...2$};
% Text Node
\draw (378,681.4) node [anchor=north west][inner sep=0.75pt]    [font=\tiny] {$1$};
% Text Node
\draw (381,721.4) node [anchor=north west][inner sep=0.75pt]  [color={rgb, 255:red, 0; green, 0; blue, 0 }  ,opacity=1 ] [font=\tiny]  {$1$};
% Text Node
\draw (238,757.4) node [anchor=north west][inner sep=0.75pt]  [color={rgb, 255:red, 208; green, 2; blue, 27 }  ,opacity=1 ]  [font=\tiny] {$1$};
% Text Node
\draw (311,774.4) node [anchor=north west][inner sep=0.75pt]   [font=\tiny]  {$2...2$};
% Text Node
\draw (377,761.4) node [anchor=north west][inner sep=0.75pt]   [font=\tiny]  {$1$};
% Text Node
\draw (378,804.4) node [anchor=north west][inner sep=0.75pt]  [color={rgb, 255:red, 0; green, 0; blue, 0 }  ,opacity=1 ]  [font=\tiny] {$1$};
% Text Node
\draw (508,241) node [anchor=north west][inner sep=0.75pt]  [font=\tiny]  [align=left]  [font=\tiny] {SU(2)};
% Text Node
\draw (580,241.62) node [anchor=north west][inner sep=0.75pt]  [font=\tiny]   {$K+3$};
% Text Node
\draw (390,240.62) node [anchor=north west][inner sep=0.75pt]    [font=\tiny] {$\ ( A_{1} ,\ D_{4})$};
% Text Node
\draw (519,366) node [anchor=north west][inner sep=0.75pt] [font=\tiny]   [align=left] {SU(2)};
% Text Node
\draw (590,366.62) node [anchor=north west][inner sep=0.75pt]   [font=\tiny]  {$K+3$};
% Text Node
\draw (411,365.62) node [anchor=north west][inner sep=0.65pt]  [font=\tiny]   {$\ ( A_{1} ,\ A_{3})$};
% Text Node

% Text Node
\draw (519,406) node [anchor=north west][inner sep=0.75pt] [font=\tiny]   [align=left] [font=\tiny]  {SU(2)};
% Text Node
\draw (590,406.62) node [anchor=north west][inner sep=0.75pt]   [font=\tiny]  {$-K+2$};
% Text Node
\draw (421,405.62) node [anchor=north west][inner sep=0.75pt]  [font=\tiny]   {$2-SU(2)-$};

\draw (468,619) node [anchor=north west][inner sep=0.75pt]  [font=\tiny]  [align=left] {Sp(4)};
% Text Node
\draw (553,618.62) node [anchor=north west][inner sep=0.75pt]  [font=\tiny]   {$6+K$};
% Text Node
\draw (340,643.62) node [anchor=north west][inner sep=0.75pt]  [font=\tiny]   {$K$};
% Text Node
\draw (532,735) node [anchor=north west][inner sep=0.75pt]    [font=\tiny] {SU(2)};
% Text Node
\draw (598,738.62) node [anchor=north west][inner sep=0.75pt]  [font=\tiny]  {$K_{2} +2$};
% Text Node
\draw (320,735.62) node [anchor=north west][inner sep=0.75pt]    [font=\tiny] {$K_{1}$};
% Text Node
\draw (317,806.62) node [anchor=north west][inner sep=0.75pt]   [font=\tiny]  {$K_{2}$};
% Text Node
\draw (466,736) node [anchor=north west][inner sep=0.75pt]  [font=\tiny]  {SU(2)};
% Text Node
\draw (405,736.62) node [anchor=north west][inner sep=0.75pt]  [font=\tiny]  {$K_{1} +2$};
% Text Node
\draw (146,878.62) node [anchor=north west][inner sep=0.75pt]  [font=\tiny]   {$0$};
% Text Node
\draw (228,882.62) node [anchor=north west][inner sep=0.75pt]  [font=\tiny]   {$0$};
% Text Node
\draw (188,864.62) node [anchor=north west][inner sep=0.75pt]  [font=\tiny]   {$3$};
% Text Node
\draw (26,947.62) node [anchor=north west][inner sep=0.75pt]   [font=\tiny]  {$f_{1} =f_{2} =\frac{1}{3} +\frac{2}{3} +1$};
% Text Node
\draw (301,968.4) node [anchor=north west][inner sep=0.75pt]   [font=\tiny]  {$3$};
% Text Node
\draw (252,941.4) node [anchor=north west][inner sep=0.75pt]   [font=\tiny]  {$1,2$};
% Text Node
\draw (262,998.4) node [anchor=north west][inner sep=0.75pt]   [font=\tiny]  {$1$};
% Text Node
\draw (332,967.4) node [anchor=north west][inner sep=0.75pt]  [font=\tiny]   {$3...3$};
% Text Node
\draw (384,938.4) node [anchor=north west][inner sep=0.75pt]   [font=\tiny]  {$2,1$};
% Text Node
\draw (386,989.4) node [anchor=north west][inner sep=0.75pt]  [color={rgb, 255:red, 0; green, 0; blue, 0 }  ,opacity=1 ] [font=\tiny]  {$1$};
% Text Node
\draw (316,1003.62) node [anchor=north west][inner sep=0.75pt]   [font=\tiny]  {$K+1$};
% Text Node
\draw (462,967) node [anchor=north west][inner sep=0.75pt]   [align=left][font=\tiny]  {SU(3)};
% Text Node
\draw (547,966.62) node [anchor=north west][inner sep=0.75pt]   [font=\tiny]  {$6+K$};

\end{tikzpicture}

\end{center}
\caption{The dual graph for non-abelian gauge theory. The lower bound for $K$ is determined so that the theory is IR free.}
\label{IRfree}
\end{figure}

By looking at all the weighted graphs in figure. \ref{weightedgenustwo}, one find the following possibilities: 
1) weighted graph 3a), and the cut is taken to be non-amphidrome; the gauge group is $SU(2)$ and is coupled with 
a rank one SCFT.  2):  weighted graph 4a), and the gauge group is ($Sp(4)$ (weight 2), and the gauge group is $SU(2)\times SU(2)$ (for weight $(1,1)$). 3) weighted graph 5e), and the gauge group is $SU(3)$, see figure. \ref{IRfree}.

\textbf{Example}: For the weighted graph 4a in figure. \ref{weightedgenustwo}, one can get non-abelian gauge theory for following two situations. Assuming the cut curves are $C_1, C_2$: a): the action of the mapping class group is $Am(C_1), Am(C_2)$, and 
the gauge group is $SU(2)\times SU(2)$, and there is one $SU(2)$ associated with cutting; b) the action acts as $(Am(C_1, C_2))$, and the gauge group is $Sp(4)$. 

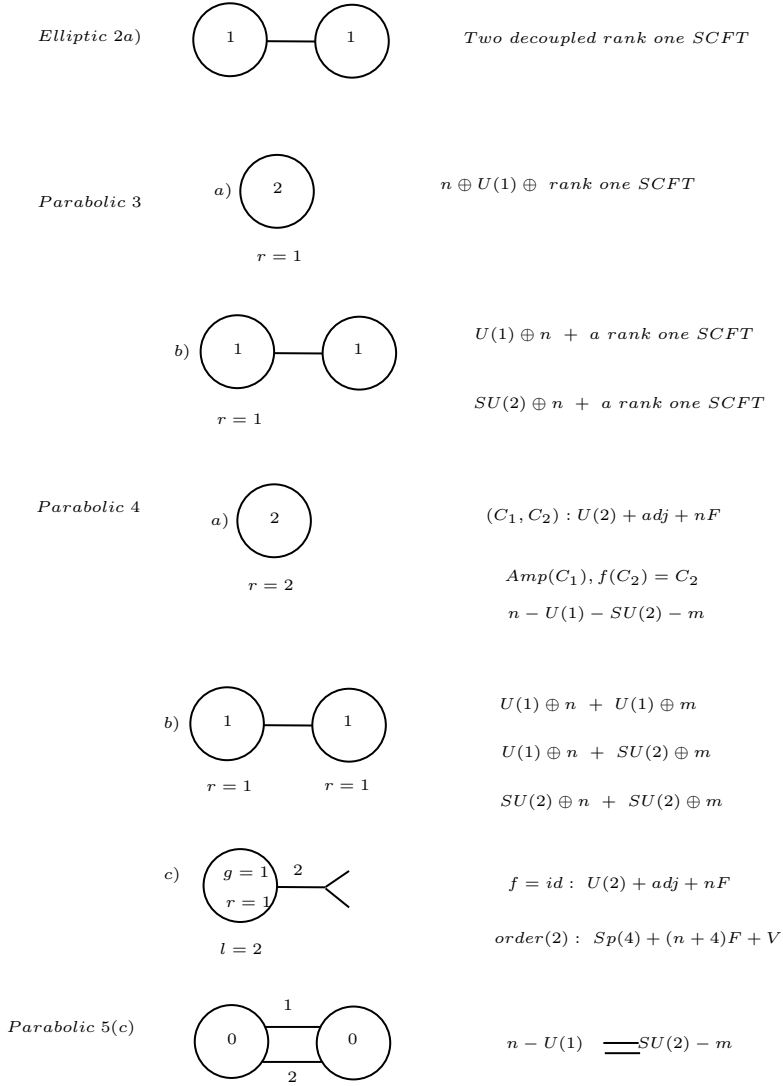
\begin{figure}[htbp]
\begin{center}

\tikzset{every picture/.style={line width=0.75pt}} %set default line width to 0.75pt        

\begin{tikzpicture}[x=0.55pt,y=0.55pt,yscale=-1,xscale=1]
%uncomment if require: \path (0,964); %set diagram left start at 0, and has height of 964

%Shape: Circle [id:dp636883219161509] 
\draw   (139,76) .. controls (139,62.19) and (150.19,51) .. (164,51) .. controls (177.81,51) and (189,62.19) .. (189,76) .. controls (189,89.81) and (177.81,101) .. (164,101) .. controls (150.19,101) and (139,89.81) .. (139,76) -- cycle ;
%Straight Lines [id:da24461399398604344] 
\draw    (189,77) -- (222,77.22) ;
%Shape: Circle [id:dp4717026108770106] 
\draw   (222,77) .. controls (222,63.19) and (233.19,52) .. (247,52) .. controls (260.81,52) and (272,63.19) .. (272,77) .. controls (272,90.81) and (260.81,102) .. (247,102) .. controls (233.19,102) and (222,90.81) .. (222,77) -- cycle ;
%Shape: Circle [id:dp3722316564037935] 
\draw   (171,180) .. controls (171,166.19) and (182.19,155) .. (196,155) .. controls (209.81,155) and (221,166.19) .. (221,180) .. controls (221,193.81) and (209.81,205) .. (196,205) .. controls (182.19,205) and (171,193.81) .. (171,180) -- cycle ;
%Shape: Circle [id:dp5653702836286971] 
\draw   (144,290) .. controls (144,276.19) and (155.19,265) .. (169,265) .. controls (182.81,265) and (194,276.19) .. (194,290) .. controls (194,303.81) and (182.81,315) .. (169,315) .. controls (155.19,315) and (144,303.81) .. (144,290) -- cycle ;
%Straight Lines [id:da2778039449414129] 
\draw    (194,291) -- (227,291.22) ;
%Shape: Circle [id:dp8931766986013527] 
\draw   (227,291) .. controls (227,277.19) and (238.19,266) .. (252,266) .. controls (265.81,266) and (277,277.19) .. (277,291) .. controls (277,304.81) and (265.81,316) .. (252,316) .. controls (238.19,316) and (227,304.81) .. (227,291) -- cycle ;
%Shape: Circle [id:dp40012599593601084] 
\draw   (169,406) .. controls (169,392.19) and (180.19,381) .. (194,381) .. controls (207.81,381) and (219,392.19) .. (219,406) .. controls (219,419.81) and (207.81,431) .. (194,431) .. controls (180.19,431) and (169,419.81) .. (169,406) -- cycle ;
%Shape: Circle [id:dp9534643807281578] 
\draw   (137,545.22) .. controls (137,531.41) and (148.19,520.22) .. (162,520.22) .. controls (175.81,520.22) and (187,531.41) .. (187,545.22) .. controls (187,559.03) and (175.81,570.22) .. (162,570.22) .. controls (148.19,570.22) and (137,559.03) .. (137,545.22) -- cycle ;
%Straight Lines [id:da3111091579963874] 
\draw    (187,546.22) -- (220,546.44) ;
%Shape: Circle [id:dp8254466545236119] 
\draw   (220,546.22) .. controls (220,532.41) and (231.19,521.22) .. (245,521.22) .. controls (258.81,521.22) and (270,532.41) .. (270,546.22) .. controls (270,560.03) and (258.81,571.22) .. (245,571.22) .. controls (231.19,571.22) and (220,560.03) .. (220,546.22) -- cycle ;
%Shape: Circle [id:dp38538650491727666] 
\draw   (146,656.22) .. controls (146,642.41) and (157.19,631.22) .. (171,631.22) .. controls (184.81,631.22) and (196,642.41) .. (196,656.22) .. controls (196,670.03) and (184.81,681.22) .. (171,681.22) .. controls (157.19,681.22) and (146,670.03) .. (146,656.22) -- cycle ;
%Straight Lines [id:da5779492473474155] 
\draw    (196,657.22) -- (229,657.44) ;
%Straight Lines [id:da37642323511947917] 
\draw    (229,657.22) -- (245,646.22) ;
%Straight Lines [id:da46982548602378715] 
\draw    (245,671.22) -- (229,658.22) ;
%Shape: Circle [id:dp21486727051921872] 
\draw   (140,763.22) .. controls (140,749.41) and (151.19,738.22) .. (165,738.22) .. controls (178.81,738.22) and (190,749.41) .. (190,763.22) .. controls (190,777.03) and (178.81,788.22) .. (165,788.22) .. controls (151.19,788.22) and (140,777.03) .. (140,763.22) -- cycle ;
%Straight Lines [id:da12601556605903586] 
\draw    (188,753.22) -- (226,753.22) ;
%Shape: Circle [id:dp912986635686043] 
\draw   (223,764.22) .. controls (223,750.41) and (234.19,739.22) .. (248,739.22) .. controls (261.81,739.22) and (273,750.41) .. (273,764.22) .. controls (273,778.03) and (261.81,789.22) .. (248,789.22) .. controls (234.19,789.22) and (223,778.03) .. (223,764.22) -- cycle ;
%Straight Lines [id:da5448636786962331] 
\draw    (186,777.22) -- (227,777.22) ;
%Straight Lines [id:da5708835974028568] 
\draw    (418,764.22) -- (442,764.22) ;
%Straight Lines [id:da24932733712365684] 
\draw    (419,771.22) -- (443,771.22) ;

\draw (159,68.4) node [anchor=north west][inner sep=0.75pt]   [font=\tiny]  {$1$};
% Text Node
\draw (241,68.4) node [anchor=north west][inner sep=0.75pt]    [font=\tiny] {$1$};
% Text Node
\draw (321,68.4) node [anchor=north west][inner sep=0.75pt]    [font=\tiny] {$Two\ decoupled\ rank\ one\ SCFT$};
% Text Node
\draw (31,66.4) node [anchor=north west][inner sep=0.75pt]  [font=\tiny]   {$Elliptic\ 2 a)$};
% Text Node
\draw (191,172.4) node [anchor=north west][inner sep=0.75pt]  [font=\tiny]   {$2$};
% Text Node
\draw (151,172.4) node [anchor=north west][inner sep=0.75pt]   [font=\tiny]  {$a)$};
% Text Node

\draw (180,218.4) node [anchor=north west][inner sep=0.75pt]    [font=\tiny] {$r=1$};
% Text Node
\draw (31,181.4) node [anchor=north west][inner sep=0.75pt]    [font=\tiny] {$Parabolic\ 3$};
% Text Node
\draw (305,168.4) node [anchor=north west][inner sep=0.75pt]   [font=\tiny]  {$n\oplus U( 1) \oplus \ rank\ one\ SCFT$};
% Text Node
\draw (164,282.4) node [anchor=north west][inner sep=0.75pt]   [font=\tiny]  {$1$};
% Text Node
\draw (124,282.4) node [anchor=north west][inner sep=0.75pt]  [font=\tiny]   {$b)$};
% Text Node

\draw (246,282.4) node [anchor=north west][inner sep=0.75pt]    [font=\tiny] {$1$};
% Text Node
\draw (153,330.4) node [anchor=north west][inner sep=0.75pt]    [font=\tiny] {$r=1$};
% Text Node
\draw (324,271.4) node [anchor=north west][inner sep=0.75pt]    [font=\tiny] {$\ U( 1) \oplus n\ +\ a\ rank\ one\ SCFT$};
% Text Node
\draw (323,318.4) node [anchor=north west][inner sep=0.75pt]    [font=\tiny] {$\ SU( 2) \oplus n\ +\ a\ rank\ one\ SCFT$};
% Text Node
\draw (30,390.4) node [anchor=north west][inner sep=0.75pt]    [font=\tiny] {$Parabolic\ 4$};
% Text Node
\draw (189,398.4) node [anchor=north west][inner sep=0.75pt]    [font=\tiny] {$2$};
% Text Node
\draw (149,398.4) node [anchor=north west][inner sep=0.75pt]   [font=\tiny]  {$a)$};
% Text Node

\draw (174,444.4) node [anchor=north west][inner sep=0.75pt]  [font=\tiny]   {$r=2$};
% Text Node
\draw (336,395.4) node [anchor=north west][inner sep=0.75pt]    [font=\tiny] {$( C_{1} ,C_{2}) :U( 2) +adj+nF$};
% Text Node
\draw (349,437.4) node [anchor=north west][inner sep=0.75pt]    [font=\tiny] {$Amp( C_{1}) ,f( C_{2}) =C_{2}$};
% Text Node
\draw (10,746.4) node [anchor=north west][inner sep=0.75pt]    [font=\tiny] {$Parabolic\ 5 (c)$};
% Text Node
\draw (157,537.62) node [anchor=north west][inner sep=0.75pt]   [font=\tiny]  {$1$};
% Text Node
\draw (117,537.62) node [anchor=north west][inner sep=0.75pt]   [font=\tiny]  {$b)$};
% Text Node

\draw (239,537.62) node [anchor=north west][inner sep=0.75pt]    [font=\tiny] {$1$};
% Text Node
\draw (146,582.4) node [anchor=north west][inner sep=0.75pt]    [font=\tiny] {$r=1$};
% Text Node
\draw (226,581.4) node [anchor=north west][inner sep=0.75pt]    [font=\tiny] {$r=1$};
% Text Node
\draw (157,641.62) node [anchor=north west][inner sep=0.75pt]  [font=\tiny]  {$g=1$};
% Text Node
\draw (117,641.62) node [anchor=north west][inner sep=0.75pt]    [font=\tiny]{$c)$};

\draw (205,639.62) node [anchor=north west][inner sep=0.75pt]   [font=\tiny]  {$2$};
% Text Node
\draw (155,693.39) node [anchor=north west][inner sep=0.75pt]  [rotate=-0.02]   [font=\tiny]{$l=2$};
% Text Node
\draw (341,524.62) node [anchor=north west][inner sep=0.75pt]   [font=\tiny]  {$\ U( 1) \oplus n\ +\ U( 1) \oplus m$};
% Text Node
\draw (342,557.62) node [anchor=north west][inner sep=0.75pt]    [font=\tiny] {$\ U( 1) \oplus n\ +\ SU( 2) \oplus m$};
% Text Node
\draw (341,591.62) node [anchor=north west][inner sep=0.75pt]    [font=\tiny] {$\ SU( 2) \oplus n\ +\ SU( 2) \oplus m$};
% Text Node
\draw (351,647.62) node [anchor=north west][inner sep=0.75pt]    [font=\tiny] {$f=id:\ U( 2) +adj+nF$};
% Text Node
\draw (159,661.62) node [anchor=north west][inner sep=0.75pt]  [font=\tiny]  {$r=1$};
% Text Node
\draw (342,685.62) node [anchor=north west][inner sep=0.75pt]   [font=\tiny]  {$order( 2) :\ Sp( 4) +( n+4) F+V$};
% Text Node
\draw (351,462.4) node [anchor=north west][inner sep=0.75pt]    [font=\tiny] {$n-U( 1) -SU( 2) -m$};
% Text Node
\draw (160,755.62) node [anchor=north west][inner sep=0.75pt]   [font=\tiny]  {$0$};
% Text Node
\draw (242,755.62) node [anchor=north west][inner sep=0.75pt]    [font=\tiny] {$0$};
% Text Node
\draw (201,781.62) node [anchor=north west][inner sep=0.75pt]    [font=\tiny] {$2$};
% Text Node
\draw (198,732.62) node [anchor=north west][inner sep=0.75pt]   [font=\tiny]  {$1$};
% Text Node
\draw (350,757.62) node [anchor=north west][inner sep=0.75pt]   [font=\tiny]  {$n-U( 1) \ \ \ \ \ \ \ \  SU( 2) -m$};

\end{tikzpicture}

\end{center}
\caption{Weighted graph and periodic map for IR free theories involving abelian gauge group.}
\label{abelian}
\end{figure}

\newpage
\subsubsection{Other cases: abelian gauge theory}
The IR theory for other cases involves decoupled systems: if the edge in the weighted graph has multiplicity one, the gauge group is $SU(1)$ and so 
the two adjacent system is decoupled. The appearance  of abelian gauge group: this happens if the internal cut is non-amphidrome: for 
an internal cut with multiplicity $n$, the gauge group is $U(n)$. The full list is:
\begin{itemize}
\item Weighted graph 2a): two decoupled rank one SCFTs, plus possible uncharged hypermultiplets;
\item  Weighted graph 3a) and the cut is taken to be non-amphidrome. The IR theory is $U(1)$ gauge group coupled with $n$ free hypermultiplets plus 
a rank one SCFT.
\item  Weighted graph 3b). If the cut is amphidrome, the IR theory is $SU(2)$ gauge group coupled with $n$ fundamental hypermultiplets plus a rank one SCFT;
If the cut is non-amphidrome, the IR theory is $U(1)$ gauge group coupled with $n$ free hypermultipelts plus a rank one SCFT.
\item The IR theory for weighted graph 4a), 4b), 4c) and 5c) are listed in figure. \ref{abelian}.
\end{itemize}

\begin{figure}
\begin{center}

\tikzset{every picture/.style={line width=0.75pt}} %set default line width to 0.75pt        

\begin{tikzpicture}[x=0.55pt,y=0.55pt,yscale=-1,xscale=1]
%uncomment if require: \path (0,787); %set diagram left start at 0, and has height of 787

%Straight Lines [id:da832523046094823] 
\draw    (265.5,159) -- (289.5,159) ;
%Straight Lines [id:da5063382583363892] 
\draw    (457.5,158) -- (481.5,158) ;
%Straight Lines [id:da9308211085521897] 
\draw    (305,173) -- (322.5,185) ;
%Straight Lines [id:da9060601394827821] 
\draw    (426,190) -- (438.5,171) ;
%Straight Lines [id:da5149131758033297] 
\draw    (304.5,147) -- (315.5,135) ;
%Straight Lines [id:da2958330543282617] 
\draw    (439,147) -- (421.5,134) ;
%Shape: Circle [id:dp022868615517726787] 
\draw  [fill={rgb, 255:red, 0; green, 0; blue, 0 }  ,fill opacity=1 ] (339,131.25) .. controls (339,129.46) and (340.46,128) .. (342.25,128) .. controls (344.04,128) and (345.5,129.46) .. (345.5,131.25) .. controls (345.5,133.04) and (344.04,134.5) .. (342.25,134.5) .. controls (340.46,134.5) and (339,133.04) .. (339,131.25) -- cycle ;
%Shape: Circle [id:dp5516726637561605] 
\draw  [fill={rgb, 255:red, 0; green, 0; blue, 0 }  ,fill opacity=1 ] (365,131.25) .. controls (365,129.46) and (366.46,128) .. (368.25,128) .. controls (370.04,128) and (371.5,129.46) .. (371.5,131.25) .. controls (371.5,133.04) and (370.04,134.5) .. (368.25,134.5) .. controls (366.46,134.5) and (365,133.04) .. (365,131.25) -- cycle ;
%Shape: Circle [id:dp4085397305412104] 
\draw  [fill={rgb, 255:red, 0; green, 0; blue, 0 }  ,fill opacity=1 ] (388,132.25) .. controls (388,130.46) and (389.46,129) .. (391.25,129) .. controls (393.04,129) and (394.5,130.46) .. (394.5,132.25) .. controls (394.5,134.04) and (393.04,135.5) .. (391.25,135.5) .. controls (389.46,135.5) and (388,134.04) .. (388,132.25) -- cycle ;

%Shape: Circle [id:dp976600046619709] 
\draw  [fill={rgb, 255:red, 0; green, 0; blue, 0 }  ,fill opacity=1 ] (347,194.25) .. controls (347,192.46) and (348.46,191) .. (350.25,191) .. controls (352.04,191) and (353.5,192.46) .. (353.5,194.25) .. controls (353.5,196.04) and (352.04,197.5) .. (350.25,197.5) .. controls (348.46,197.5) and (347,196.04) .. (347,194.25) -- cycle ;
%Shape: Circle [id:dp07855524084059229] 
\draw  [fill={rgb, 255:red, 0; green, 0; blue, 0 }  ,fill opacity=1 ] (373,194.25) .. controls (373,192.46) and (374.46,191) .. (376.25,191) .. controls (378.04,191) and (379.5,192.46) .. (379.5,194.25) .. controls (379.5,196.04) and (378.04,197.5) .. (376.25,197.5) .. controls (374.46,197.5) and (373,196.04) .. (373,194.25) -- cycle ;
%Shape: Circle [id:dp0792058970319095] 
\draw  [fill={rgb, 255:red, 0; green, 0; blue, 0 }  ,fill opacity=1 ] (396,195.25) .. controls (396,193.46) and (397.46,192) .. (399.25,192) .. controls (401.04,192) and (402.5,193.46) .. (402.5,195.25) .. controls (402.5,197.04) and (401.04,198.5) .. (399.25,198.5) .. controls (397.46,198.5) and (396,197.04) .. (396,195.25) -- cycle ;

%Shape: Ellipse [id:dp4869993558271203] 
\draw   (725,175) .. controls (725,161.19) and (760.93,150) .. (805.25,150) .. controls (849.57,150) and (885.5,161.19) .. (885.5,175) .. controls (885.5,188.81) and (849.57,200) .. (805.25,200) .. controls (760.93,200) and (725,188.81) .. (725,175) -- cycle ;
%Straight Lines [id:da35756348399478055] 
\draw    (724.25,122) -- (724.13,148.99) -- (723.75,232) ;
%Straight Lines [id:da659238342104159] 
\draw    (886.25,125) -- (886.13,151.99) -- (885.75,235) ;
%Curve Lines [id:da6775403200656526] 
\draw    (724.5,187) .. controls (727,209) and (867.5,224) .. (886.5,188) ;
%Straight Lines [id:da6009862206045898] 
\draw [color={rgb, 255:red, 208; green, 2; blue, 27 }  ,draw opacity=1 ] [dash pattern={on 4.5pt off 4.5pt}]  (878,124) -- (879.5,247) ;
%Straight Lines [id:da05057611490604508] 
\draw [color={rgb, 255:red, 208; green, 2; blue, 27 }  ,draw opacity=1 ] [dash pattern={on 4.5pt off 4.5pt}]  (737,124) -- (738.5,247) ;
%Straight Lines [id:da7855231648336732] 
\draw    (771.5,120) -- (771.75,185) ;
%Straight Lines [id:da3258191019695259] 
\draw    (801.5,120) -- (801.75,185) ;
%Straight Lines [id:da7324539154549131] 
\draw    (834.5,120) -- (834.75,185) ;
%Straight Lines [id:da9743170642205393] 
\draw    (757.5,177) -- (757.75,242) ;
%Straight Lines [id:da9963723822581341] 
\draw    (796.5,188) -- (796.75,253) ;
%Straight Lines [id:da426539482861912] 
\draw    (852.5,184) -- (852.75,249) ;
%Shape: Ellipse [id:dp5583514768031362] 
\draw   (500,432) .. controls (500,418.19) and (535.93,407) .. (580.25,407) .. controls (624.57,407) and (660.5,418.19) .. (660.5,432) .. controls (660.5,445.81) and (624.57,457) .. (580.25,457) .. controls (535.93,457) and (500,445.81) .. (500,432) -- cycle ;
%Straight Lines [id:da28378896701109324] 
\draw  [dash pattern={on 4.5pt off 4.5pt}]  (499.25,379) -- (499.13,405.99) -- (498.75,489) ;
%Straight Lines [id:da5506725125823438] 
\draw  [dash pattern={on 4.5pt off 4.5pt}]  (661.25,382) -- (661.13,408.99) -- (660.75,492) ;
%Curve Lines [id:da8981675966765574] 
\draw    (499.5,444) .. controls (502,466) and (642.5,481) .. (661.5,445) ;
%Straight Lines [id:da5709693110201751] 
\draw [color={rgb, 255:red, 208; green, 2; blue, 27 }  ,draw opacity=1 ]   (648,383) -- (649.5,506) ;
%Straight Lines [id:da012503351848467625] 
\draw [color={rgb, 255:red, 208; green, 2; blue, 27 }  ,draw opacity=1 ]   (512,381) -- (513.5,504) ;
%Straight Lines [id:da24438098891591398] 
\draw  [dash pattern={on 4.5pt off 4.5pt}]  (546.5,377) -- (546.75,442) ;
%Straight Lines [id:da016793078672437622] 
\draw  [dash pattern={on 4.5pt off 4.5pt}]  (576.5,377) -- (576.75,442) ;
%Straight Lines [id:da5445670639130603] 
\draw  [dash pattern={on 4.5pt off 4.5pt}]  (609.5,377) -- (609.75,442) ;
%Straight Lines [id:da0033322967394343594] 
\draw  [dash pattern={on 4.5pt off 4.5pt}]  (531.5,434) -- (531.75,499) ;
%Straight Lines [id:da08859945014191639] 
\draw  [dash pattern={on 4.5pt off 4.5pt}]  (571.5,445) -- (571.75,510) ;
%Straight Lines [id:da2255578126728145] 
\draw  [dash pattern={on 4.5pt off 4.5pt}]  (627.5,441) -- (627.75,506) ;
%Straight Lines [id:da40827980988904833] 
\draw    (790,284) -- (714.12,338.83) ;
\draw [shift={(712.5,340)}, rotate = 324.15] [color={rgb, 255:red, 0; green, 0; blue, 0 }  ][line width=0.75]    (10.93,-3.29) .. controls (6.95,-1.4) and (3.31,-0.3) .. (0,0) .. controls (3.31,0.3) and (6.95,1.4) .. (10.93,3.29)   ;

% Text Node
\draw (294,149.4) node [anchor=north west][inner sep=0.75pt]    {$2$};
% Text Node
\draw (441,150.4) node [anchor=north west][inner sep=0.75pt]    {$2$};
% Text Node
\draw (251,150.4) node [anchor=north west][inner sep=0.75pt]    {$1$};
% Text Node
\draw (485,150.4) node [anchor=north west][inner sep=0.75pt]    {$1$};
% Text Node
\draw (329,184.4) node [anchor=north west][inner sep=0.75pt]    {$2$};
% Text Node
\draw (409,185.4) node [anchor=north west][inner sep=0.75pt]    {$2$};
% Text Node
\draw (319,122.4) node [anchor=north west][inner sep=0.75pt]    {$1$};
% Text Node
\draw (405,122.4) node [anchor=north west][inner sep=0.75pt]    {$1$};
% Text Node
\draw (687,112.4) node [anchor=north west][inner sep=0.75pt]  [font=\footnotesize]  {$NS5$};
% Text Node
\draw (881,104.4) node [anchor=north west][inner sep=0.75pt]  [font=\footnotesize]  {$NS5$};
% Text Node
\draw (726,254.4) node [anchor=north west][inner sep=0.75pt]  [font=\footnotesize]  {$D5$};
% Text Node
\draw (496,514.4) node [anchor=north west][inner sep=0.75pt]  [font=\footnotesize]  {$NS5$};
% Text Node
\draw (656,361.4) node [anchor=north west][inner sep=0.75pt]  [font=\footnotesize]  {$D5$};
% Text Node
\draw (484,347.4) node [anchor=north west][inner sep=0.75pt]  [font=\footnotesize]  {$D5$};
% Text Node
\draw (677,296.4) node [anchor=north west][inner sep=0.75pt]    [font=\footnotesize]  {$S\ duality$};
% Text Node
\draw (868,255.4) node [anchor=north west][inner sep=0.75pt]  [font=\footnotesize]  {$D5$};
% Text Node
\draw (639,517.4) node [anchor=north west][inner sep=0.75pt]  [font=\footnotesize]  {$NS5$};

\end{tikzpicture}

\end{center}
\caption{Left: the dual graph of a genus two degeneration determined by weighted graph 5c) in figure. \ref{weightedgenustwo}. Right: we represent the quiver on the left side by a type IIB configuration \cite{Hanany:1996ie}, notice here one of the $U(1)$ gauge group is ungauged; 
We then do the S duality to find the 3d mirror, and the gauge group is $U(2)\times U(1)$. To get the mirror of the left side quiver, one need to change the $U(2)$ gauge group to $SU(2)$ gauge group.  }
\label{iib}
\end{figure}
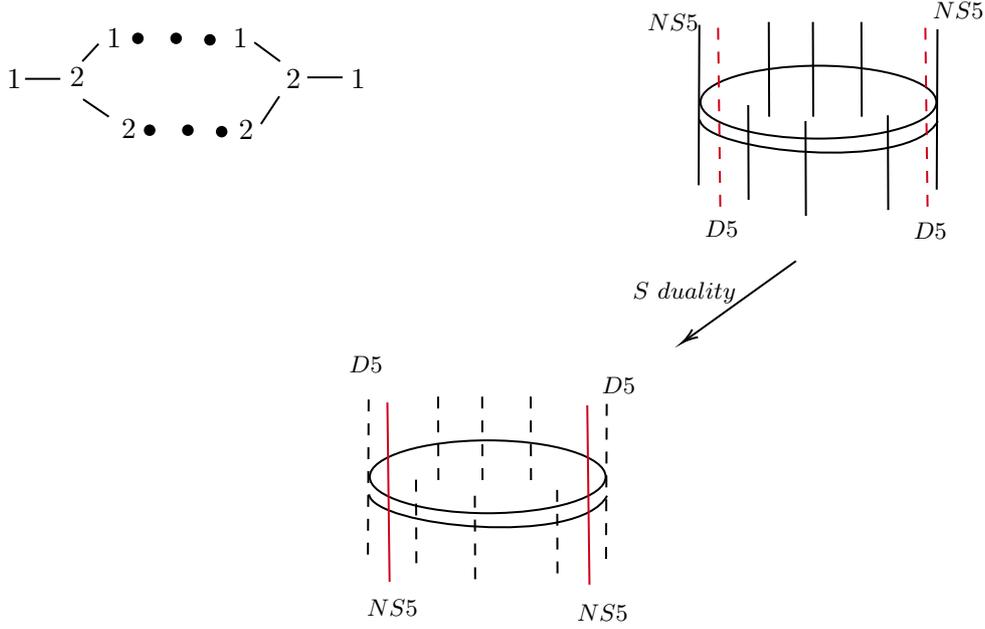

\textbf{Example}:  Let's consider the last item in figure. \ref{abelian}.  The periodic map on each genus zero component should be $(\frac{1}{2})+(1)+\frac{1}{2}$, and 
the dual graph is shown in figure. \ref{iib}. To find the IR theory, we try to find the 3d mirror of the dual graph, which can be achieved using S duality of type IIB string theory.

\section{Global SW geometry}
Let's now study the global SW geometry for rank one and rank two theory from the perspective of mapping class group. A one dimensional slice of the Coulomb branch is taken, and there are singular fibers where the low energy physics is different from the generic point. 
We add a point at $\infty$ to get a compact base, and this introduces a singular fiber at $\infty$ which determines the UV theory. 
We first study the generic one dimensional slice, so that the bulk singularity is either $I_1$ or $\tilde{I}_1$ type, see figure. \ref{intro1} 
for the illustration.
The physical understanding of these singularities are: a) The vanishing cycle associated with $I_1$ singularity is non-separating and its homology class is non-trivial; the low energy theory is  a massless hypermultiplet charged with 
a $U(1)$ gauge group (in proper duality frame), plus other free vector multiplets; b) There is a separating vanishing cycle associated with $\tilde{I}_1$ singularity, and there is an extra massless hypermultiplet which is not 
charged under the low energy gauge groups.

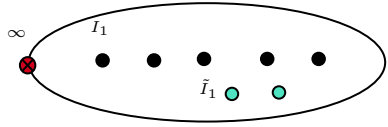
\begin{figure}[htbp]
\begin{center}

\tikzset{every picture/.style={line width=0.75pt}} %set default line width to 0.75pt        

\begin{tikzpicture}[x=0.55pt,y=0.55pt,yscale=-1,xscale=1]
%uncomment if require: \path (0,509); %set diagram left start at 0, and has height of 509

%Shape: Ellipse [id:dp2839587160648307] 
\draw   (185,138.61) .. controls (185,116.18) and (239.29,98) .. (306.25,98) .. controls (373.21,98) and (427.5,116.18) .. (427.5,138.61) .. controls (427.5,161.04) and (373.21,179.22) .. (306.25,179.22) .. controls (239.29,179.22) and (185,161.04) .. (185,138.61) -- cycle ;
%Flowchart: Summing Junction [id:dp22577707529233892] 
\draw  [fill={rgb, 255:red, 208; green, 2; blue, 27 }  ,fill opacity=1 ] (179,140.61) .. controls (179,137.51) and (181.35,135) .. (184.25,135) .. controls (187.15,135) and (189.5,137.51) .. (189.5,140.61) .. controls (189.5,143.71) and (187.15,146.22) .. (184.25,146.22) .. controls (181.35,146.22) and (179,143.71) .. (179,140.61) -- cycle ; \draw   (180.54,136.64) -- (187.96,144.58) ; \draw   (187.96,136.64) -- (180.54,144.58) ;
%Shape: Circle [id:dp28052247345949644] 
\draw  [fill={rgb, 255:red, 0; green, 0; blue, 0 }  ,fill opacity=1 ] (231,137.25) .. controls (231,134.9) and (232.9,133) .. (235.25,133) .. controls (237.6,133) and (239.5,134.9) .. (239.5,137.25) .. controls (239.5,139.6) and (237.6,141.5) .. (235.25,141.5) .. controls (232.9,141.5) and (231,139.6) .. (231,137.25) -- cycle ;
%Shape: Circle [id:dp3078783453518914] 
\draw  [fill={rgb, 255:red, 0; green, 0; blue, 0 }  ,fill opacity=1 ] (266,137.25) .. controls (266,134.9) and (267.9,133) .. (270.25,133) .. controls (272.6,133) and (274.5,134.9) .. (274.5,137.25) .. controls (274.5,139.6) and (272.6,141.5) .. (270.25,141.5) .. controls (267.9,141.5) and (266,139.6) .. (266,137.25) -- cycle ;
%Shape: Circle [id:dp34950396940104267] 
\draw  [fill={rgb, 255:red, 0; green, 0; blue, 0 }  ,fill opacity=1 ] (343,136.25) .. controls (343,133.9) and (344.9,132) .. (347.25,132) .. controls (349.6,132) and (351.5,133.9) .. (351.5,136.25) .. controls (351.5,138.6) and (349.6,140.5) .. (347.25,140.5) .. controls (344.9,140.5) and (343,138.6) .. (343,136.25) -- cycle ;
%Shape: Circle [id:dp2160729606961309] 
\draw  [fill={rgb, 255:red, 0; green, 0; blue, 0 }  ,fill opacity=1 ] (300,136.25) .. controls (300,133.9) and (301.9,132) .. (304.25,132) .. controls (306.6,132) and (308.5,133.9) .. (308.5,136.25) .. controls (308.5,138.6) and (306.6,140.5) .. (304.25,140.5) .. controls (301.9,140.5) and (300,138.6) .. (300,136.25) -- cycle ;
%Shape: Circle [id:dp34903805394050047] 
\draw  [fill={rgb, 255:red, 0; green, 0; blue, 0 }  ,fill opacity=1 ] (378,136.25) .. controls (378,133.9) and (379.9,132) .. (382.25,132) .. controls (384.6,132) and (386.5,133.9) .. (386.5,136.25) .. controls (386.5,138.6) and (384.6,140.5) .. (382.25,140.5) .. controls (379.9,140.5) and (378,138.6) .. (378,136.25) -- cycle ;
%Shape: Circle [id:dp1000229969087536] 
\draw  [fill={rgb, 255:red, 80; green, 227; blue, 194 }  ,fill opacity=1 ] (319,160.25) .. controls (319,157.9) and (320.9,156) .. (323.25,156) .. controls (325.6,156) and (327.5,157.9) .. (327.5,160.25) .. controls (327.5,162.6) and (325.6,164.5) .. (323.25,164.5) .. controls (320.9,164.5) and (319,162.6) .. (319,160.25) -- cycle ;
%Shape: Circle [id:dp21888496767899324] 
\draw  [fill={rgb, 255:red, 80; green, 227; blue, 194 }  ,fill opacity=1 ] (351,159.25) .. controls (351,156.9) and (352.9,155) .. (355.25,155) .. controls (357.6,155) and (359.5,156.9) .. (359.5,159.25) .. controls (359.5,161.6) and (357.6,163.5) .. (355.25,163.5) .. controls (352.9,163.5) and (351,161.6) .. (351,159.25) -- cycle ;

\draw (168,114.4) node [anchor=north west][inner sep=0.75pt] [font=\tiny]   {$\infty $};
% Text Node
\draw (225,108.4) node [anchor=north west][inner sep=0.75pt]   [font=\tiny]  {$I_{1}$};
% Text Node
\draw (299,148.4) node [anchor=north west][inner sep=0.75pt]  [font=\tiny]   {$\tilde{I}_{1}$};

\end{tikzpicture}

\end{center}
\caption{Global Coulomb  branch geometry with $I_1$ and $\tilde{I}_1$ type  singular fibers  at the bulk. The fiber at infinity gives the information for UV theory, i.e. which space-time dimension the UV theory lives.}
\label{global}
\end{figure}

The SW fiberation with just $I_1$ or $\tilde{I}_1$ singular fibers are called Lefschetz fiberation. Notice that in our case the singular fiber at $\infty$ 
could be special, however, we assume that it could also be deformed into $I_1$ and $\tilde{I}_1$ singularities, see the  discussion below. The Coulomb branch structure is shown in figure. \ref{global}.

There are following topological constraints that the global SW geometry has to satisfy: 
\begin{itemize}
\item One can define two topological invariants for a singular fiber: $d_x$ and $\delta_x$. $\delta_x$ 
can be easily computed from the dual graph and $d_x$ can be computed from holomorphic data, see \cite{Xie:2022aad} for those numbers. The difference of $d_x$ and $\delta_x$ reflects the number of $\tilde{I}_1$ singularity in the generic deformation. The first topological constraint is 
 \begin{equation*}
 \sum d_x = 2,
 \end{equation*}
 which reflects the fact the total space of SW fiberation is a rational surface.
The topological data for $I_1$ singularity is $d_x=\delta_x=1$, while the data for $\tilde{I}_1$ singularity is  $d_x=2, \delta_x=1$. The above topological constraint comes from 
the assumption that the total space of the genus two fiberation is rational.  The above constraint implies that
the choice of $I_1, \tilde{I}_1$ singularities are $(20,0), (18,1), (16,2)$, etc. 
\item The second topological constraint is due to the compactness of the base of the fiberation. By choosing a generic point on the moduli space, there is one mapping class group element $M_i$
for each singular fiber by choosing a path, and the product of them should be trivial:
\begin{equation*}
M_1\ldots M_s=I.
\end{equation*}
Therefore, the global SW geometry has a simple topological meaning: a factorization of identity element in terms of mapping class group elements associated with $I_1$ and $\tilde{I}_1$ singularity.
\end{itemize}

Therefore we need to solve two problems in mapping class group: a) Find the factorization of the UV  MCG element in terms of the MCG element of $I_1$ and $\tilde{I}_1$ singularity; b) Find the factorization 
of the identity element in terms of suitable number MCG element of $I_1$ and $\tilde{I}_1$ singularity.

\subsection{Dehn twist}
\begin{figure}[htbp]
\begin{center}

\tikzset{every picture/.style={line width=0.75pt}} %set default line width to 0.75pt        

\begin{tikzpicture}[x=0.55pt,y=0.55pt,yscale=-1,xscale=1]
%uncomment if require: \path (0,716); %set diagram left start at 0, and has height of 716

%Shape: Ellipse [id:dp39470531482587234] 
\draw   (197,147.86) .. controls (197,131.92) and (202.37,119) .. (209,119) .. controls (215.63,119) and (221,131.92) .. (221,147.86) .. controls (221,163.8) and (215.63,176.72) .. (209,176.72) .. controls (202.37,176.72) and (197,163.8) .. (197,147.86) -- cycle ;
%Straight Lines [id:da30941371719315613] 
\draw    (209,119) -- (345,119.72) ;
%Shape: Ellipse [id:dp25002346076500515] 
\draw   (333,148.58) .. controls (333,132.64) and (338.37,119.72) .. (345,119.72) .. controls (351.63,119.72) and (357,132.64) .. (357,148.58) .. controls (357,164.52) and (351.63,177.44) .. (345,177.44) .. controls (338.37,177.44) and (333,164.52) .. (333,148.58) -- cycle ;
%Straight Lines [id:da9270765581832079] 
\draw    (209,176.72) -- (345,177.44) ;
%Straight Lines [id:da3428406252223697] 
\draw [color={rgb, 255:red, 80; green, 227; blue, 194 }  ,draw opacity=1 ]   (221,147.86) -- (333,148.58) ;
%Shape: Ellipse [id:dp3075194187901156] 
\draw  [color={rgb, 255:red, 208; green, 2; blue, 27 }  ,draw opacity=1 ] (267,148.22) .. controls (267,131.55) and (269.24,118.04) .. (272,118.04) .. controls (274.76,118.04) and (277,131.55) .. (277,148.22) .. controls (277,164.89) and (274.76,178.4) .. (272,178.4) .. controls (269.24,178.4) and (267,164.89) .. (267,148.22) -- cycle ;
%Straight Lines [id:da9692300441071451] 
\draw    (266,200) -- (266,259.72) ;
\draw [shift={(266,261.72)}, rotate = 270] [color={rgb, 255:red, 0; green, 0; blue, 0 }  ][line width=0.75]    (10.93,-3.29) .. controls (6.95,-1.4) and (3.31,-0.3) .. (0,0) .. controls (3.31,0.3) and (6.95,1.4) .. (10.93,3.29)   ;
%Shape: Ellipse [id:dp9360821030040072] 
\draw   (199,310.86) .. controls (199,294.92) and (204.37,282) .. (211,282) .. controls (217.63,282) and (223,294.92) .. (223,310.86) .. controls (223,326.8) and (217.63,339.72) .. (211,339.72) .. controls (204.37,339.72) and (199,326.8) .. (199,310.86) -- cycle ;
%Straight Lines [id:da5702575079961398] 
\draw    (211,282) -- (347,282.72) ;
%Shape: Ellipse [id:dp6915614378329047] 
\draw   (335,311.58) .. controls (335,295.64) and (340.37,282.72) .. (347,282.72) .. controls (353.63,282.72) and (359,295.64) .. (359,311.58) .. controls (359,327.52) and (353.63,340.44) .. (347,340.44) .. controls (340.37,340.44) and (335,327.52) .. (335,311.58) -- cycle ;
%Straight Lines [id:da20814860879798758] 
\draw    (211,339.72) -- (347,340.44) ;
%Curve Lines [id:da14371336552430458] 
\draw [color={rgb, 255:red, 80; green, 227; blue, 194 }  ,draw opacity=1 ]   (223,310.86) .. controls (264,321.72) and (241,283.72) .. (268,282.72) ;
%Curve Lines [id:da0031877152034414102] 
\draw [color={rgb, 255:red, 80; green, 227; blue, 194 }  ,draw opacity=1 ] [dash pattern={on 0.84pt off 2.51pt}]  (268,282.72) .. controls (290,293.72) and (289,319.72) .. (289,338.72) ;
%Curve Lines [id:da8555193324834744] 
\draw [color={rgb, 255:red, 80; green, 227; blue, 194 }  ,draw opacity=1 ]   (289,338.72) .. controls (304,294.72) and (308,312.58) .. (335,311.58) ;

\draw (289,215) node [anchor=north west][inner sep=0.75pt]   [align=left] {Dehn twist};
% Text Node
\draw (264,87.4) node [anchor=north west][inner sep=0.75pt]    {$C$};

\end{tikzpicture}

\end{center}
\caption{Dehn twist along curve $C$, and this action is taken to be negative around the indicated direction.}
\label{Dehn}
\end{figure}
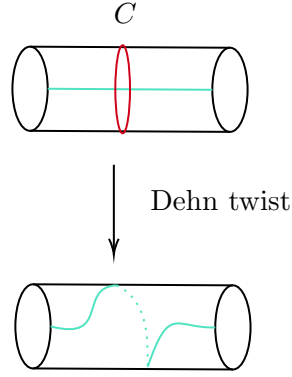

As we discussed at the beginning of this section, the global SW geometry has a meaning in terms of factorization of mapping class group (MCG) element  into  $I_1$ or $\tilde{I}_1$ singularity. 
There is a vanishing cycle for each $I_1$ or $\tilde{I}_1$ singularity, and one can have an associated MCG element called Dehn twist, see figure. \ref{Dehn}.

The following are the basic relations for Dehn twist associated with two cycles $a,b$:
\begin{enumerate}
\item If $a,b$ is disjoint, then $T_aT_b=T_bT_a$.
\item If $b=h(a)$ with $h$ an element of mapping class group, then $T_b=hT_ah^{-1}$.
\item If $(a,b)=1$, then $T_a T_b T_a=T_b T_a T_b$.
\item If $(a,b)>1$, then there are no relations between $T_a$ and $T_b$.
\end{enumerate}
The first and third relations are called braid relations, which would play crucial roles later.

\subsection{Rank one theory}
Let's now apply our classification strategy to rank one theory, and we'd like to recover the classification of rank one theory.

\begin{figure}[H]
 
 \begin{center}

\tikzset{every picture/.style={line width=0.75pt}} %set default line width to 0.75pt        

\begin{tikzpicture}[x=0.55pt,y=0.55pt,yscale=-1,xscale=1]
%uncomment if require: \path (0,964); %set diagram left start at 0, and has height of 964

%Shape: Ellipse [id:dp2732109334277837] 
\draw   (179,130.11) .. controls (179,98.57) and (231.61,73) .. (296.5,73) .. controls (361.39,73) and (414,98.57) .. (414,130.11) .. controls (414,161.65) and (361.39,187.22) .. (296.5,187.22) .. controls (231.61,187.22) and (179,161.65) .. (179,130.11) -- cycle ;
%Curve Lines [id:da1050906401575642] 
\draw    (263,117) .. controls (279,88.22) and (310,100.22) .. (315,118.22) ;
%Curve Lines [id:da3122526712301241] 
\draw    (262,114.22) .. controls (272,132.44) and (297,142.22) .. (317,115.22) ;
%Curve Lines [id:da9586326305138149] 
\draw    (293,133) .. controls (283,137.22) and (280,180.22) .. (295,186.22) ;
\draw [shift={(284.86,153.93)}, rotate = 92.21] [fill={rgb, 255:red, 0; green, 0; blue, 0 }  ][line width=0.08]  [draw opacity=0] (8.93,-4.29) -- (0,0) -- (8.93,4.29) -- cycle    ;
%Curve Lines [id:da26719682909459563] 
\draw    (293,133) .. controls (303,149.22) and (305,163.22) .. (296,185.22) ;
%Curve Lines [id:da0413251291039709] 
\draw    (227,122) .. controls (216.11,148.95) and (333.61,186.46) .. (389.33,127.05) ;
\draw [shift={(391,125.22)}, rotate = 131.58] [fill={rgb, 255:red, 0; green, 0; blue, 0 }  ][line width=0.08]  [draw opacity=0] (10.72,-5.15) -- (0,0) -- (10.72,5.15) -- (7.12,0) -- cycle    ;
%Curve Lines [id:da10673818437048421] 
\draw    (226,123) .. controls (237,72.22) and (387,75.22) .. (390,125.22) ;

\draw (340,125.4) node [anchor=north west][inner sep=0.75pt]    {$a_{1}$};
% Text Node
\draw (269,134.4) node [anchor=north west][inner sep=0.75pt]    {$b_{1}$};

\end{tikzpicture}
 
 \end{center}
 \caption{The two cycles generate the homology group of genus one curve.}
 \label{genusone}
 
 \end{figure}
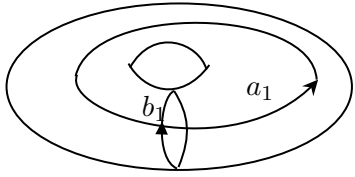

The genus one MCG is just $SL(2,Z)$ group, and it  is generated by the Dehn twist around two cycles $a_1, b_1$, see 
figure. \ref{genusone}. The generators are denoted as $\tau_1, \tau_2$, and there is just one non-trivial relation:
\begin{equation*}
(\tau_1\tau_2)^6=1.
\end{equation*} 
We also have the  braid relation $\tau_1\tau_2\tau_1=\tau_2\tau_1\tau_2$. 
The intersection form is $(a_1, b_1)=1$, and so the action of the Dehn twist on homology classes $[a_1], [b_1]$ is
\begin{align*}
&T_1(a_1)=a_1+(a_1,a_1)a_1=a_1,~~~~~~~~T_1(b_1)=b_1+(b_1,a_1)a_1=-a_1+b_1 \nonumber\\
&T_2(a_1)=a_1+(a_1,b_1)b_1=a_1+b_1,~~T_2(b_1)=b_1+(b_1,b_1)b_1=b_1
\end{align*} 
So the representation matrix on the basis of homology groups  $([a_1], [b_1])$ is given as 
 \begin{equation*}
 \tau_1=\left(\begin{array}{cc}1&-1 \\ 0&1 \end{array} \right),~~ \tau_2=\left(\begin{array} {cc}1&0 \\ 1&1 \end{array} \right)
 \end{equation*}
 which gives the standard representation for the generators of $SL(2,Z)$ group.
 
\textbf{Factorization for MCG element of singular fibers}:
The classification of rank one IR theory is given by the degeneration of elliptic curve, and it is the same as the classification of 
 the conjugacy class of $M_1$ whose homology representation $M$
satisfying  $Tr(M)\leq 2$.  We will find the factorization of these conjugacy class by requiring:
\begin{enumerate}
\item It is given by a product of positive Dehn twist, namely it involves only the generators $\tau_1,\tau_2$, but not the inverse of them.
\item The number of elements in the factorization is the same as the Euler number, which is also equal to the number of $I_1$ singularities under the generic deformation of the singularity.
\end{enumerate}
The factorizations for genus one degeneration is shown in table. \ref{genus1mgp}.

\begin{table}
\begin{center}
  \begin{tabular}{ |c|c|c|c|  }
    \hline
   Mapping class group element& Data & Name & Euler number\\ \hline
    $\zeta_1=\tau_1\tau_2$& $\frac{1}{6}+\frac{1}{3}+\frac{1}{2}$ & II & 2  \\     \hline
   $\zeta_1^2$& $\frac{1}{3}+\frac{1}{3}+\frac{1}{3}$  & IV & 4\\     \hline
      $\zeta_1^3$& $\frac{1}{2}+\frac{1}{2}+\frac{1}{2}+\frac{1}{2}$ &$I_0^*$ & 6  \\     \hline
   $\zeta_1^4$& $\frac{2}{3}+\frac{2}{3}+\frac{2}{3}$ & $IV^*$ & 8 \\     \hline
   $\zeta_1^{5}$&$\frac{5}{6}+\frac{2}{3}+\frac{1}{2}$ & $II^*$ & 10 \\ \hline
      $\zeta_1^{6}=1$&$1$ &$I_0$ & \\ \hline
  \end{tabular}
  \quad
 \begin{tabular}{ |c|c|c|c|  }
    \hline
   Mapping class group element & Data & Name & Euler number \\ \hline
    $\zeta_2=\tau_1\tau_2\tau_1$& $\frac{1}{4}+\frac{1}{4}+\frac{1}{2}$ & $III$ &3  \\     \hline
   $\zeta_2^2$& $\frac{1}{2}+\frac{1}{2}+\frac{1}{2}+\frac{1}{2}$ &$I_0^*$ &6   \\     \hline
      $\zeta_2^3$& $\frac{3}{4}+\frac{3}{4}+\frac{1}{2}$ & $ III^*$ & 9 \\     \hline
   $\zeta_2^4$& $1$  &$I_0$ & \\     \hline
  \end{tabular}
  \quad
   \begin{tabular}{ |c|c|c|  }
    \hline
   Mapping class group element  & Name & Euler number \\ \hline
    $\tau_2^{k} (\tau_1\tau_2)^3$ & $I_k^*$ & $k+6$  \\     \hline
   $\tau_1^k$ &$I_k$  & $k$\\     \hline
  \end{tabular}
  \caption{The factorization of mapping class group element associated with genus one degeneration.}
  \label{genus1mgp}
\end{center}
\end{table}

Given the representation of MCG element in table. \ref{genus1mgp}, one can then easily find the possible IR configurations: one simply rearrange the 
word by using the braid relation and the Hurwitz move
\begin{align}
& \tau_1\tau_2\tau_1=\tau_2\tau_1\tau_2, \nonumber\\
&(\tau_1 \ldots \underbrace{\tau_i\tau_{i+1}}\ldots\tau_r)=(\tau_1\ldots\underbrace{(\tau_i\tau_{i+1}\tau_i^{-1})\tau_i}\ldots\tau_r)
\end{align}
Since $\tau_i\tau_{i+1}\tau_i^{-1}= \tau_{i(i+1)}$ is also a positive Dehn twist around the curve $i(i+1)$ (which is the resulting curve of  the Dehn twist $\tau_i$ 
acting on the curve $(i+1)$.). The above two moves would also give the positive factorization. One can find the IR configuration by looking at the possible 
collapsing of the $I_1$ singularities.

\textit{Example 1}: Let's  look at  $E_8$ SCFT which is represented by the word $(\tau_1\tau_2)^5$, and it is easy to find various singular fiber combinations. Here we just give several simple examples.
The simplest one is the configuration with five type II singularities:
\begin{equation*}
(\tau_1 \tau_2) (\tau _1 \tau_2) (\tau_1 \tau_2) (\tau_1 \tau_2) (\tau_1 \tau_2)
\end{equation*}
We use bracket to indicate that the singularities inside it is collapsed. 
The next one involves a $I_0^*$ singularity 
\begin{equation*}
(\tau_1\tau_2 \tau_1 \tau_2 \tau_1\tau_2) \tau_1\tau_2 \tau_1\tau_2
\end{equation*}
Finally one can find a configuration with a $I_3^*$ singularity
\begin{equation*}
(\tau_1\tau_2 \tau_1) \tau_2 \tau_1\tau_2 \tau_1 \tau_2 \tau_1\tau_2 \to (\tau_2 \tau_1) \tau_2 \tau_2 \tau_1\tau_2 \tau_1 \tau_2 \tau_1\tau_2 \to \tau_{2(1)} (\tau_2^3 ( \tau_1\tau_2)^3)
\end{equation*}
Here one use the braid relation in the first step, and use the Hurwitz move in the second step, and  finally one get a $I_3^*$ singularity. The interested reader can work out all 
the other configurations  listed in \cite{persson1990configurations, Xie:2022lcm}. 

\textbf{Global SW geometry}:  The total mapping class group around all the singularities on the compactified Coulomb branch should be trivial, which implies
that the ordered product of Dehn twists should be trivial. We conjecture that the corresponding elliptic fibered surface should be 
a rational surface, which implies that the total number of $I_1$ singularities should be 12 \cite{persson1990configurations,Xie:2022lcm}. So one should find a positive factorization of identity element with length 12, and the only choice is
\begin{equation*}
(\tau_1\tau_2)^{6}=(\tau_1\tau_2\tau_1)^4=1.
\end{equation*}
We can then find the global SW geometry by using braid move and Hurwitz move to rearrange the above letters, and get the configuration for the factorization of the UV fiber, see table. \ref{genus1mgp}.

\textit{ 4d SCFT}: The configuration for SCFT is quite simple, we have 
\begin{align}
&((\tau_1\tau_2)^i, (\tau_1\tau_2)^{6-i}), \nonumber\\
&((\tau_1\tau_2 \tau_1)^i, (\tau_1\tau_2\tau)^{4-i}).
\end{align}
So one can get a pair of SCFT by putting one  configuration at $\infty$, and others are bulk singularities which can be moved freely.

\textit{ 4d asymptotical free theory}:  The configuration for asymptotical free theory is  
\begin{equation}
(I_{bulk}, I_\infty)=( \tau_1 \tau_{22(1)} \tau_2^{4-k},\tau_2^k(\tau_1 \tau_2)^3).
\label{rankoneaf}
\end{equation} 
This is the SW geometry for $SU(2)$ with $4-k$ fundamental hypermultiplets. The configuration is found by doing the braid move and Hurwitz move on the fundamental factorization of identity, so that one can form a $I_k^*$ singularity:
\begin{equation*}
1=\tau_1\tau_2 (\tau_1 \tau_2 \tau_1) \tau_2 (\tau_1\tau_2)^3 = \tau_1(\tau_2 \tau_2 \tau_1) \tau_2 \tau_2 (\tau_1\tau_2)^3 =  \tau_1 \tau_{22(1)} \tau_2^{4} (\tau_1\tau_2)^3.
\end{equation*}
One do the braid move for the first step, and do the Hurwitz move for the second step. The notation $\tau_{22(1)}$ means the positive Dehn twist along the curve $22(1)$, which is the 
curve derived by first doing  Dehn twist along curve 2 on curve $1$ to get a curve $2(1)$, and then do the Dehn twist along curve 2 on curve $2(1)$. Using the action of Dehn twist on homology, one has 
\begin{equation*}
22([1])=2([1]+(1,2)[2])=[1]+(1,2)[2]+(1,2)[2]=[1]+2[2].
\end{equation*}
Here one use the intersection number $(1,2)=1$. For the pure $SU(2)$ theory $(k=4)$ in \ref{rankoneaf}, the bulk singularity is $\tau_1 \tau_{22(1)}$; So there is 
one vanishing cycle with homology $[1]$,  and another vanishing cycle with homology $[1]+2[2]$.  The intersection pair of these two vanishing cycles are $2$, which indeed gives 
the BPS quiver of this theory. 

Notice that the singular fiber at $\infty$ is $I_k^*$ singular fiber, which is represented by $\tau_2^k(\tau_1 \tau_2)^3$, and is the singular fiber at $\infty$ for 4d asymptotical free theory.

\textit{ 5d KK theory}: One simply use the braid move and Hurwitz move to get a $I_k$ singularity on the right:
\begin{align*}
& \tau_1(\tau_2 \tau_1 \tau_2) \tau_1 (\tau _2 \tau_1\tau_2) \tau_1 (\tau_2 \tau_1\tau_2) \to \tau_1 \tau_1\tau_2 \tau_1 \tau_1 \tau_1\tau_2\tau_1 \tau_1 \tau_1 \tau_2 \tau_1 \nonumber\\
& \to \tau_{1^2(2)} \tau_{1^5(2)} \tau_{1^8(2)} \tau_1^9.
\end{align*}
and one can get a $I_9$ singularity. Different type of 5d theory is found by putting  $I_k$ singularity at $\infty$.

\textit{ 6d KK theory}: This one is the simplest one as there is no singularity at $\infty$, so the bulk singularity is just $(\tau_1\tau_2)^6$.

\textit{Non-deformable singularity}: To get theory with non-deformable singularity, one simply use braid move and Hurwitz move on the word for a SCFT to get the non-deformable singularity type 
(such as $I_{n}, I_k^*, II^*, III^*, IV^*$) on the bulk.  

\textbf{ $\mathcal{N}=2^*$ theory}: SW geometry of $\mathcal{N}=2^*$ theory is found by $SU(2)$ with $N_f=4$ word, which is a $I_0^*$ singularity. It can be written as $I_4I_1^2$ configuration, 
which can be found as follows:
\begin{equation*}
\tau_1(\tau_2\tau_1\tau_2)\tau_1\tau_2=(\tau_1^2\tau_2)\tau_1^2\tau_2=\tau_{1^2(2)}(\tau_1^4) \tau_2.
\end{equation*}
Here we used braid move and the Hurwitz move to get a $I_4$ singularity. Another realization is given by $I_2I_2I_2$ configuration, which can be found from $I_0^*$ word as follows:
\begin{equation*}
\tau_1(\tau_2\tau_1 \tau_2)\tau_1\tau_2=\tau_1^2 (\tau_2\tau_1\tau_1)\tau_2=(\tau_1^2) (\tau_{2(1)}^2) (\tau_2^2).
\end{equation*}

\subsection{Rank two theory}

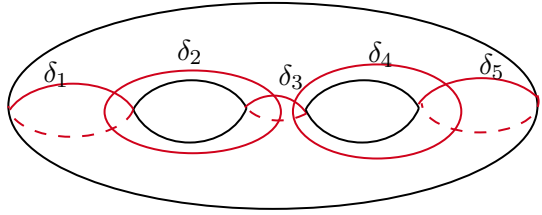
\begin{figure}[H]
\begin{center}

\tikzset{every picture/.style={line width=0.75pt}} %set default line width to 0.75pt        

\begin{tikzpicture}[x=0.50pt,y=0.50pt,yscale=-1,xscale=1]
%uncomment if require: \path (0,799); %set diagram left start at 0, and has height of 799

%Shape: Ellipse [id:dp3246353739327341] 
\draw   (99,144.72) .. controls (99,101.8) and (187.65,67) .. (297,67) .. controls (406.35,67) and (495,101.8) .. (495,144.72) .. controls (495,187.64) and (406.35,222.44) .. (297,222.44) .. controls (187.65,222.44) and (99,187.64) .. (99,144.72) -- cycle ;
%Shape: Arc [id:dp945263110503583] 
\draw  [draw opacity=0] (277.27,146.39) .. controls (271.42,161.37) and (254.73,172.24) .. (234.97,172.38) .. controls (215.31,172.51) and (198.54,161.97) .. (192.42,147.2) -- (234.71,134.71) -- cycle ; \draw   (277.27,146.39) .. controls (271.42,161.37) and (254.73,172.24) .. (234.97,172.38) .. controls (215.31,172.51) and (198.54,161.97) .. (192.42,147.2) ;  
%Shape: Arc [id:dp10315564058872828] 
\draw  [draw opacity=0] (192.62,147.33) .. controls (199.1,134.53) and (216.38,125.32) .. (236.25,125.39) .. controls (255.47,125.45) and (271.47,134.18) .. (277.27,146.39) -- (234.75,157.51) -- cycle ; \draw   (192.62,147.33) .. controls (199.1,134.53) and (216.38,125.32) .. (236.25,125.39) .. controls (255.47,125.45) and (271.47,134.18) .. (277.27,146.39) ;  

%Shape: Arc [id:dp9630053518576597] 
\draw  [draw opacity=0] (100.09,147.71) .. controls (107.61,136.11) and (126.16,127.89) .. (147.48,127.96) .. controls (168.38,128.03) and (185.88,136.05) .. (192.62,147.33) -- (146.06,158.41) -- cycle ; \draw  [color={rgb, 255:red, 208; green, 2; blue, 27 }  ,draw opacity=1 ] (100.09,147.71) .. controls (107.61,136.11) and (126.16,127.89) .. (147.48,127.96) .. controls (168.38,128.03) and (185.88,136.05) .. (192.62,147.33) ;  
%Shape: Arc [id:dp22624272135371137] 
\draw  [draw opacity=0][dash pattern={on 4.5pt off 4.5pt}] (192.38,147.75) .. controls (188.37,159.14) and (169.26,167.87) .. (146.23,168.02) .. controls (120.78,168.2) and (99.97,157.84) .. (99,144.72) -- (146.06,143.53) -- cycle ; \draw  [color={rgb, 255:red, 208; green, 2; blue, 27 }  ,draw opacity=1 ][dash pattern={on 4.5pt off 4.5pt}] (192.38,147.75) .. controls (188.37,159.14) and (169.26,167.87) .. (146.23,168.02) .. controls (120.78,168.2) and (99.97,157.84) .. (99,144.72) ;  
%Shape: Arc [id:dp31581519217696874] 
\draw  [draw opacity=0] (275.99,145.93) .. controls (281.85,140.36) and (289.53,136.96) .. (297.79,136.98) .. controls (307.99,137.02) and (316.81,142.26) .. (322.16,150.36) -- (296.23,170.54) -- cycle ; \draw  [color={rgb, 255:red, 208; green, 2; blue, 27 }  ,draw opacity=1 ] (275.99,145.93) .. controls (281.85,140.36) and (289.53,136.96) .. (297.79,136.98) .. controls (307.99,137.02) and (316.81,142.26) .. (322.16,150.36) ;  
%Shape: Arc [id:dp48675276391190625] 
\draw  [draw opacity=0] (405.98,143.37) .. controls (413.52,131.86) and (432.05,123.71) .. (453.33,123.78) .. controls (471,123.84) and (486.24,129.55) .. (494.6,138.09) -- (451.92,154.07) -- cycle ; \draw  [color={rgb, 255:red, 208; green, 2; blue, 27 }  ,draw opacity=1 ] (405.98,143.37) .. controls (413.52,131.86) and (432.05,123.71) .. (453.33,123.78) .. controls (471,123.84) and (486.24,129.55) .. (494.6,138.09) ;  
%Shape: Arc [id:dp7310887060564322] 
\draw  [draw opacity=0][dash pattern={on 4.5pt off 4.5pt}] (494.94,136.72) .. controls (495.52,138.21) and (495.83,139.76) .. (495.84,141.34) .. controls (495.93,154.1) and (476.41,164.57) .. (452.24,164.74) .. controls (428.59,164.9) and (409.25,155.12) .. (408.36,142.74) -- (452.09,141.64) -- cycle ; \draw  [color={rgb, 255:red, 208; green, 2; blue, 27 }  ,draw opacity=1 ][dash pattern={on 4.5pt off 4.5pt}] (494.94,136.72) .. controls (495.52,138.21) and (495.83,139.76) .. (495.84,141.34) .. controls (495.93,154.1) and (476.41,164.57) .. (452.24,164.74) .. controls (428.59,164.9) and (409.25,155.12) .. (408.36,142.74) ;  
%Shape: Arc [id:dp583037899689498] 
\draw  [draw opacity=0] (406.27,146.39) .. controls (400.42,161.37) and (383.73,172.24) .. (363.97,172.38) .. controls (344.31,172.51) and (327.54,161.97) .. (321.42,147.2) -- (363.71,134.71) -- cycle ; \draw   (406.27,146.39) .. controls (400.42,161.37) and (383.73,172.24) .. (363.97,172.38) .. controls (344.31,172.51) and (327.54,161.97) .. (321.42,147.2) ;  
%Shape: Arc [id:dp11078762413610221] 
\draw  [draw opacity=0] (321.62,147.33) .. controls (328.1,134.53) and (345.38,125.32) .. (365.25,125.39) .. controls (384.47,125.45) and (400.47,134.18) .. (406.27,146.39) -- (363.75,157.51) -- cycle ; \draw   (321.62,147.33) .. controls (328.1,134.53) and (345.38,125.32) .. (365.25,125.39) .. controls (384.47,125.45) and (400.47,134.18) .. (406.27,146.39) ;  

%Shape: Arc [id:dp624454660334151] 
\draw  [draw opacity=0][dash pattern={on 4.5pt off 4.5pt}] (322.16,150.36) .. controls (317.49,152.9) and (312.17,154.68) .. (306.5,155.4) .. controls (294.36,156.93) and (283.57,153.24) .. (277.27,146.39) -- (305.84,127.79) -- cycle ; \draw  [color={rgb, 255:red, 208; green, 2; blue, 27 }  ,draw opacity=1 ][dash pattern={on 4.5pt off 4.5pt}] (322.16,150.36) .. controls (317.49,152.9) and (312.17,154.68) .. (306.5,155.4) .. controls (294.36,156.93) and (283.57,153.24) .. (277.27,146.39) ;  
%Shape: Ellipse [id:dp12455276841343743] 
\draw  [color={rgb, 255:red, 208; green, 2; blue, 27 }  ,draw opacity=1 ] (171,149.22) .. controls (171,131.98) and (200.55,118) .. (237,118) .. controls (273.45,118) and (303,131.98) .. (303,149.22) .. controls (303,166.46) and (273.45,180.44) .. (237,180.44) .. controls (200.55,180.44) and (171,166.46) .. (171,149.22) -- cycle ;
%Shape: Ellipse [id:dp4506166378819504] 
\draw  [color={rgb, 255:red, 208; green, 2; blue, 27 }  ,draw opacity=1 ] (312,148.94) .. controls (312,129.33) and (340.21,113.44) .. (375,113.44) .. controls (409.79,113.44) and (438,129.33) .. (438,148.94) .. controls (438,168.54) and (409.79,184.44) .. (375,184.44) .. controls (340.21,184.44) and (312,168.54) .. (312,148.94) -- cycle ;

\draw (123,109.4) node [anchor=north west][inner sep=0.75pt]    {$\delta _{1}$};
% Text Node
\draw (223,92.4) node [anchor=north west][inner sep=0.75pt]    {$\delta _{2}$};
% Text Node
\draw (299,111.4) node [anchor=north west][inner sep=0.75pt]    {$\delta _{3}$};
% Text Node
\draw (366,94.4) node [anchor=north west][inner sep=0.75pt]    {$\delta _{4}$};
% Text Node
\draw (449,103.4) node [anchor=north west][inner sep=0.75pt]    {$\delta _{5}$};

\end{tikzpicture}

\end{center}
\caption{The Dehn twists along curves $\delta_i$ generate the genus two mapping class group.}
\label{genustwodehn}
\end{figure}

The mapping class group  $M_2$ of genus two curve  is generated by Dehn twists associated with the five curves $\delta_i$ shown in figure. \ref{genustwodehn}. The generators are labeled as
\begin{equation}
\tau_1,\tau_2,\tau_3, \tau_4,\tau_5
\end{equation}
and the relations are
\begin{align}
& \tau_j \tau_{j+1}\tau_j=\tau_{j+1}\tau_j \tau_{j+1} \nonumber\\
& \tau_i\tau_j=\tau_j \tau_i,~~if~~|i-j|>1 \nonumber \\
& I \tau_j=\tau_j I \nonumber\\
& I^2=1 \nonumber\\
& (\tau_1 \tau_2 \tau_3 \tau_4 \tau_5)^6=1
\end{align}
Here $I=\tau_1\tau_2\tau_3\tau_4\tau_5^2\tau_4\tau_3\tau_2\tau_1$, and is the hyperelliptic involution. 
The first two relations are called braid relations.  

An element in $M_2$ can be represented by a positive product of generators; However, the representation 
is far from unique due to the braid relations and  other relations in $M_2$. Furthermore, we also need to impose following two equivalence relations:
\begin{itemize}
\item The Hurwitz equivalence:
\begin{equation*}
(\tau_1 \ldots \underbrace{\tau_i\tau_{i+1}}\ldots\tau_r)=(\tau_1\ldots\underbrace{(\tau_i\tau_{i+1}\tau_i^{-1})\tau_i}\ldots\tau_r)=(\tau_1\ldots \tau_{i(i+1)}\tau_i \ldots\tau_r).
\end{equation*}
\item Global conjugacy:
\begin{equation*}
\phi(\tau_1\ldots \tau_r)\phi^{-1}=(\phi \tau_1 \phi^{-1} \ldots \phi \tau_r \phi^{-1}).
\end{equation*}
This is due to the fact the degeneration is given by the conjugacy class of mapping class group.
\end{itemize}
The goal for finding a special factorization for the degeneration is following: a): It should be given by a positive factorization, namely, 
the word consists of only Dehn twist, but not its inverse; b): The number of generators are determined by the local invariant $d_x$ and $\delta_x$:
\begin{align}
&\# \tau= 2\delta_x-d_x, \nonumber\\
&\# \sigma= d_x-\delta_x. 
\end{align}
 Here $\# \tau$ is the number of Dehn twist along non-separating curve, and $\# \sigma$ is the Dehn twist along the separating curve $\sigma$.
The task of finding above factorization is a difficult one, and we will solve it for the SCFT and many other IR theories in this paper.

Let's now summarize some important relations regarding group $M_2$, which will be quite useful for our later studies.
\begin{enumerate}
\item Let $\zeta_{a,b}=\prod_{i=a}^b \tau_i$, then 
\begin{equation}
\tau_i \zeta_{a,b}=\zeta_{a,b} \tau_{i-1}
\label{fundamental}
\end{equation}
Here $a<i\leq b$. This equation can be proven using the braid relation:

\textbf{Proof}:   
\begin{align*}
&\tau_i (\tau_a \tau_{a+1}\ldots \tau_b)= \tau_a \ldots (\tau_i \tau_{i-1} \tau_{i}) \tau_{i+1}\ldots  \tau_b=\tau_a \ldots \tau_{i-1} \tau_{i} (\tau_{i-1}\tau_{i+1})\ldots  \tau_b \nonumber\\
&=(\tau_a\ldots \tau_b) \tau_{i-1}
\end{align*}

\item Let $\zeta=\tau_1\tau_2\tau_3\tau_4\tau_5,\eta=\tau_1\tau_2\tau_3\tau_4$, we have the relation
\begin{align}
&\tau_i \zeta^j=\zeta^j\tau_{i-j},~~~~~ (i\neq j)~~  \nonumber\\
&\tau_i \zeta^i=\zeta^{i+1} \eta^{-1},~~(\eta=\zeta^{-i}\tau_i^{-1}\zeta^{i+1})
\label{eta}
\end{align}
These two equations are derived using the  relation \ref{fundamental}. From the second relation, one find that the generators can be expressed in terms of $\zeta, \eta$:
\begin{equation}
\tau_i= \zeta^{i+1} \eta^{-1}\zeta^{-i}
\label{basic}
\end{equation}
So the mapping class group $M_2$ is generated by $\zeta, \eta$ subject to relation $\zeta^6=\eta^{10}=1$. 

\item  Let $\epsilon=\tau_1^2\tau_2\tau_3\tau_4, \eta=\tau_1\tau_2\tau_3\tau_4$. We  have  the relation $\epsilon^4=\eta^5$, and  $\eta^5$ is conjugate with $I$.

\textbf{Proof}: Let's first do the computation for $\eta^5$. Using formula. \ref{eta}, we have
\begin{align*}
&\eta^5=(\zeta^{-1}\tau_1^{-1}\zeta^2) (\zeta^{-2} \tau_2^{-1} \zeta^{3})\ldots (\zeta^{-5}\tau_5^{-1}\zeta^{6})= \nonumber\\
&=\zeta^{-1}\tau_1^{-1}\tau_2^{-1}\tau_3^{-1}\tau_4^{-1}\tau_5^{-1}=\nonumber\\
&=\tau_5^{-1}\tau_4^{-1}\tau_3^{-1}\tau_2^{-1}\tau_1^{-1}\tau_1^{-1}\tau_2^{-1}\tau_3^{-1}\tau_4^{-1}\tau_5^{-1}=\nonumber \\
&=\tau_5\tau_4\tau_3\tau_2\tau_1\tau_1\tau_2\tau_3\tau_4\tau_5=\tilde{I}
\end{align*}
The last expression is named $\tilde{I}$ which is conjugate to $I$, and the corresponding conjugate element is $\phi=\tau_1\tau_2\tau_3\tau_4\tau_5$. On the other hand, we have $\epsilon=\tau_1\eta$ (definition), and $\tau_{i+1}\eta=\eta\tau_i$ 
(see \ref{fundamental}), we have
\begin{align*}
&\epsilon^4=\tau_1(\eta\tau_1)\eta \tau_1\eta\tau_1\eta=\tau_1 \tau_2 \eta (\eta \tau_1)\eta\tau_1\eta=\tau_1\tau_2 (\eta \tau_2) \eta \eta\tau_1\eta \nonumber\\
&=\tau_1\tau_2 \tau_3 \eta \eta (\eta\tau_1)\eta=\tau_1\tau_2 \tau_3 \eta \eta \tau_2 \eta \eta =\ldots=\tau_1\tau_2\tau_3\tau_4 \eta^4=\eta^5.
\end{align*}

\end{enumerate}

In the following, we'd like to list some useful conjugation transformation.
Let $\zeta_{a,b}=\prod_{i=a}^b \tau_i$, then we have
\begin{enumerate}
\item For any $a,b$ such that $1\leq a<b\leq 5$, we have 
 \begin{equation*}
C(\zeta_{a,b})(\tau_i)=\begin{cases} \tau_i,~~~~if~i<a-1 \\ \tau_{i+1},~~if~a\leq i<b,\\ \tau_i,~~if~i>b+1
\end{cases}
\end{equation*}
\item For any $a,b$ such that $1\leq a<b\leq 5$, let $\eta_{a,b}=\prod_{c=0}^{b-a} \tau_{a,b-c}$, then
 \begin{equation}
C(\eta_{a,b})(\zeta_i)=\begin{cases} \tau_i,~~~~if~i<a-1 \\ \tau_{a+b-i},~~if~a\leq i \leq b,\\ \tau_i,~~if~i>b+1
\end{cases}
\label{conju}
\end{equation}
In particular, one can take $a=1, b=5$, and so its action on the index would be $5\to 1, 4\to 2, 3\to3$.
\item $C(\zeta_{a,b}^2)(\tau_b)=\tau_a$.
\item There is following conjugacy condition 
\begin{equation*}
\zeta_1^{t_1}\zeta_2^{t_2}\ldots \zeta_5^{t_5}\sim \zeta_{s(1)}^{t_{s(1)}} \zeta_{s(1)}^{t_{s(1)}} \ldots \zeta_{s(5)}^{s(t_5)}
\end{equation*}
Here $s$ is a permutation.
\end{enumerate}
One can prove above conjugacy equation by using braid relations.

\subsubsection{Mapping class elements for 4d SCFT}
\begin{figure}[H]
\begin{center}

\tikzset{every picture/.style={line width=0.75pt}} %set default line width to 0.75pt        

\begin{tikzpicture}[x=0.55pt,y=0.55pt,yscale=-1,xscale=1]
%uncomment if require: \path (0,964); %set diagram left start at 0, and has height of 964

%Shape: Ellipse [id:dp2732109334277837] 
\draw   (182,124.61) .. controls (182,105.5) and (222.29,90) .. (272,90) .. controls (321.71,90) and (362,105.5) .. (362,124.61) .. controls (362,143.72) and (321.71,159.22) .. (272,159.22) .. controls (222.29,159.22) and (182,143.72) .. (182,124.61) -- cycle ;
%Curve Lines [id:da1050906401575642] 
\draw    (216,126) .. controls (232,97.22) and (263,109.22) .. (268,127.22) ;
%Curve Lines [id:da3122526712301241] 
\draw    (215,123.22) .. controls (225,141.44) and (250,151.22) .. (270,124.22) ;
%Curve Lines [id:da8178070484931126] 
\draw    (287,127) .. controls (303,98.22) and (334,110.22) .. (339,128.22) ;
%Curve Lines [id:da31112229370476707] 
\draw    (287,115) .. controls (297,133.22) and (319,143.22) .. (339,116.22) ;
%Shape: Ellipse [id:dp0788033866716038] 
\draw  [color={rgb, 255:red, 80; green, 227; blue, 194 }  ,draw opacity=1 ] (273,124.11) .. controls (273,105.39) and (274.57,90.22) .. (276.5,90.22) .. controls (278.43,90.22) and (280,105.39) .. (280,124.11) .. controls (280,142.83) and (278.43,158) .. (276.5,158) .. controls (274.57,158) and (273,142.83) .. (273,124.11) -- cycle ;

\draw (270,55.4) node [anchor=north west][inner sep=0.75pt]    {$\sigma $};

\end{tikzpicture}

\end{center}
\caption{The separating curve on genus two curve.}
\label{separate}
\end{figure}
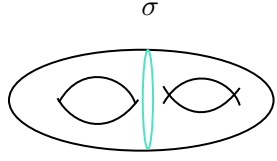

We are going to list the mapping class elements for rank two $\mathcal{N}=2$ SCFT realized as the periodic map of genus two curve. We'd like to find a factorization of 
the mapping class element so that the number of $I_1$ and $\tilde{I}_1$ singularities are given as (see next section for the derivation):
\begin{align}
&\# I_1= 2\delta_x-d_x \nonumber\\
&\# \tilde{I}_1= d_x-\delta_x 
\end{align}
Here the $I_1$ singularity is represented by the Dehn twist along any closed curve whose homology class is nontrivial, and 
$\tilde{I}_1$ singularity is given by the Dehn twist along the cycle with trivial homology class.
This problem has been studied in \cite{hirose2010presentations, nakamura2018generation, dhanwani2023factoring, sakalli2023singular} and the results are listed in table. [\ref{fac1}, \ref{fac2}, \ref{fac3}, \ref{fac4}]. One of the 
crucial relation is the following representation of the Dehn twist along the separating curve:
\begin{equation}
\boxed{(\tau_1\tau_2)^6=\sigma}
\end{equation}
Here $\sigma$ is the Dehn twist along the non-separating curve, see figure. \ref{separate}.

\begin{table}[H]
\begin{center}
  \begin{tabular}{ |c|c|c|c| c| }
    \hline
   Monodromy & Data & Singularity &Scaling dimension & $(d_x, \delta_x)$\\ \hline
    $\eta=\tau_1\tau_2\tau_3\tau_4$& $\frac{1}{10}+\frac{2}{5}+\frac{1}{2}$ & $y^2=x^5+t$ &$(\frac{10}{7}, \frac{8}{7})$  & $(4,4)$ \\     \hline
   $\eta^2$& $\frac{1}{5}+\frac{2}{5}+\frac{2}{5}$  &$y^2=x^5+t^2$ &$(\frac{5}{2}, \frac{3}{2})$ & $(8,8)$ \\     \hline
      $\eta^3$& $\frac{4}{5}+\frac{7}{10}+\frac{1}{2}$ & $y^2=x^5+t^3$ &$(10,4)$ & $(12,12)$ \\     \hline
   $\phi_{\tilde{4}}=\eta^4= \tau_5^2\tau_4\tau_3\tau_2\tau_1$& $\frac{1}{5}+\frac{1}{5}+\frac{3}{5}$ &&$(\frac{5}{3},\frac{4}{3})$ & $(6,6)$  \\     \hline
   $\eta^5=\tilde{\text{I}}$& $\frac{1}{2}+\frac{1}{2}+\frac{1}{2}+\frac{1}{2}$ &&$(2,2)$ & $(10,10)$  \\     \hline
   $\eta\tilde{\text{I}}$& $\frac{4}{5}+\frac{4}{5}+\frac{2}{5}$  && $(5,4)$ & $(14,14)$\\     \hline
      $\eta^7$& $\frac{1}{5}+\frac{3}{10}+\frac{1}{2}$ && $(\frac{10}{4},\frac{4}{3})$ & $(8,7)$ \\     \hline
   $\phi_{\tilde{4}}\phi_{\tilde{4}}$& $\frac{4}{5}+\frac{3}{5}+\frac{3}{5}$ &&$(5,3)$ & $(12,12)$ \\     \hline
   $\tilde{\text{I}}\phi_{\tilde{4}}$& $\frac{9}{10}+\frac{3}{5}+\frac{1}{2}$ & &$(10,8)$ & $(16,16)$ \\     \hline
   $\eta^{10}=1$&~& & &\\ \hline
  \end{tabular}
\end{center}
\caption{Factorization for periodic maps. This family is generated by the factorization $\eta$.}
 \label{fac1}
\end{table}

\begin{table}[H]
\begin{center}
  \begin{tabular}{ |c|c|c|c|c|  }
    \hline
   Monodromy & Data & Singularity & Scaling dimension & $(d_x, \delta_x)$)\\ \hline
    $\epsilon=\tau_1^2\tau_2\tau_3\tau_4$& $\frac{1}{8}+\frac{3}{8}+\frac{1}{2}$ & $y^2=x(x^4+t)$ & $(\frac{8}{5},\frac{6}{5}) $ & $(5,5)$\\     \hline
   $\epsilon^2$& $\frac{1}{2}+\frac{1}{2}+\frac{1}{4}+\frac{3}{4}$  &$y^2=x(x^4+t^2)$ &$(4,2)$ & $(10,10)$\\     \hline
      $\epsilon^3 = \tau_5^2\tau_4\tau_3\tau_2$& $\frac{1}{8}+\frac{3}{8}+\frac{1}{2}$  &$y^2=x(x^4+t^3)$ &$(\frac{8}{5},\frac{6}{5}) $ & $(5,5)$ \\     \hline
      $\epsilon^4=\tilde{I}$& $\frac{1}{2}+\frac{1}{2}+\frac{1}{2}+\frac{1}{2}$ &&$(2,2)$ & $(10,10)$ \\     \hline
   $\epsilon \tilde{I}$& $\frac{7}{8}+\frac{5}{8}+\frac{1}{2}$ && $(8,6)$ & $(15,15)$ \\     \hline
      $(\tau_5^2\tau_4\tau_3\tau_2)^2$& $\frac{1}{2}+\frac{1}{2}+\frac{3}{4}+\frac{1}{4}$ && $(4,2)$ & $(10,10)$\\     \hline
            $ \tau_5^2\tau_4\tau_3\tau_2 \tilde{I}$& $\frac{7}{8}+\frac{5}{8}+\frac{1}{2}$ &&$(8,6)$ &  $(15,15)$ \\     \hline
      $\epsilon^{8}=1$&~&& &\\ \hline
  \end{tabular}
\end{center}
 \caption{Factorization for periodic maps. This family is generated by the factorization $\epsilon$.}
  \label{fac2}
  \end{table}

\begin{table}[H]
\begin{center}
  \begin{tabular}{ |c|c|c| c| c|}
    \hline
   Monodromy & Data & Singularity &  Scaling dimension & $(d_x, \delta_x)$ \\ \hline
    $\zeta=\tau_1\tau_2\tau_3\tau_4\tau_5$& $\frac{1}{6}+\frac{1}{6}+\frac{2}{3}$ & $y^2=x^6+t$ &$(\frac{3}{2},\frac{5}{4})$ & $(5,5)$  \\     \hline
   $\zeta^2$& $\frac{1}{3}+\frac{2}{3}+\frac{2}{3}+\frac{1}{3}$ & $y^2=x^6+t^2$ & $(3,2)$& $(10,10)$  \\     \hline
      $\zeta^3=\sigma 1_{32}2_{43} 3_{54}$& $(n=2,g^{'}=1, \frac{1}{2}+\frac{1}{2})$ & $y^2=x^6+t^3$ &$(2,1)$ & $(5,4)$   \\     \hline
      $\zeta^4=\phi_A=(\tau_5\tau_4\tau_3\tau_2\tau_1)^2$& $\frac{1}{3}+\frac{2}{3}+\frac{2}{3}+\frac{1}{3}$ & &$(3,2)$ & $(10,10)$  \\     \hline
   $\phi_A\zeta$& $\frac{5}{6}+\frac{5}{6}+\frac{1}{3}$ &&$(6,5)$& $(15,15)$  \\     \hline
      $\zeta^{6}=1$&~&&& $~$ \\ \hline
  \end{tabular}
\end{center}
  \caption{Factorization for periodic maps. This family is generated by the factorization $\zeta$.}
  \label{fac3}
\end{table}

\begin{table}[H]
\begin{center}
  \begin{tabular}{ |c|c| c| c|}
    \hline
   Monodromy & Data & Scaling dimension & $(d_x, \delta_x)$ \\ \hline
     $\phi_4=\tau_1\tau_2\tau_4^{-1}\tau_5^{-1}$& $\frac{1}{2}+\frac{1}{2}+\frac{2}{3}+\frac{1}{3}$ &$ (6,2)$ & $(10,9)$   \\     \hline
    $\phi_4^2$  & $\frac{1}{3}+\frac{1}{3}+\frac{2}{3}+\frac{2}{3}$  & $(3,2)$ & $(10,10)$\\ \hline
    $\phi_4^3=I$ & $\frac{1}{2}+\frac{1}{2}+\frac{1}{2}+\frac{1}{2}$ & $(2,2)$ & $(10,10)$\\ \hline
        $\phi_4^4\sim \phi_4^2$ & $\frac{1}{3}+\frac{1}{3}+\frac{2}{3}+\frac{2}{3}$ & $(3,2)$ & $(10,10)$\\ \hline
    $\phi_4^5 \sim \phi_4 $ & $\frac{1}{2}+\frac{1}{2}+\frac{2}{3}+\frac{1}{3}$ & $(6,2)$ &$(10,9)$\\ \hline
        $\phi_4^6 =1 $ & $~$&& \\ \hline
  \end{tabular}
\end{center}
  \caption{Factorization for periodic maps. This family is generated by the factorization $\phi_4$.}
  \label{fac4}
\end{table}

%\begin{center}
  %\begin{tabular}{ |c|c| c| c|}
    %\hline
  % Monodromy & Data & Singularity & Scaling dimension\\ \hline
  %$\phi_5=\tau_1^2\tau_2\tau_3\tau_4\tau_5$& $\frac{1}{5}+\frac{1}{5}+\frac{3}{5}$ & $y^2=x(x^5+t)$ & $(\frac{5}{3},\frac{4}{3})$ \\     \hline
    %$\phi_5^2$  & $\frac{4}{5}+\frac{3}{5}+\frac{3}{5}$ &$y^2=x(x^5+t^2)$  & $(5,3)$ \\ \hline
    %$\phi_5^3 \sim \tau_4\tau_{a_2}\tau_3\tau_5\tau_2\tau_4\tau_{b_2} \tau_{b_2^{'}} $ & $\frac{1}{5}+\frac{2}{5}+\frac{2}{5}$ && $(\frac{5}{2},\frac{3}{2})$ \\ \hline
      %  $\phi_5^4$ & $\frac{4}{5}+\frac{4}{5}+\frac{2}{5}$ &&$(5,4)$ \\ \hline
    %$\phi_5^5 =1$ & $~$ \\ \hline
  %\end{tabular}
%\end{center}

Let's verify some of the results in  table. [\ref{fac1}, \ref{fac2}, \ref{fac3}, \ref{fac4}]:
\begin{enumerate}
\item The first one would be the factorization of $\eta^4$:
\begin{align*}
\eta^4=(\zeta^{-1}\tau_1^{-1}\zeta^2)(\zeta^{-2} \tau_2^{-1} \zeta^3)\ldots (\zeta^{-4}\tau_4^{-1} \zeta^{5})=\zeta^{-1}\tau_1^{-1}\tau_2^{-1}\tau_3^{-1}\tau_4^{-1}\zeta^{-1}
\end{align*}
Here we used equation $\eta=\zeta^{-i}\tau_i^{-1}\zeta^{i+1}$, and $\zeta^5=\zeta^{-1}$.  We then use the relation $I^2=1$ to express above expression 
in terms of positive product  of generators:
\begin{align*}
&1=I^2=\tau_1\tau_2\tau_3\tau_4\tau_5 (\tau_5\tau_4\tau_3\tau_2\tau_1 \tau_1\tau_2\tau_3\tau_4\tau_5) \textcolor{red}{\tau_5}\tau_4\tau_3\tau_2\tau_1 \nonumber\\
&=(\tau_1\tau_2\tau_3\tau_4\tau_5) \tau_5 \tau_5\tau_4\tau_3\tau_2\tau_1 \tau_1\tau_2\tau_3\tau_4\tau_5 \tau_4\tau_3\tau_2\tau_1 \nonumber\\
&=\tau_5^2 \tau_4\tau_3\tau_2\tau_1( \tau_1\tau_2\tau_3\tau_4\tau_5 \tau_4\tau_3\tau_2\tau_1 \tau_1\tau_2\tau_3\tau_4\tau_5) \nonumber\\
&=\tau_5^2 \tau_4\tau_3\tau_2\tau_1 (\zeta \tau_4\tau_3\tau_2\tau_1 \zeta)\to \nonumber\\
&\zeta^{-1}\tau_1^{-1}\tau_2^{-1}\tau_3^{-1}\tau_4^{-1}\zeta^{-1}=\tau_5^2 \tau_4\tau_3\tau_2\tau_1
\end{align*}
Here in the first step we used the fact $\tilde{I}=\tau_5\tau_4\tau_3\tau_2\tau_1 \tau_1\tau_2\tau_3\tau_4\tau_5$ commutes with  all the generators (in particular $\tau_5$),
and in the second step one used the cyclic equivalence by moving letters $\tau_1\tau_2\tau_3\tau_4\tau_5$ to the end of the word.
\item Second, let's compute $\epsilon^3$. Since $\epsilon^4=\tilde{I}$, and so
\begin{equation*}
\epsilon^3=\epsilon^{-1}\tilde{I}^{-1}
\end{equation*}
Next, we'd like to use the relation 
\begin{align*}
&1=\tilde{I}^2=\tilde{I} \tau_5\tau_4\tau_3\tau_2\tau_1 \tau_1 \tau_2\tau_3\tau_4 \tau_5=\tau_5^2\tau_4\tau_3\tau_2\tilde{I} \tau_1^2\tau_2\tau_3\tau_4=\tau_5^2\tau_4\tau_3\tau_2 \tilde{I} \epsilon \nonumber\\
&\to \epsilon^{-1}\tilde{I}^{-1}=\tau_5^2\tau_4\tau_3\tau_2
\end{align*}
Here in the first step we used the fact that $\tilde{I}$ commutes with any generators, and in the second step one used conjugation to move $\tau_5$ from the end to the beginning (cyclic relation).
\item Thirdly, we'd like to compute the word $\zeta^3$.
\begin{align*}
&\zeta^3=12(345 123)45 12345=1213243(545)12345=12132(434)5412345 \nonumber\\
&=121(323)435412345=12123243(5412)345=(12)^23243 12 5(434)5=(12)^23243 12 5(343)5 \nonumber\\
&=(12)^23243 12 3(545)3=(12)^23241(3 2 3)4543=(12)^23(212) (43 4)2543=(12)^2 1(323)1 43 2543 \nonumber\\
&=(12)^212321432543=(12)^3321432543=(12)^6 \overline{12}^3321432543
\end{align*}
We then need to compute following word:
\begin{align*}
&\overline{12}^3321432543=\bar{2}\bar{1}\bar{2}\bar{1}\bar{2}\bar{1}321432543=\bar{2}\bar{1}\bar{2}\bar{1}\bar{2}3(\bar{1}21)432543 =\bar{2}\bar{1}\bar{2}\bar{1}(\bar{2}32)14(\bar{2}32)543\nonumber\\
&=\bar{2}\bar{1}\bar{2}3(\bar{1}21)(\bar{3}43)25(\bar{3}43)=\bar{2}\bar{1}\bar{2}3(21\bar{2})(43\bar{4})25(43\bar{4})=\bar{2}\bar{1}(\bar{2}32)14(\bar{2}32)(\bar{4}54)3\bar{4} \nonumber\\
&=\bar{2}\bar{1}(32\bar{3})14(32\bar{3})(54\bar{5})3\bar{4}=\bar{2}3(\bar{1}21)\bar{3}4325(\bar{3}43)\bar{5}\bar{4}=(\bar{2}32)1\bar{2}\bar{3}432543(\bar{4}\bar{5}\bar{4})\nonumber\\
&=321(\bar{3}\bar{2}\bar{3})432(54\bar{5})3\bar{4}\bar{5}=321\bar{2}\bar{3}(\bar{2}432\bar{4})543\bar{4}\bar{5}=1_{32} \cdot (\bar{2}432\bar{4}) \cdot 3_{54} \nonumber\\
&=1_{32} \cdot (432 \bar{3}\bar{4}) \cdot1_{45}=1_{32} \cdot 2_{43} \cdot 3_{54}
\end{align*}
In the process of the computation, we  used the following braid relation:
\begin{equation*}
\bar{\tau}_{i+1} \tau_i \tau_{i+1}=\tau_{i} \tau_{i+1} \bar{\tau}_{i}
\end{equation*}
which is easily derived from the standard braid relation. So finally one has the following important result
\begin{equation}
\zeta^3=\sigma \cdot 1_{32} \cdot 2_{43} \cdot 3_{54}
\label{crucial}
\end{equation}

\item We'd like to compute $\eta^7$:
\begin{align*}
&\eta^7=\tilde{I} \eta^2=\tau_5\tau_4\tau_3\tau_2\tau_1 \tau_1\tau_2\tau_3\tau_4\tau_5 \tau_1\tau_2\tau_3\tau_4 \tau_1\tau_2\tau_3\tau_4 = \nonumber\\
&\tau_5\tau_4\tau_3\tau_2\tau_1\tau_1\tau_2\tau_3\tau_4\tau_5 \tau_1\tau_2\tau_3\tau_4 (\tau_5 \bar{\tau}_5) \tau_1\tau_2\tau_3\tau_4 (\tau_5\bar{\tau}_5)= \nonumber\\
&\tau_5\tau_4\tau_3\tau_2\tau_1(\tau_1\tau_2\tau_3\tau_4\tau_5)^2 \tau_1\tau_2\tau_3(\bar{\tau}_5  \tau_4 \tau_5)\bar{\tau}_5 = \nonumber \\
&\tau_5\tau_4\tau_3\tau_2\tau_1(\tau_1\tau_2\tau_3\tau_4\tau_5)^2 \tau_1\tau_2\tau_3 \tau_4 \tau_5\bar{\tau}_4\bar{\tau}_5= \nonumber\\
&\tau_5\tau_4[\tau_3\tau_2\tau_1(\tau_1\tau_2\tau_3\tau_4\tau_5)^3 ]\bar{\tau}_4\bar{\tau}_5
\end{align*}
So $\eta^7$ is conjugate to the element $\tau_3\tau_2\tau_1(\tau_1\tau_2\tau_3\tau_4\tau_5)^3= \tau_3\tau_2\tau_1 \zeta^3$, and so it has a factorization with $6$ $I_1$ singularities and one $\tilde{I}_1$ singularity by 
using the factorization of $\zeta^3$. This is also consistent with the local invariant of degeneration $(\frac{3}{10}+\frac{1}{5}+\frac{1}{2})$: $(d_x=8, \delta_x=7)$.

\item Let's now verify $\phi_4^3=I$. First, we have trivial relation
\begin{equation*}
\tau_1^{-1}\tau_2^{-1}\tau_3^{-1}\tau_4^{-1}\tau_5^{-1}=\zeta I \to \tau_4^{-1}\tau_5^{-1}=\tau_3\tau_2\tau_1 \zeta I
\end{equation*}
and so 
\begin{align*}
& \phi_4=(\tau_1\tau_2)(\tau_4^{-1}\tau^{-1}_5)=(\zeta \tau_5^{-1}\tau_4^{-1}\tau_3^{-1})(\tau_3\tau_2\tau_1 \zeta I)=\zeta \tau_5^{-1}\tau_4^{-1} \tau_2\tau_1 \zeta I \nonumber\\
&=(\zeta \tau_2) \tau_5^{-1} \tau_1 (\tau_4^{-1} \zeta) I=\tau_3 (\zeta \tau_5^{-1}) (\tau_1 \zeta) \tau_3^{-1} I=\tau_3\eta \zeta^2 \eta^{-1} \tau_3^{-1} I \nonumber\\
&=(\tau_3 \eta) \zeta^2(\tau_3\eta)^{-1} I
\end{align*}
here we used the relation $\tau_1\zeta=\zeta^2\eta^{-1}$. So $\phi_4^3=I$ (notice that $I$ is the central element of the mapping class group), by using $\zeta^6=1, I^2=1$.  On the other hand, using the above formula, we find that
\begin{align}
&\phi_4=(\tau_3 \eta) \zeta^2(\tau_3\eta)^{-1} I=(\tau_3 \eta) \zeta^3 \zeta^{-1} I (\tau_3 \eta) ^{-1} \nonumber\\
&=(\tau_3 \eta) \zeta^3 (\tau_5\tau_4 \tau_3 \tau_2 \tau_1) (\tau_3 \eta) ^{-1} 
\end{align}
Now use the fact that $\zeta^3$ can be factorized into a $\tilde{I}_1$ singularity and three $I_1$ singularity, one see from above formula that $\phi_4$ can be factorized into a $\tilde{I}_1$ singularity and eight $I_1$ singularity.
This agrees with the result from the local invariant ($d_x=10, \delta_x=9$).

\end{enumerate}

\textbf{Singular configurations for SCFT}: Once we find out the desired factorization for mapping class group elements of SCFT, one can find various singular configurations of them by doing braid moves and Hurwitz moves.
Here let's just give several simple examples.

\textit{Example 1}: Consider the theory whose word is $\zeta=\tau_1\tau_2\tau_3\tau_4\tau_5$, one can have following singular configuration: a): $(\tau_1\tau_2\tau_3\tau_4) \tau_5$, namely 
there is a AD theory represented by $\eta=(\tau_1\tau_2\tau_3\tau_4)$, and a $I_1$ singularity; b):  $(\tau_1\tau_2)\tau_3\tau_4 \tau_5$, namely there is a rank one AD theory represented by $\tau_1\tau_2$, 
and three $I_1$ singularities.

\textit{Example 2}: Consider the theory whose word is $(\tau_1\tau_2\tau_3\tau_4\tau_5)^2$, one can have following singular configuration: a): $(\tau_1\tau_2\tau_3\tau_4 \tau_5)(\tau_1 \tau_2\tau_3 \tau_4 \tau_5)$, namely 
there are two  rank two AD theories represented by the word $\zeta$; b): $(\tau_1\tau_2\tau_3\tau_4) \tau_5 (\tau_1 \tau_2\tau_3 \tau_4) \tau_5$, namely there are two AD theory represented by the word $\eta$, and 
two extra $I_1$ singularities.

It is possible to use the braid move and Hurwitz move to get undeformable singularities, see following examples for $I_n$ type. Here we use braid moves 
to give the configuration with four identical letter, and then one can use Hurwitz moves to move those letters together (moving the letters from left to right). The $I_4$ series 
corresponding to scaling dimension $(5,3)$ and $(4,2)$ were discussed in \cite{Xie:2023zxn}.
\begin{align*}
&(8,6):~~\bm{\tau_1^2}\tau_2\tau_3\tau_4(\tau_5\tau_4\tau_3\tau_2\bm{\tau_1^2}\tau_2\tau_3\tau_4\tau_5) \nonumber\\
&(10,4):~~\tau_1\tau_2\tau_3\tau_4\tau_1\tau_2\tau_3(\tau_4 \tau_1\tau_2\tau_3\tau_4)=\tau_1\tau_2 \bm{\tau_3}\tau_4\tau_1\tau_2 \bm{\tau_3} \tau_1\tau_2 \bm{\tau_3}\tau_4 \bm{\tau_3}  \nonumber\\
%&~~~=\tau_1 \tau_5\tau_4\tau_3\tau_2\tau_1^2\tau_2\tau_3(\tau_4\tau_5 \tau_1\tau_2\tau_3\tau_4)= \tau_1 \tau_5\tau_4\tau_3\tau_2\tau_1^2\tau_2\tau_3 \tau_1\tau_2\tau_3\tau_4(\tau_5\tau_3\tau_4\bar{\tau}_5)\nonumber\\
%&~~~~=\tau_1 \tau_5\tau_4\bm{\tau_3}\tau_2\tau_1^2\tau_2\bm{\tau_3} \tau_1\tau_2\bm{\tau_3}\tau_4 \bm{\tau_3}\tau_{5(4)} \nonumber\\
%&(5,4):~~(\tau_5\tau_4\tau_3\tau_2\tau_1^2\tau_2\tau_3\tau_4\tau_5)\tau_1\tau_2\tau_3\tau_4= \tau_5\tau_4\tau_3\tau_2\bm{\tau_1^2 \tau_1}\tau_2\tau_3\tau_4 \bm{\tau_1}\tau_2\tau_3\tau_{5(4)} \nonumber\\
&(5,3):~~(\bm{\tau_1^2}\tau_2\tau_3\tau_4\tau_5 )(\bm{\tau_1^2}\tau_2\tau_3\tau_4\tau_5) \nonumber\\
%&(3,2):~~\tau_1(\tau_2\tau_3\tau_4\tau_5 \tau_1\tau_2\tau_3\tau_4\tau_5)=\tau_1^2(\tau_2\tau_3\tau_4\tau_5 \tau_1\tau_2\tau_3\tau_4)=\nonumber\\
%&~~~~~~~~=\bm{\tau_1^2}\tau_2\tau_3\tau_4\tau_5 \bm{\tau_1}\tau_2\tau_3\tau_4 \bm{ \tau_1}\tau_2\tau_3 \tau_{5(4)} \nonumber\\
& (4,2):~~~ (\bm{\tau_1^2}\tau_2\tau_3\tau_4) (\bm{\tau_1^2}\tau_2\tau_3\tau_4)    \nonumber\\
&(2,2):~~\tau_1\tau_2\tau_3\tau_4\tau_5 \tau_5 \tau_4\tau_3\tau_2\tau_1 \sim \tau_2 \tau_1^2 \tau_2\tau_3\tau_4\tau_5^2  \tau_4\tau_3=\tau_{2(1)}^2 \tau_2^2\tau_3\tau_4 \tau_5^2 \tau_4 \tau_3 \nonumber \\
&=\tau_{2(1)}^2 \tau_2^2\tau_3 \tau_{4(5)}^2 \tau_4^2 \tau_3=\tau_{2(1)}^2 \tau_2^2 \tau_{34(5)}^2 \tau_{3(4)}^2 \tau_3^2
\end{align*}
There are also underformable singularities of $Z_2$ type \cite{Xie:2023out}, which is given by mapping class group element $I$. 
By looking at table. [\ref{fac1}, \ref{fac2}, \ref{fac3}, \ref{fac4}], i.e. theory with scaling dimension $(10,4)$, $(8,6)$ and $(5,4)$.

\subsubsection{Global SW geometry}
The global SW geometry is given by the positive factorization of the identity element in mapping class group. 
Since $I^2=1$, the first choice is 
\begin{equation}
(\tau_1\tau_2\tau_3\tau_4 \tau_5^2 \tau_4 \tau_3\tau_2\tau_1)^2=1.
\label{type1}
\end{equation}
This is the $(20,0)$ type as there are a total of 20 $I_1$ singularities. The second choice would be $\eta^{10}=\zeta^6=1$, and since one need to have the topological constraint on the number of $I_1$ and $\tilde{I_1}$ singularities, we use the equivalent factorization
\begin{equation}
 \eta^7 \eta^3=[\tau_5\tau_4[\tau_3\tau_2\tau_1(\tau_1\tau_2\tau_3\tau_4\tau_5)^3 ]\bar{\tau}_4\bar{\tau}_5] \eta^3=
[\tau_5\tau_4[\tau_3\tau_2\tau_1\sigma \cdot 1_{32} \cdot 2_{43} \cdot 3_{54}] \bar{\tau}_4\bar{\tau}_5] \eta^3=1.
\label{type2}
\end{equation}
This is the $(18,1)$ type. Finally, we have the relation $\phi_4^6=1$, and so $\phi_4^{5}=\phi_4^{-1}=\tau_5\tau_4\tau_2^{-1}\tau_1^{-1}$, 
and one can find an element in mapping class group so that its action would change the index as $5\to 1,~4\to2$ (see \ref{conju}, and take $a+b=6$),
so $\phi_4^{-1}$ is conjugate with $\phi_4$, and its factorization involves a Dehn twist along $\sigma$ and eight Dehn twists along non-separating curves. So the factorization is just
\begin{equation}
\phi_4 \phi_4^{-1}=1.
\label{type3}
\end{equation}
There are now two $\sigma$ Dehn twists in above factorization, and so
it gives the $(16,2)$ type.

To find the global configuration for the SCFT, one need to rearrange the above  configuration to get a sensible singular fiber 
at $\infty$. Using the data in table. [\ref{fac1},\ref{fac2},\ref{fac3},\ref{fac4}], one can easily find the results, see table. \ref{gl1}.

\begin{table}
\begin{center}
  \begin{tabular}{ |c|c| }
    \hline
   Factorization & Theory\\ \hline
            $(\tau_1^2\tau_2\tau_3\tau_4, \tau_1^2\tau_2\tau_3\tau_4(\tau_5^2\tau_4\tau_3\tau_2)^2)$ & $((\frac{8}{5},\frac{6}{5}),(8,6))$  \\ \hline
  $(\tau_1\tau_2\tau_3\tau_4, \tau_5^2\tau_4\tau_3\tau_2\tau_1I)$ & $((\frac{8}{7},\frac{10}{7}),(10,8))$  \\ \hline
    $(\tau_1\tau_2\tau_3\tau_4\tau_5, \tau_5\tau_4\tau_3\tau_2\tau_1I)$ & $((\frac{3}{2},\frac{5}{4}),(6,5))$  \\ \hline
        $(\tau_5^2\tau_4\tau_3\tau_2\tau_1, \tau_1\tau_2\tau_3\tau_4\tau_5^2\tau_4\tau_3\tau_2\tau_1 \tau_1\tau_2\tau_3\tau_4)$ & $((\frac{3}{2},\frac{5}{4}),(5,4))$  \\ \hline
       $(\tau_1\tau_2\tau_3\tau_4 \tau_5^2 \tau_4 \tau_3\tau_2\tau_1,\tau_1\tau_2\tau_3\tau_4 \tau_5^2 \tau_4 \tau_3\tau_2\tau_1)$ & $((2,2),(2,2))$  \\ \hline
      $((\tau_1\tau_2\tau_3\tau_4\tau_5)^2, (\tau_5\tau_4\tau_3\tau_2\tau_1)^2)$ & $((3,2),(3,2))$ \\ \hline
          $((\tau_1^2\tau_2\tau_3\tau_4)^2, (\tau_5^2\tau_4\tau_3\tau_2)^2)$ & $((4,2),(4,2))$  \\ \hline
        $((\tau_1\tau_2\tau_3\tau_4)^2, (\tau_5^2\tau_4\tau_3\tau_2\tau_1)^2)$ & $((\frac{5}{2},\frac{3}{2}),(5,3))$  \\ \hline
                $(\eta^7, (\tau_1\tau_2\tau_3\tau_4)^3)$ & $((\frac{10}{3},\frac{4}{3}),(10,4))$  \\ \hline
                    $(\phi_4,\phi_4^{-1})$ & $((6,2),(6,2))$  \\ \hline
  \end{tabular}
  \end{center}
  \caption{The global SW geometry for a pair of rank two SCFTs. We use scaling dimension to denote the theory. }
  \label{gl1}
  \end{table}

Let's show some moves to derive the equivalent factorization of identity which is used to derive the result in table. \ref{gl1}. 
\begin{enumerate}
\item First, we have the following equivalent factorization:
\begin{equation*}
1=I^2=\tau_1\tau_2\tau_3\tau_4 \tau_5^2 \tau_4 \tau_3\tau_2\tau_1 I=\tau_1\tau_2\tau_3\tau_4 \tau_5 I \tau_5 \tau_4\tau_3 \tau_2 \tau_1=(\tau_1\tau_2\tau_3\tau_4\tau_5)^2(\tau_5\tau_4\tau_3\tau_2\tau_1)^2.
\end{equation*}
The fact $I$ commuting with all the generators is used.
\item Secondly, we have the following factorization (see table. \ref{fac2}):
\begin{equation*}
1=\epsilon^8=\epsilon^2 (\epsilon^3)^2=(\tau_1^2\tau_2\tau_3\tau_4)^2(\tau_5^2\tau_4\tau_3\tau_2)^2.
\end{equation*}
\end{enumerate}

\textbf{Other 4d $\mathcal{N}=2$ SCFTs}:
Let's now give the global SW geometry of all rank two $\mathcal{N}=2$ SCFT discussed in section 2.
We have found the configuration for theories engineered using periodic maps. Here 
we'd list the global SW geometry for other choices, see table. \ref{gl2}. The important difference is that 
the fiber at $\infty$ is no longer given by  the periodic map. We first use the following factorization of identity 
\begin{equation*}
\zeta^6=1,
\end{equation*}
here $\zeta=\tau_1\tau_2\tau_3\tau_4\tau_5$. We then use the factorization $\zeta^3=\sigma \cdot 1_{32} \cdot 2_{43} \cdot 3_{54}$, 
and finally the fact $\sigma=(\tau_1\tau_2)^6$, and so the factorization of identity becomes 
\begin{equation*}
\zeta^3(\zeta)^3=\sigma \cdot 1_{32} \cdot 2_{43} \cdot 3_{54}\cdot \sigma 1_{32}2_{43} 3_{54}= \sigma \cdot 1_{32} \cdot 2_{43} \cdot 3_{54}\cdot (\tau_1\tau_2)^6 \cdot 1_{32} \cdot 2_{43} \cdot 3_{54}
\end{equation*}
We then split the middle factor $(\tau_1\tau_2)^6$ to the bulk and infinity.

\begin{table}[H]
\begin{center}
  \begin{tabular}{ |c|l| }
    \hline
   Scaling dimension & $[I_\infty, bulk]$\\ \hline
   $(4,3)$ &$[\tau_1\tau_2\tau_3\tau_4\tau_5\tau_1\tau_2\tau_3,~\tau_4\tau_5(\tau_5\tau_4\tau_3\tau_2\tau_1)^2]$ \\ \hline
   $(6,4)$ & $[\tau_1\tau_2\tau_3\tau_4\tau_1\tau_2\tau_3,~\tau_5\tau_4\tau_5(\tau_5\tau_4\tau_3\tau_2\tau_1)^2$ \\ \hline
   $(12,6)$ &$[\zeta^3\tau_1\tau_2,~(\tau_1\tau_2)^5 \zeta_5]$  \\ \hline
   $(8,4)$ & $[\zeta^3\tau_1\tau_2\tau_1,~\tau_2(\tau_1\tau_2)^4 \zeta_5]$ \\ \hline
   $(6,3)$ & $[\zeta^3(\tau_1\tau_2)^2,~(\tau_1\tau_2)^4 \zeta_5]$ \\ \hline
   $(4,2)$ & $[\zeta^3(\tau_1\tau_2)^3,~(\tau_1\tau_2)^3 \zeta_5]$ \\\hline
 $(3,\frac{3}{2})$ &$[\zeta^3(\tau_1\tau_2)^4,~(\tau_1\tau_2)^2 \zeta_5]$ \\ \hline
$ (\frac{8}{3},\frac{4}{3})$ & $[\zeta^3(\tau_1\tau_2\tau_1)^3,~\tau_2(\tau_1\tau_2) \zeta_5]$ \\ \hline
$(\frac{12}{5},\frac{6}{5})$ & $[\zeta^3(\tau_1\tau_2)^5,~(\tau_1\tau_2)\zeta_5]$ \\ \hline
  \end{tabular}
  \end{center}
  \caption{Here $\zeta_5=1_{32}2_{43} 3_{54}$, and one use the factorization of $\zeta^3=\sigma \cdot 1_{32} \cdot 2_{43} \cdot 3_{54}$ so that the type would be $(18,1)$. We used the fact $\sigma=(\tau_1\tau_2)^6$.}
  \label{gl2}
  \end{table}

\section{Representation of mapping class group of genus two}
To have a complete understanding of the mapping class group elements for rank two theory, one need to have 
several useful representation for $M_2$. In this section, we are going to discuss three important representation: a): the action on homology groups; 
b): Jones representation; c): the signature function. We then use these representations to discuss the candidate configuration for 4d asymptotical free theories, 
5d and 6d KK theories.

\subsection{Symplectic representation}
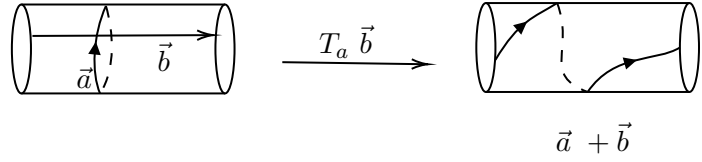
\begin{figure}[H]
\begin{center}

\tikzset{every picture/.style={line width=0.75pt}} %set default line width to 0.75pt        

\begin{tikzpicture}[x=0.55pt,y=0.55pt,yscale=-1,xscale=1]
%uncomment if require: \path (0,799); %set diagram left start at 0, and has height of 799

%Shape: Ellipse [id:dp6330496235211077] 
\draw   (55,130.94) .. controls (55,114.09) and (57.91,100.44) .. (61.5,100.44) .. controls (65.09,100.44) and (68,114.09) .. (68,130.94) .. controls (68,147.78) and (65.09,161.44) .. (61.5,161.44) .. controls (57.91,161.44) and (55,147.78) .. (55,130.94) -- cycle ;
%Straight Lines [id:da4370056106496113] 
\draw    (61.5,100.44) -- (200,100.44) ;
%Shape: Ellipse [id:dp8876209519114929] 
\draw   (193.5,130.94) .. controls (193.5,114.09) and (196.41,100.44) .. (200,100.44) .. controls (203.59,100.44) and (206.5,114.09) .. (206.5,130.94) .. controls (206.5,147.78) and (203.59,161.44) .. (200,161.44) .. controls (196.41,161.44) and (193.5,147.78) .. (193.5,130.94) -- cycle ;
%Straight Lines [id:da4363853671209744] 
\draw    (61.5,161.44) -- (200,161.44) ;
%Curve Lines [id:da7410773815768243] 
\draw    (115,161.44) .. controls (109,147.44) and (110,121.44) .. (119,101.44) ;
\draw [shift={(112.15,125.85)}, rotate = 96.13] [fill={rgb, 255:red, 0; green, 0; blue, 0 }  ][line width=0.08]  [draw opacity=0] (8.93,-4.29) -- (0,0) -- (8.93,4.29) -- cycle    ;
%Curve Lines [id:da7103396528731534] 
\draw  [dash pattern={on 4.5pt off 4.5pt}]  (119,101.44) .. controls (124,118.44) and (126,144.44) .. (115,161.44) ;
%Straight Lines [id:da025243460914341043] 
\draw    (69,122) -- (190,121.45) ;
\draw [shift={(192,121.44)}, rotate = 179.74] [color={rgb, 255:red, 0; green, 0; blue, 0 }  ][line width=0.75]    (10.93,-3.29) .. controls (6.95,-1.4) and (3.31,-0.3) .. (0,0) .. controls (3.31,0.3) and (6.95,1.4) .. (10.93,3.29)   ;
%Straight Lines [id:da9193747828284777] 
\draw    (239,140) -- (336,141.41) ;
\draw [shift={(338,141.44)}, rotate = 180.83] [color={rgb, 255:red, 0; green, 0; blue, 0 }  ][line width=0.75]    (10.93,-3.29) .. controls (6.95,-1.4) and (3.31,-0.3) .. (0,0) .. controls (3.31,0.3) and (6.95,1.4) .. (10.93,3.29)   ;
%Shape: Ellipse [id:dp6241386171435456] 
\draw   (371,129.94) .. controls (371,113.09) and (373.91,99.44) .. (377.5,99.44) .. controls (381.09,99.44) and (384,113.09) .. (384,129.94) .. controls (384,146.78) and (381.09,160.44) .. (377.5,160.44) .. controls (373.91,160.44) and (371,146.78) .. (371,129.94) -- cycle ;
%Straight Lines [id:da5937346877778042] 
\draw    (377.5,99.44) -- (516,99.44) ;
%Shape: Ellipse [id:dp7908429061544529] 
\draw   (509.5,129.94) .. controls (509.5,113.09) and (512.41,99.44) .. (516,99.44) .. controls (519.59,99.44) and (522.5,113.09) .. (522.5,129.94) .. controls (522.5,146.78) and (519.59,160.44) .. (516,160.44) .. controls (512.41,160.44) and (509.5,146.78) .. (509.5,129.94) -- cycle ;
%Straight Lines [id:da19770659517803146] 
\draw    (377.5,160.44) -- (516,160.44) ;
%Curve Lines [id:da03598684194411361] 
\draw    (384,138.44) .. controls (400,111.44) and (410,109.44) .. (426,99.44) ;
\draw [shift={(405.59,112.11)}, rotate = 139.66] [fill={rgb, 255:red, 0; green, 0; blue, 0 }  ][line width=0.08]  [draw opacity=0] (8.93,-4.29) -- (0,0) -- (8.93,4.29) -- cycle    ;
%Curve Lines [id:da45816073435832005] 
\draw  [dash pattern={on 4.5pt off 4.5pt}]  (426,99.44) .. controls (438,131.44) and (415,148.44) .. (446.75,160.44) ;
%Curve Lines [id:da6195620051158065] 
\draw    (446.75,160.44) .. controls (465,133.44) and (493.5,139.94) .. (509.5,129.94) ;
\draw [shift={(480.34,138.46)}, rotate = 164.26] [fill={rgb, 255:red, 0; green, 0; blue, 0 }  ][line width=0.08]  [draw opacity=0] (8.93,-4.29) -- (0,0) -- (8.93,4.29) -- cycle    ;

\draw (97,141.4) node [anchor=north west][inner sep=0.75pt]    {$\vec{a}$};
% Text Node
\draw (152,124.4) node [anchor=north west][inner sep=0.75pt]    {$\vec{b}$};
% Text Node
\draw (262,112.4) node [anchor=north west][inner sep=0.75pt]    {$T_{a} \ \vec{b}$};
% Text Node
\draw (423,179.4) node [anchor=north west][inner sep=0.75pt]    {$\vec{a} \ +\vec{b}$};

\end{tikzpicture}

\end{center}
\caption{The action of Dehn twist on oriented curve.}
\label{twist}
\end{figure}

The Dehn twist action on homology represented by oriented curves is given as:
\begin{equation*}
(T_b)([\vec{a}])=[\vec{a}]+(\vec{a},\vec{b})[\vec{b}].
\end{equation*}
Here $(\vec{a},\vec{b}))$ is the intersection number, see figure. \ref{twist} for the illustration.

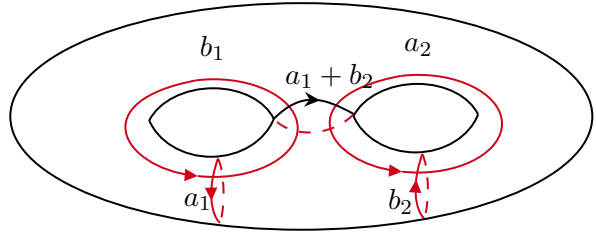
\begin{figure}
\begin{center}

\tikzset{every picture/.style={line width=0.75pt}} %set default line width to 0.75pt        

\begin{tikzpicture}[x=0.55pt,y=0.55pt,yscale=-1,xscale=1]
%uncomment if require: \path (0,799); %set diagram left start at 0, and has height of 799

%Shape: Ellipse [id:dp8019341011682583] 
\draw   (106,159.72) .. controls (106,116.8) and (194.65,82) .. (304,82) .. controls (413.35,82) and (502,116.8) .. (502,159.72) .. controls (502,202.64) and (413.35,237.44) .. (304,237.44) .. controls (194.65,237.44) and (106,202.64) .. (106,159.72) -- cycle ;
%Shape: Arc [id:dp41471600745117865] 
\draw  [draw opacity=0] (285.27,161.39) .. controls (279.42,176.37) and (262.73,187.24) .. (242.97,187.38) .. controls (223.31,187.51) and (206.54,176.97) .. (200.42,162.2) -- (242.71,149.71) -- cycle ; \draw   (285.27,161.39) .. controls (279.42,176.37) and (262.73,187.24) .. (242.97,187.38) .. controls (223.31,187.51) and (206.54,176.97) .. (200.42,162.2) ;  
%Shape: Arc [id:dp12104823499043726] 
\draw  [draw opacity=0] (200.62,162.33) .. controls (207.1,149.53) and (224.38,140.32) .. (244.25,140.39) .. controls (263.47,140.45) and (279.47,149.18) .. (285.27,161.39) -- (242.75,172.51) -- cycle ; \draw   (200.62,162.33) .. controls (207.1,149.53) and (224.38,140.32) .. (244.25,140.39) .. controls (263.47,140.45) and (279.47,149.18) .. (285.27,161.39) ;  

%Shape: Arc [id:dp19913644129561625] 
\draw  [draw opacity=0] (233.59,200.6) .. controls (237.13,201) and (240.8,201.21) .. (244.54,201.21) .. controls (276.02,201.21) and (301.54,186.22) .. (301.54,167.71) .. controls (301.54,149.21) and (276.02,134.21) .. (244.54,134.21) .. controls (240.45,134.21) and (236.46,134.47) .. (232.62,134.95) -- (244.54,167.71) -- cycle ; \draw  [color={rgb, 255:red, 208; green, 2; blue, 27 }  ,draw opacity=1 ] (233.59,200.6) .. controls (237.13,201) and (240.8,201.21) .. (244.54,201.21) .. controls (276.02,201.21) and (301.54,186.22) .. (301.54,167.71) .. controls (301.54,149.21) and (276.02,134.21) .. (244.54,134.21) .. controls (240.45,134.21) and (236.46,134.47) .. (232.62,134.95) ;  
%Curve Lines [id:da5273746829471798] 
\draw [color={rgb, 255:red, 208; green, 2; blue, 27 }  ,draw opacity=1 ]   (230.56,200.29) .. controls (181.63,194.47) and (156.55,149.15) .. (232.62,134.95) ;
\draw [shift={(233.59,200.6)}, rotate = 184.78] [fill={rgb, 255:red, 208; green, 2; blue, 27 }  ,fill opacity=1 ][line width=0.08]  [draw opacity=0] (8.93,-4.29) -- (0,0) -- (8.93,4.29) -- cycle    ;
%Shape: Arc [id:dp8076722606148747] 
\draw  [draw opacity=0][dash pattern={on 4.5pt off 4.5pt}] (247.28,188.63) .. controls (249.46,192.56) and (251.02,201.52) .. (251.09,211.95) .. controls (251.15,220.39) and (250.22,227.88) .. (248.74,232.53) -- (244.98,212) -- cycle ; \draw  [color={rgb, 255:red, 208; green, 2; blue, 27 }  ,draw opacity=1 ][dash pattern={on 4.5pt off 4.5pt}] (247.28,188.63) .. controls (249.46,192.56) and (251.02,201.52) .. (251.09,211.95) .. controls (251.15,220.39) and (250.22,227.88) .. (248.74,232.53) ;  
%Curve Lines [id:da6025500388731361] 
\draw [color={rgb, 255:red, 208; green, 2; blue, 27 }  ,draw opacity=1 ]   (248.74,232.53) .. controls (247,232.44) and (237,218.44) .. (247.28,188.63) ;
\draw [shift={(242.89,217.45)}, rotate = 268.33] [fill={rgb, 255:red, 208; green, 2; blue, 27 }  ,fill opacity=1 ][line width=0.08]  [draw opacity=0] (8.93,-4.29) -- (0,0) -- (8.93,4.29) -- cycle    ;
%Shape: Arc [id:dp9471129061119897] 
\draw  [draw opacity=0] (424.27,158.39) .. controls (418.42,173.37) and (401.73,184.24) .. (381.97,184.38) .. controls (362.31,184.51) and (345.54,173.97) .. (339.42,159.2) -- (381.71,146.71) -- cycle ; \draw   (424.27,158.39) .. controls (418.42,173.37) and (401.73,184.24) .. (381.97,184.38) .. controls (362.31,184.51) and (345.54,173.97) .. (339.42,159.2) ;  
%Shape: Arc [id:dp8214057883130208] 
\draw  [draw opacity=0] (339.62,159.33) .. controls (346.1,146.53) and (363.38,137.32) .. (383.25,137.39) .. controls (402.47,137.45) and (418.47,146.18) .. (424.27,158.39) -- (381.75,169.51) -- cycle ; \draw   (339.62,159.33) .. controls (346.1,146.53) and (363.38,137.32) .. (383.25,137.39) .. controls (402.47,137.45) and (418.47,146.18) .. (424.27,158.39) ;  

%Shape: Arc [id:dp014587236953269023] 
\draw  [draw opacity=0] (372.59,197.6) .. controls (376.13,198) and (379.8,198.21) .. (383.54,198.21) .. controls (415.02,198.21) and (440.54,183.22) .. (440.54,164.71) .. controls (440.54,146.21) and (415.02,131.21) .. (383.54,131.21) .. controls (379.45,131.21) and (375.46,131.47) .. (371.62,131.95) -- (383.54,164.71) -- cycle ; \draw  [color={rgb, 255:red, 208; green, 2; blue, 27 }  ,draw opacity=1 ] (372.59,197.6) .. controls (376.13,198) and (379.8,198.21) .. (383.54,198.21) .. controls (415.02,198.21) and (440.54,183.22) .. (440.54,164.71) .. controls (440.54,146.21) and (415.02,131.21) .. (383.54,131.21) .. controls (379.45,131.21) and (375.46,131.47) .. (371.62,131.95) ;  
%Curve Lines [id:da46472880513102854] 
\draw [color={rgb, 255:red, 208; green, 2; blue, 27 }  ,draw opacity=1 ]   (369.56,197.29) .. controls (320.63,191.47) and (295.55,146.15) .. (371.62,131.95) ;
\draw [shift={(372.59,197.6)}, rotate = 184.78] [fill={rgb, 255:red, 208; green, 2; blue, 27 }  ,fill opacity=1 ][line width=0.08]  [draw opacity=0] (8.93,-4.29) -- (0,0) -- (8.93,4.29) -- cycle    ;
%Shape: Arc [id:dp0819086496345166] 
\draw  [draw opacity=0][dash pattern={on 4.5pt off 4.5pt}] (386.28,185.63) .. controls (388.46,189.56) and (390.02,198.52) .. (390.09,208.95) .. controls (390.15,217.39) and (389.22,224.88) .. (387.74,229.53) -- (383.98,209) -- cycle ; \draw  [color={rgb, 255:red, 208; green, 2; blue, 27 }  ,draw opacity=1 ][dash pattern={on 4.5pt off 4.5pt}] (386.28,185.63) .. controls (388.46,189.56) and (390.02,198.52) .. (390.09,208.95) .. controls (390.15,217.39) and (389.22,224.88) .. (387.74,229.53) ;  
%Curve Lines [id:da2649009368330174] 
\draw [color={rgb, 255:red, 208; green, 2; blue, 27 }  ,draw opacity=1 ]   (387.74,229.53) .. controls (386,229.44) and (376,215.44) .. (386.28,185.63) ;
\draw [shift={(382.13,203.03)}, rotate = 94.12] [fill={rgb, 255:red, 208; green, 2; blue, 27 }  ,fill opacity=1 ][line width=0.08]  [draw opacity=0] (8.93,-4.29) -- (0,0) -- (8.93,4.29) -- cycle    ;
%Shape: Arc [id:dp7640645477153458] 
\draw  [draw opacity=0][dash pattern={on 4.5pt off 4.5pt}] (339.76,158.12) .. controls (333.6,164.48) and (324.6,169.12) .. (314.5,170.4) .. controls (302.36,171.93) and (291.57,168.24) .. (285.27,161.39) -- (313.84,142.79) -- cycle ; \draw  [color={rgb, 255:red, 208; green, 2; blue, 27 }  ,draw opacity=1 ][dash pattern={on 4.5pt off 4.5pt}] (339.76,158.12) .. controls (333.6,164.48) and (324.6,169.12) .. (314.5,170.4) .. controls (302.36,171.93) and (291.57,168.24) .. (285.27,161.39) ;  
%Curve Lines [id:da02238355576274642] 
\draw    (285.27,161.39) .. controls (304,142.44) and (319,146.44) .. (339.76,158.12) ;
\draw [shift={(316.74,148.72)}, rotate = 184.23] [fill={rgb, 255:red, 0; green, 0; blue, 0 }  ][line width=0.08]  [draw opacity=0] (10.72,-5.15) -- (0,0) -- (10.72,5.15) -- (7.12,0) -- cycle    ;

\draw (372,104.4) node [anchor=north west][inner sep=0.75pt]    {$a_{2}$};
% Text Node
\draw (361,207.4) node [anchor=north west][inner sep=0.75pt]    {$b_{2}$};
% Text Node
\draw (233,102.4) node [anchor=north west][inner sep=0.75pt]    {$b_{1}$};
% Text Node
\draw (222,210.4) node [anchor=north west][inner sep=0.75pt]    {$a_{1}$};
% Text Node
\draw (291,121.4) node [anchor=north west][inner sep=0.75pt]    {$a_{1} +b_{2}$};

\end{tikzpicture}

\end{center}
\caption{The oriented curves on genus two Riemann surface.}
\label{twohomology}
\end{figure}

For genus two, the basis for homology group is $a_1,b_1, a_2, b_2$ and the intersection form is $(a_1,b_1)=1, (a_2,b_2)=1,~\delta_3=a_1+b_2$, see figure. \ref{twohomology}.
we have the matrix representation for those generators $\tau_i$:
\begin{align*}
& \tau_1=\left(\begin{array}{cccc} 1&-1&0&0 \\ 0&1&0&0 \\ 0&0&1&0 \\ 0&0&0&1 \end{array} \right),~\tau_2=\left(\begin{array}{cccc} 1&0&0&0\\1&1&0&0 \\ 0&0&1&0 \\ 0&0&0&1 \end{array} \right),\tau_3=\left(\begin{array}{cccc} 1&-1&1&0\\0&1&0&0 \\ 0&0&1&0 \\ 0&-1&1&1 \end{array} \right), \nonumber\\
&\tau_4=\left(\begin{array}{cccc} 1&0&0&0\\0&1&0&0 \\ 0&0&1&-1 \\ 0&0&0&1 \end{array} \right),~\tau_5=\left(\begin{array}{cccc} 1&0&0&0\\0&1&0&0 \\ 0&0&1&0 \\ 0&0&1&1 \end{array} \right).
\end{align*}
The representation matrix for some mapping class groups  are then:
\begin{align*}
& I=\left(\begin{array}{cccc} -1&0&0&0 \\ 0&-1&0&0 \\ 0&0&-1&0 \\ 0&0&0&-1 \end{array} \right),~\zeta=\left(\begin{array}{cccc} 0&-1&0&0\\1&0&0&-1 \\ 0&0&0&-1 \\ 0&-1&1&0 \end{array} \right),\phi_4=\left(\begin{array}{cccc} 0&-1&0&0\\1&1&0&0 \\ 0&0&0&1 \\ 0&0&-1&1 \end{array} \right), \nonumber\\
&\epsilon=\left(\begin{array}{cccc} -1&-1&-1&1\\1&0&1&-1 \\ 0&0&1&-1 \\ 0&-1&1&0 \end{array} \right),~\eta=\left(\begin{array}{cccc} 0&-1&0&0\\1&0&1&-1 \\ 0&0&1&-1 \\ 0&-1&1&0 \end{array} \right).
\end{align*}

Given an element $\alpha$ if mapping class group $M_2$,  the characteristic polynomial of its symplectic representation $Sp(\alpha)$ is 
\begin{equation*}
Det(y I_4-Sp(\alpha))=y^4+i_1(\alpha) y^3+i_2(\alpha) y^2+i_1(\alpha) y+1.
\end{equation*}

\subsection{Jones representation}
Jones \cite{jones1987hecke} gave another representation for $M_2$ which is given by  $M_{5\times 5}[q]$, namely the representation 
is $5\times 5$ matrix with the entries in polynomial of $q$, and the explicit form for the generators are shown here:
\begin{align*}
&\tau_1=\left(\begin{array}{ccccc} -1&0&0&0 &q\\ 0&-1&1&0&0 \\ 0&0&q&0&0 \\ 0&0&1&-1 &0 \\ 0&0&0&0&q \end{array} \right),~\tau_2=\left(\begin{array}{ccccc} -1&0&0&0 &q\\ 0&-1&1&0&0 \\ 0&0&q&0&0 \\ 0&0&1&-1 &0 \\ 0&0&0&0&q \end{array} \right),~ \tau_3=\left(\begin{array}{ccccc} -1&0&0&q&0\\ 0&-1&1&0&0 \\ 0&0&0&q&0 \\ 0&0&1&-1 &0 \\ 0&0&1&0&-1 \end{array} \right)\nonumber\\
& \tau_4=\left(\begin{array}{ccccc} q&0&0&0&0\\ 1&-1&0&0&0 \\ 0&0&-1&0&q \\ 1&0&0&-1 &0 \\ 0&0&0&0&q \end{array} \right), ~\tau_5=\left(\begin{array}{ccccc} -1&q&0&0&0\\ 0&q&0&0&0 \\ 0&0&q&0&0 \\ 0&0&1&-1 &0 \\ 0&0&1&0&-1 \end{array} \right).
\end{align*}
Unlike homology representation where different mapping class group would give the same monodromy group, Jones's presentation is more unique.

\subsection{Signature funcation}
\label{signature}
One can define a Meyer's function  \cite{endo2000meyer} for a mapping class group as follows. First, given two element $\alpha, \beta \in M_2$, we have two $4\times 4$ matrices $A=Sp(\alpha)$ and $B=Sp(\beta)$
by using the homology representation. Now define a real vector space by following equation
\begin{equation*}
V_{A,B}=\{(x,y) \in R^{4}\times R^4|(I_{4}-A^{-1})x+(I_{4}-B)y=0\};
\end{equation*}
The above equation defines a linear subspace inside $R^4\times R^4$, since one can solve some components of $x$ and $y$ coordinates in terms of other components linearly.
The dimension actually depends on the specific form of $A$ and $B$. Then define the following quadratic form on $V_{A,B}$ as follows
\begin{equation*}
\psi_{A,B}((x_1, y_1),(x_2, y_2))=(x_1+y_1)^tJ(I_4-B)y_2;
\end{equation*}
Here $J=\left(\begin{array}{cc} 0& I_2 \\ -I_2 & 0 \end{array}\right)$. One then define a cocycle as follows
\begin{equation}
\tau_2(\alpha, \beta):=sgn \psi_{A,B};
\label{sigequation}
\end{equation}
Here $sgn$ counts the difference of the positive and negative eigenvalues of the quadratic form $\psi_{A,B}$.

For a word, one can define a Meyer's function for it as follows: First for the generators: $\phi_2(\tau_i)=\frac{3}{5}$, and secondly for a word $(\tau_1\tau_2\ldots \tau_r)$, its Meyer's function is
\begin{equation*}
\phi_2(\tau_1\tau_2\ldots \tau_r)=\frac{3}{5}r-\sum_{j=1}^{r-1} \tau_2(\tau_1\ldots \tau_j,\tau_{j+1});
\end{equation*}
Here $\phi_2$ is the function defined in \ref{sigequation}.  In particular, the Meyer's function for $I_1$ and $\tilde{I}_1$ singularity are $\phi_2(I_1)=\frac{3}{5},~~\phi_2(\tilde{I}_1)=\frac{4}{5}$.
The Meyer's function satisfies the equation:
\begin{align*}
&\phi_2(1)=0, \nonumber\\
&\phi_2(\alpha^{-1})=-\phi_2(\alpha) ,\nonumber\\
&\phi_2(\beta \alpha \beta^{-1})=\phi_2(\alpha).
\end{align*}
In particular, the value of the signature depends only on the conjugacy class. The Meyer's function also satisfies the equation
\begin{equation*}
\phi_2(B)-\phi_2(AB)+\phi_2(A)=\tau_2(A,B).
\end{equation*}

One can then define a signature function for a singular fiber $F$ as follows:
\begin{equation*}
\sigma(F)=-\phi_2(F)+sgn(f^{-1}(D)),
\end{equation*}
and $D$ is the small disk around the singularity. On the other hand, the local signature can also be computed using the local invariant $d_x, \delta_x$ as follows:
\begin{equation}
\sigma(F)=\frac{2}{5} d_x -\delta_x=-\frac{3}{5}(2\delta_x-d_x)-\frac{1}{5}(d_x-\delta_x).
\label{signature1}
\end{equation}
So one can compute the signature for a local singularity  from local invariants. In particular,
The signature value for the $I_1$ and $\tilde{I}_1$ singularity is 
\begin{equation*}
\sigma(I_1)=-\frac{3}{5},~~\sigma(\tilde{I}_1)=-\frac{1}{5}.
\end{equation*}

The signature function is conserved under the deformation of a degeneration. 
 if a degeneration $F$  is split into several singularities $F_i$, then the  signatures of the corresponding mapping class group satisfies the important condition \cite{matsumoto2011pseudo}:
\begin{equation*}
\sigma(F)=\sum_i \sigma(F_i).
\end{equation*}
The above fact is  useful to find the number of terms in the factorization of the mapping class group element (see formula \ref{signature1}):
\begin{align}
&\# I_1= 2\delta_x-d_x, \nonumber\\
&\# \tilde{I}_1= d_x-\delta_x.
\end{align}
which is derived by the conservation of the signature function.

\subsection{Asymptotical free theory}
To find the configuration for asymptotical free theory, one might use the method similar to rank one theory: namely one start with 
the conformal theory and then move the bulk $I_1$ singularities to the fiber at $\infty$. 

\textbf{$SU(3)$ gauge theory}: 
The factorization for $SU(3)$ with $N_f=6$ fundamental flavor is
\begin{equation*}
(I_{bulk}, I_{\infty})=((\tau_5\tau_4\tau_3\tau_2\tau_1)^2, (\tau_1\tau_2\tau_3\tau_4\tau_5)^2).
\end{equation*}
To get the configuration for $SU(3)$ gauge theory with $N_f<6$, we'd like to move some letter from the bulk to $\infty$.
One of the constraint is that the characteristic polynomial of the corresponding monodromy group at $\infty$ is 
\begin{equation*}
y^4+2x^3+3x^2+2x+1=0.
\end{equation*}

\begin{figure}
\begin{center}

\tikzset{every picture/.style={line width=0.75pt}} %set default line width to 0.75pt        

\begin{tikzpicture}[x=0.45pt,y=0.45pt,yscale=-1,xscale=1]
%uncomment if require: \path (0,787); %set diagram left start at 0, and has height of 787

%Straight Lines [id:da4349425829610677] 
\draw    (475,135) -- (418,135) ;
\draw [shift={(415,135)}, rotate = 360] [fill={rgb, 255:red, 0; green, 0; blue, 0 }  ][line width=0.08]  [draw opacity=0] (8.93,-4.29) -- (0,0) -- (8.93,4.29) -- cycle    ;
%Straight Lines [id:da5444448760477274] 
\draw    (404.03,154) -- (404.5,205) ;
\draw [shift={(404,151)}, rotate = 89.47] [fill={rgb, 255:red, 0; green, 0; blue, 0 }  ][line width=0.08]  [draw opacity=0] (8.93,-4.29) -- (0,0) -- (8.93,4.29) -- cycle    ;
%Straight Lines [id:da3495292979146546] 
\draw    (474,144) -- (417,144) ;
\draw [shift={(414,144)}, rotate = 360] [fill={rgb, 255:red, 0; green, 0; blue, 0 }  ][line width=0.08]  [draw opacity=0] (8.93,-4.29) -- (0,0) -- (8.93,4.29) -- cycle    ;
%Straight Lines [id:da47131767373509625] 
\draw    (410.5,156) -- (479.13,209.16) ;
\draw [shift={(481.5,211)}, rotate = 217.76] [fill={rgb, 255:red, 0; green, 0; blue, 0 }  ][line width=0.08]  [draw opacity=0] (8.93,-4.29) -- (0,0) -- (8.93,4.29) -- cycle    ;
%Straight Lines [id:da9443198041697536] 
\draw    (417,150) -- (482.14,201.15) ;
\draw [shift={(484.5,203)}, rotate = 218.14] [fill={rgb, 255:red, 0; green, 0; blue, 0 }  ][line width=0.08]  [draw opacity=0] (8.93,-4.29) -- (0,0) -- (8.93,4.29) -- cycle    ;
%Straight Lines [id:da28404066757425384] 
\draw    (491,201) -- (491.47,154) ;
\draw [shift={(491.5,151)}, rotate = 90.57] [fill={rgb, 255:red, 0; green, 0; blue, 0 }  ][line width=0.08]  [draw opacity=0] (8.93,-4.29) -- (0,0) -- (8.93,4.29) -- cycle    ;
%Straight Lines [id:da9379267852357341] 
\draw    (475,213) -- (418,213) ;
\draw [shift={(415,213)}, rotate = 360] [fill={rgb, 255:red, 0; green, 0; blue, 0 }  ][line width=0.08]  [draw opacity=0] (8.93,-4.29) -- (0,0) -- (8.93,4.29) -- cycle    ;
%Straight Lines [id:da35155277821430575] 
\draw    (475,222) -- (418,222) ;
\draw [shift={(415,222)}, rotate = 360] [fill={rgb, 255:red, 0; green, 0; blue, 0 }  ][line width=0.08]  [draw opacity=0] (8.93,-4.29) -- (0,0) -- (8.93,4.29) -- cycle    ;

% Text Node
\draw (399,129.4) node [anchor=north west][inner sep=0.75pt]    {$a$};
% Text Node
\draw (482,208.4) node [anchor=north west][inner sep=0.75pt]    {$d$};
% Text Node
\draw (483,127.4) node [anchor=north west][inner sep=0.75pt]    {$c$};
% Text Node
\draw (399,209.4) node [anchor=north west][inner sep=0.75pt]    {$b$};

\end{tikzpicture}

\end{center}
\caption{The BPS quiver for the pure $SU(3)$ theory. The charges are $(-a, -b, c, d)$, see \ref{chargevec} for the charge vectors.}
\label{purebps}
\end{figure}
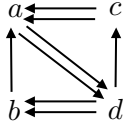

One possible choice is to move the letter $\tau_1, \tau_3, \tau_5$ for the conformal case to $\infty$ and the end result is
\begin{equation}
(\tau_{5(4)}\tau_{3(2)} \tau_{35^2(4)}\tau_{13^2(2)}, \tau_3\tau_5\tau_1\tau_3\tau_5\tau_1 (\tau_1\tau_2\tau_3\tau_4\tau_5)^2).
\label{su3}
\end{equation} 
We'd like to conjecture that the SW geometry for the SU(3) with $N_f=k$ fundamental flavor is just
\begin{equation*}
(\tau_{5(4)}\tau_{3(2)} \tau_{35^2(4)}\tau_{13^2(2)} I_k, I_{6-k} (\tau_1\tau_2\tau_3\tau_4\tau_5)^2).
\end{equation*} 
Here $I_k$ involves $k$ letter of $1,3,5$. Since $1,3,5$ commute with each other, 
the ordering in $I_k$ is not important. An interesting check is that when $k=0$, the BPS quiver defined from 
the BPS particle associated with the vanishing cycle is given in figure. \ref{purebps}, which is exactly the same as found earlier.
We have bulk word $\tau_a\tau_b\tau_c\tau_d$, and the symplectic pairing as 
\begin{equation*}
(a,b)=-1,~~(a,c)=2,~~(a,d)=-2,~~(b,c)=0,~~(b,d)=2,~~(c,d)=-1.
\end{equation*}
Here 
\begin{align}
&a=5(4)=[4]+[5], \nonumber\\
 &b=3(2)=[2]-[3]=[2]-[1]-[5],  \nonumber\\
&c=35^2(4)=[4]+[3]+2[5]=[4]+[1]+3[5],  \nonumber\\
&d=13^2(2)=[2]-[1]-2[3]=[2]-3[1]-2[5].
\label{chargevec}
\end{align}
We used the relation $[3]=[1]+[5]$. The basis for BPS quiver is $(-a, -b,c,d)$, which gives the BPS quiver.

\textbf{$SU(2)\times SU(2)$ gauge theory}:  The factorization for the conformal theory is 
\begin{equation}
(I_{bulk}, I_{\infty})=(\tau_1\tau_2\tau_3\tau_4\tau_5\tau_5\tau_4\tau_3\tau_2\tau_1, \tau_1\tau_2\tau_3\tau_4\tau_5\tau_5\tau_4\tau_3\tau_2\tau_1).
\end{equation}
This time one should move the letter $\tau_2, \tau_4$ of the bulk letter to $\infty$, and there are just four possibilities, which 
agree with the field theory expectation.
The above choice might be understood from the cut system for the corresponding mapping class group element.

\subsection{5d and 6d  KK theory}
Let's now discuss the global SW geometry for 5d $\mathcal{N}=1$ KK theory. The main  difficulty here is to determine the singular fiber at $\infty$. 
There are several clues about the monodromy group that would be useful:  firstly the eigenvalues of the monodromy group at $\infty$ 
for many 5d theory engineered using toric singularity are $(1,1,-1,-1)$ \footnote{More generally, one pair of eigenvalues can be one, and the other pair is $(\exp(2\pi i/3),\exp(4\pi/3))$.}; 
secondly the monodromy group is not periodic; thirdly the number of Dehn twist is determined by the local invariants $d_x, \delta_x$.

 We'd like to find 
the SW geometry for 5d theory whose flavor symmetry is $SO(20)$ \cite{Jefferson:2018irk,Apruzzi:2019opn}, and other cases should be found in a similar way. Our basic idea is following: one can find the UV singular fiber of the 5d theory by using the 
singular fiber of the  corresponding 4d theory with same flavor symmetry, namely one need to move a $I_1$ factor of the 4d UV fiber to the bulk.  
The reason is following: the dimension of the charge lattice of the 5d theory is one bigger than that of the 4d theory, as the BPS particle of the KK theory could carry the winding mode charge. 
Now the $I_\infty$ for the 4d theory with $SO(20)$ theory is $\tau_1^2\tau_2\tau_3\tau_4$, and by moving $\tau_4$ to the bulk, we get the word:
\begin{equation*}
\tau_1\tau_1\tau_2\tau_3,
\end{equation*}
which gives the required eigenvalues: $(1,1,-(-1)^{1/3},(-1)^{2/3})$, and certainly this is not a periodic map. The global SW geometry is then 
\begin{equation*}
(I_\infty, I_{bulk})=(\tau_1\tau_1\tau_2\tau_3,~\tau_4\tau_1^2\tau_2\tau_3\tau_4(\tau_5^2\tau_4\tau_3\tau_2)^2).
\end{equation*}

\begin{figure}
\begin{center}

\tikzset{every picture/.style={line width=0.75pt}} %set default line width to 0.75pt        

\begin{tikzpicture}[x=0.45pt,y=0.45pt,yscale=-1,xscale=1]
%uncomment if require: \path (0,964); %set diagram left start at 0, and has height of 964

%Curve Lines [id:da5751583450883184] 
\draw    (143,157) .. controls (176.5,78.22) and (332.5,84.22) .. (379.5,160.22) ;
%Shape: Ellipse [id:dp4351103044806457] 
\draw   (143,159.22) .. controls (143,154.8) and (153.19,151.22) .. (165.75,151.22) .. controls (178.31,151.22) and (188.5,154.8) .. (188.5,159.22) .. controls (188.5,163.64) and (178.31,167.22) .. (165.75,167.22) .. controls (153.19,167.22) and (143,163.64) .. (143,159.22) -- cycle ;
%Shape: Ellipse [id:dp9183978506888526] 
\draw   (334,161.22) .. controls (334,156.8) and (344.19,153.22) .. (356.75,153.22) .. controls (369.31,153.22) and (379.5,156.8) .. (379.5,161.22) .. controls (379.5,165.64) and (369.31,169.22) .. (356.75,169.22) .. controls (344.19,169.22) and (334,165.64) .. (334,161.22) -- cycle ;
%Shape: Ellipse [id:dp6691056137418792] 
\draw   (242,162.22) .. controls (242,157.8) and (252.19,154.22) .. (264.75,154.22) .. controls (277.31,154.22) and (287.5,157.8) .. (287.5,162.22) .. controls (287.5,166.64) and (277.31,170.22) .. (264.75,170.22) .. controls (252.19,170.22) and (242,166.64) .. (242,162.22) -- cycle ;
%Curve Lines [id:da4077649132341221] 
\draw    (188,161) .. controls (197.5,134.22) and (232.5,137.22) .. (241.5,161.22) ;
%Curve Lines [id:da29404391019586607] 
\draw    (288,162) .. controls (298.5,138.22) and (322.5,139.22) .. (332.5,161.22) ;
%Curve Lines [id:da9930942219213963] 
\draw    (144,207) .. controls (165.5,293.22) and (362.5,282.22) .. (386.5,209.22) ;
%Shape: Ellipse [id:dp5449821763218394] 
\draw   (145,205.22) .. controls (145,200.8) and (155.19,197.22) .. (167.75,197.22) .. controls (180.31,197.22) and (190.5,200.8) .. (190.5,205.22) .. controls (190.5,209.64) and (180.31,213.22) .. (167.75,213.22) .. controls (155.19,213.22) and (145,209.64) .. (145,205.22) -- cycle ;
%Shape: Ellipse [id:dp652452440607338] 
\draw   (243,206.22) .. controls (243,201.8) and (253.19,198.22) .. (265.75,198.22) .. controls (278.31,198.22) and (288.5,201.8) .. (288.5,206.22) .. controls (288.5,210.64) and (278.31,214.22) .. (265.75,214.22) .. controls (253.19,214.22) and (243,210.64) .. (243,206.22) -- cycle ;
%Shape: Ellipse [id:dp4855559280813462] 
\draw   (341,207.22) .. controls (341,202.8) and (351.19,199.22) .. (363.75,199.22) .. controls (376.31,199.22) and (386.5,202.8) .. (386.5,207.22) .. controls (386.5,211.64) and (376.31,215.22) .. (363.75,215.22) .. controls (351.19,215.22) and (341,211.64) .. (341,207.22) -- cycle ;
%Curve Lines [id:da3313515897116315] 
\draw    (191,207) .. controls (195.5,230.22) and (239.5,228.22) .. (242.5,206.22) ;
%Curve Lines [id:da13492656073440012] 
\draw    (290,208) .. controls (294.5,231.22) and (338.5,229.22) .. (341.5,207.22) ;

\end{tikzpicture}

\end{center}
\caption{The cutting system for the $I_{n-p-q}$ singularity.}
\label{6dkk}
\end{figure}
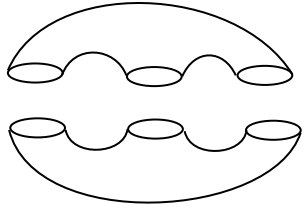

For the fiber at $\infty$ for 6d $(1,0)$ KK theory, one of the possible choice is the so-called $I_{n-p-q}$ singularity: the eigenvalues of the monodromy group is $(1,1,1,1)$. 
In the MM's theory, the cut curves are shown in figure. \ref{6dkk}. The periodic map on two genus zero component is just trivial. Since
the mapping group action associated with each cut curve is a Dehn twist, it is natural to identify the factorization as 
\begin{equation*}
\tau_1^n \tau_3^p \tau_5^q.
\end{equation*}
When $n=p=q=1$, this should give the $F_\infty$ so that the theory has $SO(20)$ flavor symmetry. The global SW geometry $I_{1-1-1}$ case can be found as follows:
\begin{align*}
&1=\tau_1\tau_2 \tau_3 \tau_4 \tau_5 \tau_5 \tau_4\tau_3\tau_2\tau_1 \tau_1\tau_2\tau_3\tau_4\tau_5^2\tau_4\tau_3\tau_2\tau_1= \nonumber\\
&\tau_{1(2)} \tau_1 \tau_3 \tau_4 \tau_5 \tau_5 \tau_4\tau_3\tau_2\tau_1\tau_1\tau_2\tau_3\tau_4\tau_5^2\tau_4\tau_3\tau_2\tau_1=\nonumber\\
&\tau_{1(2)} \tau_{3(4) }(\tau_1 \tau_3  \tau_5) \tau_5 \tau_4\tau_3\tau_2\tau_1\tau_1\tau_2\tau_3\tau_4\tau_5^2\tau_4\tau_3\tau_2\tau_1=\nonumber\\
& (\tau_1 \tau_3  \tau_5) \tau_5 \tau_4\tau_3\tau_2\tau_1(\tau_1\tau_2\tau_3\tau_4\tau_5^2\tau_4\tau_3\tau_2\tau_1)\tau_{1(2)} \tau_{3(4) }.
\end{align*}
In the first two steps, we used Hurwitz move to move the letter $1,3,5$ together, and finally we use cyclic equivalence. So finally, we have
\begin{equation*}
(I_\infty, I_{bulk})=(\tau_1 \tau_3  \tau_5,~\tau_5 \tau_4\tau_3\tau_2\tau_1(\tau_1\tau_2\tau_3\tau_4\tau_5^2\tau_4\tau_3\tau_2\tau_1)\tau_{1(2)} \tau_{3(4) }).
\end{equation*}

Alternatively, one can move the one letter of UV fiber of 5d KK theory with flavor symmetry $SO(20)$ to the bulk, 
as the charge lattice of 6d theory is one dimensional higher, since there is one more winding charge for 6d KK theory. To satisfy the eigenvalue condition, 
the choice is  $\tau_1^2 \tau_3$, which might be conjugate to the simple choice $\tau_1\tau_3\tau_5$.

\section{Conclusion}
We studies the global SW geometry for rank two theories with eight supercharges by using a topological approach. 
The main task is to find a factorization of a mapping class group element for the degeneration in terms of positive Dehn twist (up to conjugation, braid move, and Hurwitz move),
and a factorization of identify element. We achieved this for most local singularities, and find the global SW geometry for most 4d theories with generic deformations. 
The rank one case studied in \cite{Argyres:2015ffa} can now easily recovered using the topological approach, while the rank two case seems significantly more complicated, although 
it is still manageable. 

We take a topological approach in the classification, and one need to find the SW geometries to solve the Coulomb branch solution, i.e. the photon couplings.
The genus two fiberations are all hyperelliptic which makes the study easier and the study of those curves are under progress. The algebraic curves and 
the choice of SW differential can give us further constraints (for example, the theory with scaling dimension $(10,8)$ seem not exist although topologically we do not find any obstruction, and one may 
explain it using holomorphic constraint, i.e. no such hyperelliptic family can be written down \cite{Xie:2023zxn}).

The approach taken in this paper can be straightforwardly generalized to higher rank theory whose SW geometry can be given by
the fiberation of genus $g$ curves. The reason is that the mapping class group can also be generated by the Dehn twists. 
One can easily find the factorization for some familiar theories (such as $(A_1, A_n)$ or $(A_1, D_n)$ theories). A thorough study takes more efforts, and 
we hope to report the progress in the future.

One of the interesting result in this paper is that the mapping class group $M_2$ is generated by 
that of the simple $(A_1,A_n)$ and $(A_1, D_n)$ type SCFTs. This suggests that one may start with simple SCFTs and generate 
other SCFTs.

In this paper, we studied the theory with generic deformation (with only $I_1$ or $\tilde{I}_1$ singularities). One can find more general theories by allowing
so-called un-deformable singularities. We have already discussed how to form those un-deformable singularities by moving the letters of the Dehn twist around,
and a thorough understanding needs the study of the automorphism of the genus two fiberation, which will be left for part III of the series.

%\section*{Acklowledgement}
%DX would like to thank D.X Zhang for helpful discussions. 

\bibliographystyle{JHEP}
\bibliography{ADhigher}

\providecommand{\href}[2]{#2}\begingroup\raggedright\begin{thebibliography}{10}

\bibitem{Argyres:2015ffa}
P.~Argyres, M.~Lotito, Y.~L\"u, and M.~Martone, {\it {Geometric constraints on
  the space of $ \mathcal{N} $ = 2 SCFTs. Part I: physical constraints on
  relevant deformations}},  {\em JHEP} {\bf 02} (2018) 001,
  [\href{http://xxx.lanl.gov/abs/1505.0481}{{\tt arXiv:1505.0481}}].

\bibitem{Argyres:2022lah}
P.~C. Argyres and M.~Martone, {\it {The rank 2 classification problem I: scale
  invariant geometries}},  \href{http://xxx.lanl.gov/abs/2209.0924}{{\tt
  arXiv:2209.0924}}.

\bibitem{Argyres:2022puv}
P.~C. Argyres and M.~Martone, {\it {The rank 2 classification problem II:
  mapping scale-invariant solutions to SCFTs}},
  \href{http://xxx.lanl.gov/abs/2209.0991}{{\tt arXiv:2209.0991}}.

\bibitem{Argyres:2022fwy}
P.~C. Argyres and M.~Martone, {\it {The rank-2 classification problem III:
  curves with additional automorphisms}},
  \href{http://xxx.lanl.gov/abs/2209.1055}{{\tt arXiv:2209.1055}}.

\bibitem{Xie:2022aad}
D.~Xie, {\it {On rank two theories with eight supercharges part I: local
  singularities}},  \href{http://xxx.lanl.gov/abs/2212.0247}{{\tt
  arXiv:2212.0247}}.

\bibitem{matsumoto2011pseudo}
Y.~Matsumoto and J.~M. Montesinos-Amilibia, {\em Pseudo-periodic maps and
  degeneration of Riemann surfaces}, vol.~2030.
\newblock Springer Science \& Business Media, 2011.

\bibitem{Seiberg:1994rs}
N.~Seiberg and E.~Witten, {\it {Electric - magnetic duality, monopole
  condensation, and confinement in N=2 supersymmetric Yang-Mills theory}},
  {\em Nucl. Phys.} {\bf B426} (1994) 19--52,
  [\href{http://xxx.lanl.gov/abs/hep-th/9407087}{{\tt hep-th/9407087}}].
  [Erratum: Nucl. Phys.B430,485(1994)].

\bibitem{matsumoto1996lefschetz}
Y.~Matsumoto, {\it Lefschetz fibrations of genus two-a topological approach},
  in {\em Topology and Teichm{\"u}ller spaces}, pp.~123--148.
\newblock World Scientific, 1996.

\bibitem{farb2011primer}
B.~Farb and D.~Margalit, {\em A primer on mapping class groups (pms-49)},
  vol.~41.
\newblock Princeton university press, 2011.

\bibitem{Seiberg:1994aj}
N.~Seiberg and E.~Witten, {\it {Monopoles, duality and chiral symmetry breaking
  in N=2 supersymmetric QCD}},  {\em Nucl. Phys.} {\bf B431} (1994) 484--550,
  [\href{http://xxx.lanl.gov/abs/hep-th/9408099}{{\tt hep-th/9408099}}].

\bibitem{namikawa1973complete}
Y.~Namikawa and K.~Ueno, {\it The complete classification of fibres in pencils
  of curves of genus two},  {\em Manuscripta mathematica} {\bf 9} (1973), no.~2
  143--186.

\bibitem{Xie:2023lko}
D.~Xie, {\it {Pseudo-periodic map and classification of theories with eight
  supercharges}},  \href{http://xxx.lanl.gov/abs/2304.1366}{{\tt
  arXiv:2304.1366}}.

\bibitem{Xie:2012hs}
D.~Xie, {\it {General Argyres-Douglas Theory}},  {\em JHEP} {\bf 1301} (2013)
  100, [\href{http://xxx.lanl.gov/abs/1204.2270}{{\tt arXiv:1204.2270}}].

\bibitem{Hanany:1996ie}
A.~Hanany and E.~Witten, {\it {Type IIB superstrings, BPS monopoles, and
  three-dimensional gauge dynamics}},  {\em Nucl. Phys. B} {\bf 492} (1997)
  152--190, [\href{http://xxx.lanl.gov/abs/hep-th/9611230}{{\tt
  hep-th/9611230}}].

\bibitem{persson1990configurations}
U.~Persson, {\it Configurations of kodaira fibers on rational elliptic
  surfaces},  {\em Mathematische Zeitschrift} {\bf 205} (1990), no.~1 1--47.

\bibitem{Xie:2022lcm}
D.~Xie, {\it {Classification of rank one 5d $\mathcal{N}=1$ and 6d $(1,0)$
  SCFTs}},  \href{http://xxx.lanl.gov/abs/2210.1732}{{\tt arXiv:2210.1732}}.

\bibitem{hirose2010presentations}
S.~Hirose, {\it Presentations of periodic maps on oriented closed surfaces of
  genera up to 4},  {\em Osaka J. Math} {\bf 47} (2010) 385--421.

\bibitem{nakamura2018generation}
G.~Nakamura and T.~Nakanishi, {\it Generation of finite subgroups of the
  mapping class group of genus 2 surface by dehn twists},  {\em Journal of Pure
  and Applied Algebra} {\bf 222} (2018), no.~11 3585--3594.

\bibitem{dhanwani2023factoring}
N.~K. Dhanwani, A.~K. Nair, and K.~Rajeevsarathy, {\it Factoring periodic maps
  into dehn twists},  {\em Journal of Pure and Applied Algebra} {\bf 227}
  (2023), no.~1 107159.

\bibitem{sakalli2023singular}
S.~Sakall{\i}, V.~Horn-Morris, et~al., {\it Singular fibers in algebraic
  fibrations of genus 2 and their monodromy factorizations},  {\em arXiv
  preprint arXiv:2303.01554} (2023).

\bibitem{Xie:2023zxn}
D.~Xie and Z.~Yu, {\it {Hyperelliptic families and 4d $\mathcal{N}=2$ SCFT}},
  \href{http://xxx.lanl.gov/abs/2310.0279}{{\tt arXiv:2310.0279}}.

\bibitem{Xie:2023out}
D.~Xie, {\it {Discrete gauging and automorphism of Seiberg-Witten geometry, to
  appear}}, .

\bibitem{jones1987hecke}
V.~Jones, {\it Hecke algebra representations of braid groups and link
  polynomials},  {\em Ann. Math.} {\bf 126} (1987) 335--388.

\bibitem{endo2000meyer}
H.~Endo, {\it Meyer's signature cocycle and hyperelliptic fibrations},  {\em
  Mathematische Annalen} {\bf 316} (2000), no.~2 237--257.

\bibitem{Jefferson:2018irk}
P.~Jefferson, S.~Katz, H.-C. Kim, and C.~Vafa, {\it {On Geometric
  Classification of 5d SCFTs}},  {\em JHEP} {\bf 04} (2018) 103,
  [\href{http://xxx.lanl.gov/abs/1801.0403}{{\tt arXiv:1801.0403}}].

\bibitem{Apruzzi:2019opn}
F.~Apruzzi, C.~Lawrie, L.~Lin, S.~Sch\"afer-Nameki, and Y.-N. Wang, {\it
  {Fibers add Flavor, Part I: Classification of 5d SCFTs, Flavor Symmetries and
  BPS States}},  {\em JHEP} {\bf 11} (2019) 068,
  [\href{http://xxx.lanl.gov/abs/1907.0540}{{\tt arXiv:1907.0540}}].

\end{thebibliography}\endgroup

\end{document}